\documentclass[english,11pt,twoside,a4paper]{article}

\usepackage{multirow}
\usepackage[english]{babel}
\usepackage{fancyhdr}
\usepackage{amsmath}
\usepackage{amssymb}
\usepackage{amsfonts}
\usepackage{psfrag}
\usepackage[applemac]{inputenc}
\usepackage{graphicx,wrapfig}
\usepackage[bf,footnotesize]{caption2}
\usepackage{pstricks}
\usepackage{dsfont}
\usepackage{arydshln}
\usepackage{hhline}
\usepackage{makeidx}
\usepackage{slashed}
\usepackage{epsfig}

\newcommand{\newc}{\newcommand}

\newc{\gsim}{\lower.7ex\hbox{$\;\stackrel{\textstyle>}{\sim}\;$}}
\newc{\lsim}{\lower.7ex\hbox{$\;\stackrel{\textstyle<}{\sim}\;$}}
\newc{\sm}{{\scriptscriptstyle {SM}}}
\newc{\gsm}{g_{\scriptscriptstyle {SM}}}
\newc{\asm}{\alpha_{\scriptscriptstyle {SM}}}
\newc{\grho}{g_{\rho}}

\newcommand{\Tr}{\mathop{\rm Tr}} %inside math eq%
\newcommand{\Trs}{\mathop{\rm Tr^2}} %inside math eq%

\newcommand{\beq}{\begin{equation}}
\newcommand{\eeq}{\end{equation}}
\newcommand{\bea}{\begin{eqnarray}}
\newcommand{\eea}{\end{eqnarray}}
\newcommand{\bag}{\begin{align}}
\newcommand{\eag}{\end{align}}

\newcommand{\ie}{$\textnormal{i.e.}$ }

\newcommand{\GeV}{\,\mathrm{GeV}}
\newcommand{\TeV}{\,\mathrm{TeV}}

\newcommand{\eq}[1]{eq.~(\ref{#1})}

\newcommand{\nn}{\nonumber}

\newcommand{\SU}{\textrm{SU}}
\newcommand{\SO}{\textrm{SO}}
\newcommand{\U}{\textrm{U}}
\newcommand{\Sp}{\textrm{Sp}}

\newcommand{\cW}{c_W}

\newcommand{\ca}{c_1}
\newcommand{\cb}{c_2}
\newcommand{\cc}{c_3}
\newcommand{\cd}{c_4}
\newcommand{\ce}{c_5}

\newcommand{\uno}{
{\scriptscriptstyle{{\widehat {\mathbf1}}}}}
\newcommand{\due}{
{\scriptscriptstyle{{\widehat {\mathbf2}}}}}

\newcommand{\downangle}{\tilde\theta}

\usepackage{nicefrac}

\begin{document}

\baselineskip=18pt

\setcounter{footnote}{0}
\setcounter{figure}{0}
\setcounter{table}{0}

%%%%%%%%%%%%%%%%%%%%%%%%%%%%%%%%%%%%%%%%%%%

%\title{Title}
%\author{Javi Serra}
%\affiliation{IFAE, Universitat Aut\`onoma de Barcelona, 08193 Bellaterra, Barcelona}
%\begin{abstract}

%Abstract.

%\end{abstract}
%\maketitle

%FRONTPAGE2%%%%%%
\begin{titlepage}

\begin{flushright}
CERN-PH-TH/2011-122 \\
%May 2011
\end{flushright}
\vspace{.2in}

\begin{center}
%\vspace{5cm}
%{
%\large Version 11.0\mbox{} -- \today\mbox{} -- \currenttime
%}\\
\vspace{1cm}

{\bf\Large   The Other Natural Two Higgs Doublet Model
}

\vspace{1.2cm}

{\bf J. Mrazek$^{a}$, A. Pomarol$^{b}$,  R. Rattazzi$^{a}$, M. Redi$^{c,d}$,}
{\bf J. Serra$^{b}$, A. Wulzer$^{e}$}

\vspace{.5cm}

\centerline{$^{a}${\it Institut de Th\'eorie des Ph\'enom\`enes Physiques, EPFL,  CH--1015 Lausanne, Switzerland}}
\centerline{$^{b}${\it Departament de F\'isica, Universitat Aut\`onoma de Barcelona, 08193 Bellaterra (Barcelona), Spain}}
\centerline{$^{c}${\it CERN, Theory Division, CH--1211 Geneva 23, Switzerland}}
\centerline{$^{d}${\it INFN, 50019 Sesto F., Firenze, Italy}}
\centerline{$^{e}${\it ETH Zurich, 8093 Zurich, Switzerland}}

\end{center}
\vspace{.8cm}

\begin{abstract}
\medskip
\noindent

We characterize models where electroweak symmetry breaking is driven by two light Higgs doublets arising as pseudo-Nambu-Goldstone bosons of new dynamics above the weak scale.
They  represent  the simplest  natural  two Higgs doublet  alternative to supersymmetry.
We construct their low-energy effective Lagrangian
making only  few specific assumptions about the strong sector. These concern their global symmetries, their patterns of spontaneous breaking and the sources of explicit breaking. In particular we assume that all the explicit breaking is associated with the couplings 
of the strong sector to the Standard Model fields, that is  gauge and (proto)-Yukawa interactions. 
Under those assumptions the scalar potential is determined at lowest order by very few free parameters associated to the top sector.
Another crucial property of our scenarios is the presence of a discrete symmetry, in addition to  custodial SO(4), that controls the $T$-parameter.  That can either be simple $CP$ or a $Z_2$ that distinguishes the two Higgs doublets.  Among various possibilities we study in detail models based on  SO(6)/SO(4)$\times$ SO(2), focussing on their predictions for the structure of the scalar spectrum and   the deviations of their  couplings from those
of a generic renormalizable two Higgs doublet model.

\end{abstract}

\bigskip

\end{titlepage}

%%%%%%%%%%%%%%%%%%%%%%%%%%%%%%%%%%%%%%%%

\tableofcontents

%%%%%%%%%%%%%%%%%%%%%%%%%%%%%%%%%%%%%%%%

\section{Introduction}

Uncovering  the Higgs sector at the LHC is going to be a difficult but
crucial   task needed  to  understand    the  breaking of the electroweak symmetry.
In the SM the Higgs sector  consists of only one scalar $\SU(2)_L$-doublet, but
models with a much more opulent  Higgs structure
have been extensively considered in the literature.
Among them,  two Higgs doublet models  (2HDM) have attracted a lot of attention
due to their rich phenomenology in electroweak  and flavor physics.

The Higgs sector, however, consisting  of  scalars, is   very sensitive  to UV physics,
giving rise to the well-known  hierarchy problem.
It is  expected then that  the   new physics needed to solve this  problem 
significantly affects  the Higgs sector and its properties.
This is exactly what happens in
the most popular solution to the hierarchy problem, supersymmetry.
There one learns that indeed  the Higgs sector of
the  supersymmetric Standard Model   is in fact quite restrictive,
requiring two Higgs doublets with
Yukawa and  potential terms  taking a very  specific form.
This shows that   different Higgs  scenarios, such as 2HDM,
must be  analyzed  within  frameworks
that address  at the same time the hierarchy problem.

The other natural  alternative to supersymmetry, that also addresses the hierarchy problem, is
to consider the Higgs bosons  as composite states  arising from a strong sector.
The Higgses can be lighter than the strong scale,
as  favored by electroweak precision tests,
if  they are    pseudo-Nambu-Goldstone bosons (PNGB)
of an  approximate  symmetry   $G$ spontaneously broken to $H$ \cite{Kaplan:1983fs,Agashe:2004rs}.
One very  interesting aspect of these scenarios is that the low-energy dynamics is to a large extent determined by symmetry.
For instance, the spectrum of light scalars is fixed by the coset $G/H$.
Furthermore, the effective Lagrangian is constrained by  the  $G/H$ construction \cite{CCWZ}
and by  the structure of  the $G$-breaking couplings, that is by selection rules.
In particular, the Higgs potential is fully   determined up to a few $O(1)$ coefficients by the couplings between the strong sector and the Standard Model (SM). That provides
information on
the main phenomenological properties  of these models
without the  necessity  of
a detailed knowledge of the strong sector.

In this paper we will explore  composite PNGB Higgs models involving two Higgs doublets.
Our interest is to show how these natural scenarios restrict  generic 2HDM.
The presence of two Higgs doublets, rather than just one,
rises  two main phenomenological challenges.
The first concerns the breaking of the  approximate
custodial $\SO(3)_c$  symmetry
by the vacuum  structure of the model. That  can  lead to
large  contributions to the $T$-parameter
even when the custodial symmetry is  preserved by the strong sector.
The second  concerns Higgs-mediated   Flavor Changing Neutral Currents (FCNC),
which is a well known potential problem of theories with extended Higgs sectors.
We will explain how these two problems can be overcome   in a natural way  by the use of discrete symmetries. These discrete symmetries   restrict the  form of the Higgs potential and of the Yukawa couplings, thus
leading to interesting predictions. We  are lead to considering  two classes of models.
In the first class,  like in the inert Higgs model \cite{Barbieri:2006dq},   the extra Higgs doublet  will be odd under
a certain parity, $C_2$. In the limit of exact $C_2$ the second Higgs does not couple   linearly to  the SM fields.
In the second class, an approximate $CP$ symmetry
will control the  $T$-parameter and  FCNC.
The simplest models  we could contruct in this second class, however, feature, somewhat unexpectedly,   an accidental approximate  $C_2$ parity, thus giving rise to  ``almost inert" Higgs scenarios.

The structure of the paper is as follows.
In section~\ref{TCHDM} we describe the general structure of composite 2HDM,
pointing out how discrete symmetries can help to avoid   constraints from the
$T$-parameter  and  FCNC.
In section~\ref{secExplicitModels} we  present  explicit composite 2HDM models.
We mainly concentrate in 2HDM arising from the
$\SO(6)/\SO(4)\times \SO(2)$ coset, although we will also briefly present models
with extended custodial symmetry such as those based on the ${\Sp(6)}/{\SU(2)\times \Sp(4)}$
coset.
In section~\ref{pheno} we  give the main phenomenological implications of composite 2HDM. We
focus first on model-independent features  and later concentrate on the phenomenology of two particular examples, the composite inert Higgs
and the almost composite inert Higgs.
The last section is devoted  to conclusions.

\section{Two Composite Higgs Doublets as PNGBs}
\label{TCHDM}

\subsection{General Structure}\label{genS}

\begin{figure}
\centering
\epsfig{file=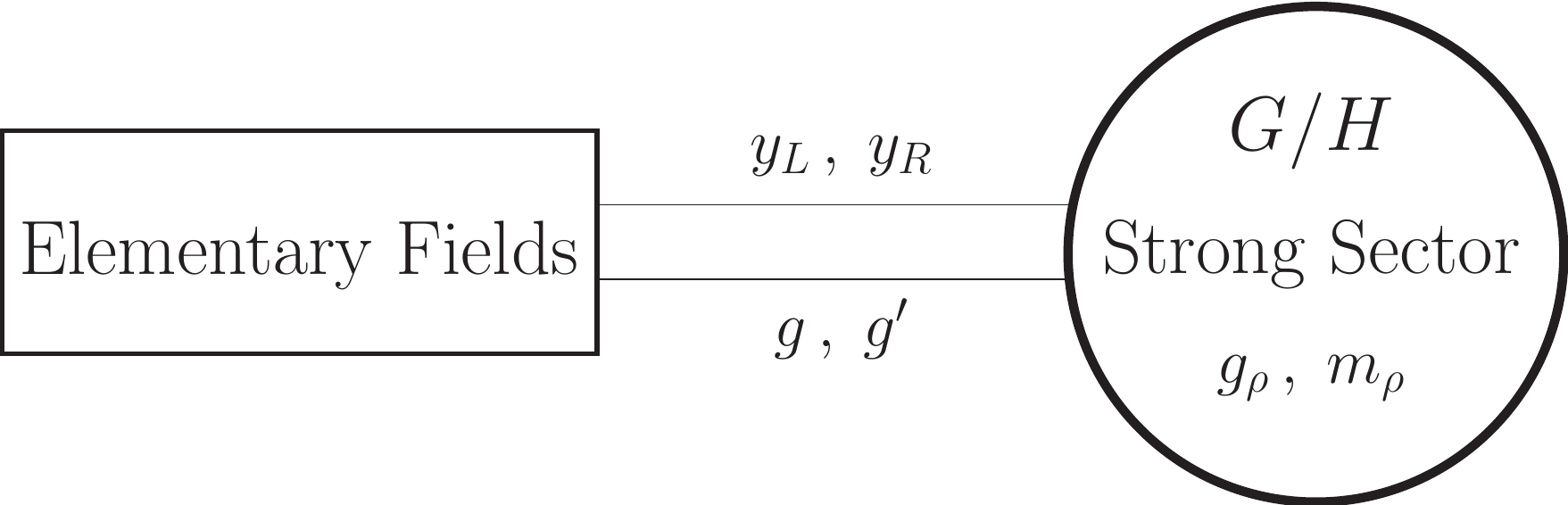,height=100pt}
\caption{Pictorial representation of our scenario.}\label{figSS}
\end{figure}
The basic structure of our composite-Higgs scenario is as follows. As depicted in 
figure~\ref{figSS}, there exists a new sector, that we denote as ``strong'', 
or ``strongly-interacting'' sector, which is endowed with a global 
group $G$ of symmetry, spontaneously broken to $H\subset G$. 
As such, the strong sector delivers a set of massless Nambu-Goldstone 
bosons (NGB). 
%As discussed in the previous  section, 
The only constraints on the choice of the $G/H$ coset 
that characterizes the strong sector 
are of phenomenological nature and they are rather mild, a 
priori. 
%For the model to have a potentially interesting phenomenology,  
The main requirement, needed to  avoid generic large contributions to the $T$-parameter,
is that the unbroken group must contain a  ``custodial'' $\SO(4)\cong \SU(2)\times\SU(2)$ symmetry, 
$H\supset \SO(4)$, and at least one Higgs $4$-plet 
({\it{i.e.}}, a $\mathbf{4}$ of $\SO(4)$) must be present. 
Compatibly with these basic requirements,  
several cosets exist. The smallest ones, chosen so that $H$ is a maximal 
subgroup of $G$, are present in table~\ref{cosets}.
\begin{table}[h!]
	\begin{center}
\begin{tabular}{cccccc}
\hline
$G$ & $H$ & $N_G$ & NGBs $\textrm{rep.}[H] = \textrm{rep.}[\textrm{SU}(2) \times \textrm{SU}(2)]$ \\

SO(5) & SO(4) & 4 & $\mathbf{4} = (\mathbf{2},\mathbf{2})$ \\

SO(6) & SO(5) & 5 & $\mathbf{5} = (\mathbf{1},\mathbf{1}) + (\mathbf{2},\mathbf{2})$ \\

SO(6) & SO(4) $\times$ SO(2) & 8 & $\mathbf{4_{+2}} + \mathbf{\bar{4}_{-2}} = 2 \times (\mathbf{2},\mathbf{2})$ \\

%SO(6) & SO(4) & 9 & $\mathbf{1} + \mathbf{4} + \mathbf{\bar{4}} = (\mathbf{1},\mathbf{1})+2 \times (\mathbf{2},\mathbf{2})$ \\

SO(7) & SO(6) & 6 & $\mathbf{6} = 2 \times (\mathbf{1},\mathbf{1}) + (\mathbf{2},\mathbf{2})$ \\

SO(7) & $\textrm{G}_2$ & 7 & $\mathbf{7} = (\mathbf{1},\mathbf{3})+(\mathbf{2},\mathbf{2})$ \\

SO(7) & SO(5) $\times$ SO(2) & 10 & $\mathbf{10_0} = (\mathbf{3},\mathbf{1})+(\mathbf{1},\mathbf{3})+(\mathbf{2},\mathbf{2})$ \\

SO(7) & $[\textrm{SO}(3)]^3$ & 12 & $(\mathbf{2},\mathbf{2},\mathbf{3}) = 3 \times (\mathbf{2},\mathbf{2})$ \\

Sp(6) & Sp(4) $\times$ SU(2) & 8 & $(\mathbf{4},\mathbf{2}) = 2 \times (\mathbf{2},\mathbf{2}), (\mathbf{2},\mathbf{2}) + 2 \times (\mathbf{2},\mathbf{1})$ \\

SU(5) & SU(4) $\times$ U(1) & 8 & $\mathbf{4}_{-5} + \mathbf{\bar{4}_{+5}} = 2 \times (\mathbf{2},\mathbf{2})$ \\

SU(5) & SO(5) & 14 & $\mathbf{14} = (\mathbf{3},\mathbf{3}) + (\mathbf{2},\mathbf{2}) + (\mathbf{1},\mathbf{1})$ \\

%SU(5) & SU(3) $\times$ SU(2) $\times$ U(1) & 12 & $(\mathbf{3},\mathbf{2})_{\mathbf{-5}} + (\mathbf{\bar{3}},\mathbf{2})_{\mathbf{+5}} = 2 \times (\mathbf{2},\mathbf{2}) + 2 \times (\mathbf{2},\mathbf{1})$ & $\mathbf{24}$, $\mathbf{?}$ \\

\hline
\end{tabular} \\
	\caption{Cosets $G/H$ from simple Lie groups, 
	with $H$ maximal subgroup of $G$. For each coset, 
	its dimension $N_G$ and the NGBs representation under $H$ and 
	$\SO(4)\simeq\SU(2)_L \times \SU(2)_R$ are reported. For $\Sp(6)/\SU(2)\times\Sp(4)$, 
	two embeddings are possible, we will be interested only in the first one, which leads 
	to two Higgs $4$-plets.}
	\label{cosets}
	\end{center}
\end{table} 
%\begin{table}[h!]
%	\begin{center}
%	\footnotesize{\input{THDcosets.tex}}
%	\caption{Set of cosets $G/H$ containing two complex Higgs doublets, and their transformation properties under $H$.}
%	\label{THDcosets}
%	\end{center}
%\end{table}
Other cosets, with non-maximal subgroups, can be obtained 
from table~\ref{cosets} in a stepwise fashion $G \rightarrow 
H \rightarrow H' $ etc.. 
The coset $\SO(6)/\SO(4)$, for instance, arises from the breaking 
$\SO(6) \rightarrow \SO(5) \rightarrow \SO(4)$. Besides two 
$(\textbf{2},\textbf{2})$ Higgs $4$-plets, this coset contains an extra scalar singlet 
$(\textbf{1},\textbf{1})$. The cosets that only contain two Higgs doublets, and 
therefore give rise to a composite Two Higgs Doublet Model (2HDM), 
are $\SO(6)/\SO(4)\times \SO(2)$, $\Sp(6)/\SU(2)\times \Sp(4)$, and $\SU(5)/\SU(4) \times \U(1)$. 
In the following, when discussing explicit realizations of the composite 2HDM scenario, 
we will mainly consider the $\SO(6)/\SO(4)\times \SO(2)$ coset, but the 
$\Sp(6)/\SU(2)\times \Sp(4)$ one will also find an interesting application, in 
section~\ref{extcust}, as an example of models with an extended custodial symmetry group.

Apart from the choice of the $G/H$ symmetry breaking pattern, very mild assumptions 
will be made on the nature of the strong sector and on its microscopic origin. 
In the spirit of \cite{Giudice:2007fh}, 
we assume its dynamics to be controlled by the smallest possible set of 
parameters: a coupling $g_\rho\leq4\pi$ 
that controls the interactions of the strong sector's resonances  
and the typical size $m_\rho$ of their masses. One possible implementation of 
this scenario could be provided by strongly-interacting confining ``QCD-like'' 
gauge theories in the large-$N$
expansion. At large-$N$, the size of all the couplings among mesonic resonances 
is fixed by 
\beq
g_\rho\simeq \frac{4\pi}{\sqrt{N}}\, ,
\label{grhon}
\eeq
while the mass $m_\rho\sim \Lambda_S$ is 
set by the confinement scale and does not depend on $N$
\footnote{
Notice that the ``universality'' of the coupling only holds in the mesonic 
sector, while resonances of different nature can interact with parametrically different 
couplings. For instance, for the glueballs in QCD, $g_G\simeq4\pi/N$. Thus if we needed to account for all classes of resonances we would not be able to  depict the strong sector just in terms of  
a single coupling $g_\rho$. We shall assume that only mesons matter and work with a single coupling. Based on 5D examples, that is not an unreasonable assumption. Moreover phenomenological constraint prefer a large $g_\rho$, in which case all distinctions disappear.}.
Other realizations of our strong sector, which are definitely easier to construct 
and to deal with, are the holographic five-dimensional models, 
discussed at length in the literature 
for the case of the ``minimal'' $\SO(5)/\SO(4)$ coset \cite{Agashe:2004rs,Contino:2006qr}.

At energies below the resonance scale 
$m_\rho$, independently of their microscopic origin, the NGB 
composite Higgses are described by the non-linear $\sigma$-model associated 
to the $G/H$ coset. At the leading two-derivatives order, the sigma-model interactions 
are dictated by the dimension-full coupling $1/f$ which, given our assumptions on the 
strong sector, has to be identified with $g_\rho/m_\rho$, leading to the relation  
$m_\rho\simeq g_\rho f$. Notice that it is only if the NGB form an 
\emph{irreducible} representation of $H$ that their two-derivative interactions 
are completely fixed, and therefore predicted, in terms of a unique parameter $f$. 
This is the case for all the cosets in table~\ref{cosets}, while for instance in 
$\SO(6)/\SO(4)$ the most general two derivative Lagrangian is described by four parameters associated to the four 
quadratic invariants which can be built out of two $4$-plets and one singlet \footnote{By performing field redefinitions one can however show that only three parameters are physically independent.}. 

There are strong phenomenological hints, some of which will be summarized in the 
following, that the observed quarks and leptons (with the possible remarkable exception 
of the right-handed top quark $t_R$) and the transverse polarizations of the EW gauge 
bosons are \emph{not} composite objects of some strongly-interacting dynamics, or 
at least that they are not \emph{entirely} composite. 
We therefore need to introduce these particles as ``elementary fields'', external to 
the strong sector, and make them communicate with the latter by a set of couplings, to be defined later,  $g,\,g',\,y_L,\,y_R$, as shown in figure~\ref{figSS}. We will generically denote these 
``elementary'' 
couplings as $\gsm$. For what concerns the SM $\SU(2)_L\times \U(1)_Y$ gauge fields, there 
is no ambiguity on how they should be coupled to the strong sector. The SM gauge group 
is identified as the appropriate subgroup of the global $\SO(4)\subset H$, and it is gauged 
with couplings $g$ and $g'$. The standard gauging basically consists in writing down a 
\emph{linear} coupling of the elementary gauge fields with the corresponding global 
currents of the strong sector. 

The fermions also need to be coupled to the strong sector, with the aim of generating 
their masses, and this could be achieved in two ways. We could 
write \emph{bilinear} terms, involving one left- and one 
right-handed fermion coupled to a bosonic strong sector operator with the 
quantum numbers of the Higgs. This is of course the standard mechanism for 
fermions mass generation in technicolor-like theories. Or, copying from what 
we just saw to happen for the gauge fields, we may adopt the ``partial compositeness'' 
paradigm \cite{Kaplan:1991dc,Agashe:2004rs} and introduce  
\emph{linear} terms, separately for the left- and right-handed components, 
which involve fermionic strong sector operators. In the present paper we will 
consider this second possibility, with $y_L$ and $y_R$ being the left- and right-handed 
fermion linear couplings, which we will denote as ``proto-Yukawa'' couplings.
Schematically, the couplings of the elementary fields to the strong sector 
can be written as
\beq
{\cal L}_{\textrm{mix}}= \gsm\cdot\Psi_\sm \cdot {\cal O}\,,
\label{mix0}
\eeq
where $\Psi_\sm=(A_\mu, f)$ collectively denotes the SM gauge fields and fermions. 
Notice that, since the elementary states do not fill complete representation of $G$, 
${\cal L}_{\textrm{mix}}$ unavoidably breaks the strong sector's global group. The 
Higgs therefore becomes a PNGB and is free to acquire a potential, 
as we will discuss below.

Because of these linear couplings, the SM fields have a  degree of mixing
\beq
\epsilon_{g}\equiv\frac{g}{g_\rho}\ ,	\quad
\epsilon_{L,R}\equiv\frac{y_{L,R}}{g_\rho}\, ,
\eeq
with the strong sector's resonances. It is only 
when this mixing is not too large that the previously-mentioned 
phenomenological bounds can be accommodated and the model made realistic \cite{Giudice:2007fh,Contino:2006nn}.
This suggests that the coupling $g_\rho$ is better taken to be large, at least 
larger than the elementary couplings $\gsm$ \footnote{As a matter of fact  $g_\rho<g_{SM}$ would not even be a radiatively stable choice.}. As in \cite{Giudice:2007fh}, we then 
restrict our parameter space to the region
\beq
\gsm\,\leq \, g_\rho\,\leq\,4\pi\,,
\label{gsm}
\eeq
where the limit of \emph{total} compositeness $\gsm\simeq g_\rho$ could  be considered 
for the $t_R$ ($y_R\simeq g_\rho$), given that phenomenological constraints on the $t_R$ 
compositeness are practically absent. Instead of taking $y_R\simeq g_\rho$, a more 
direct way to achieve total $t_R$ compositeness is not to introduce the elementary 
$t_R$ field to start with, and assume that a massless resonance with the quantum 
numbers of the $t_R$ emerges from the strong sector. 

%Due to the mixing \eq{mix0},  the operators of the strong sector
%must  transform under the full SM group.
%This implies that  the global symmetry of the strong sector, apart from the groups
%of table~\ref{cosets}, 
%must also  contain a $\SU(3)_c\times \U(1)_X$ symmetry where
%we define $Y=T_R^3+X$.

Due to the couplings in eq.~(\ref{mix0}) to the SM fermions, and in particular to 
the quarks, the strong sector must be charged under the full SM group, including the 
color $\SU(3)_c$. On top of the $G/H$ cosets discussed until now, and listed in table~\ref{cosets}, 
the strong sector must therefore also enjoy an unbroken $\SU(3)_c$ global group, 
weakly gauged with coupling $g_{\textrm {strong}}$ by elementary gluon fields. This gluon gauge 
coupling should also appear in eq.~(\ref{mix0}), but it will  be ignored since it does not play any role 
in what follows. Another unbroken symmetry of the strong sector that we have not mentioned 
  is the strong sector matter charge ${\textrm{U}}(1)_X$, which is needed to assign the correct hypercharge 
to the fermionic operators. The hypercharge is identified as $Y=T^3_R+X$, in terms of the third 
$\SU(2)_R$ generator $T^3_R$.

\subsubsection*{The Structure of the Potential}

Let us briefly recall, for future use, the general structure of the effective potential of our 
PNGB Higgs.

In general, given a strong sector, one could imagine breaking its global symmetry $G$ 
either by adding new weak interactions among the composites or by their direct (weak) 
coupling to 
external elementary fields. For instance, in QCD the chiral symmetry is broken both by fermion
masses, belonging to the first class of couplings, and by the coupling of quarks to the photon,
which belongs to the second class. In our composite Higgs scenario, as described above, 
the second class of effects is always unavoidably present,  
while the first is not. It is thus not unreasonable, and also motivated by simplicity, 
to assume all the breaking of $G$ is due to the coupling to the SM fields
in eq.~(\ref{mix0}). 
We will work under this assumption, bearing however in mind that by relaxing the latter 
 the parameter 
space of PNGB Higgs models could be significantly enlarged. 

\begin{figure}
\centering
\epsfig{file=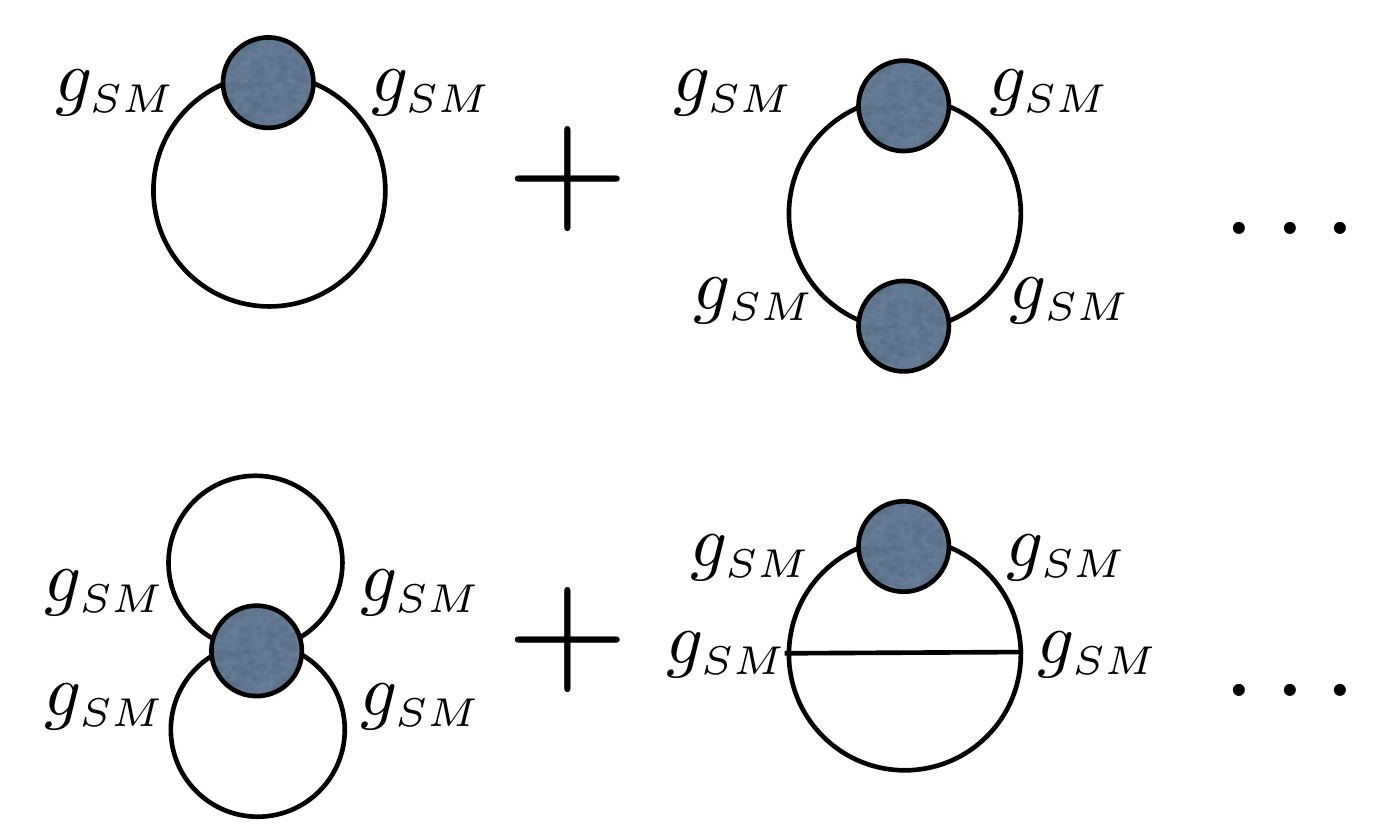,height=150pt}
\caption{Power counting for the Higgs potential.}\label{potfig}
\end{figure}

Thanks to the above assumption, the potential only originates from insertions of the $\gsm$ 
couplings of eq.~(\ref{mix0}), and much can be said on its structure. First of all, 
its size can be estimated, as  figure~\ref{potfig} shows, 
in an expansion in 
loops  and in powers of the degree of the mixing $\epsilon=\gsm/\grho$. By 
noticing that each strong sector's 
$\langle {\mathcal O }\ldots {\mathcal O } \rangle$ correlator 
(represented as a circle in   figure~\ref{potfig}) 
is proportional to $1/g_\rho^2\propto N$, the estimate reads
\beq
V(\Pi)=\frac{m_\rho^4}{16\pi^2}\left (\sum_{n=1}^{\infty}%\frac{\gsm^{2n}}{\grho^{2n}}
\epsilon^{2n}F_{1\,n}(\Pi/f)\,+\,\frac{\grho^2}{16\pi^2}\sum_{n=1}^{\infty}%\frac{\gsm^{2n}}{\grho^{2n}}
\epsilon^{2n}F_{2\,n}(\Pi/f)\,+\,{\rm higher \,\, loops}\right)\,.
\label{estimate}
\eeq
The Higgs bosons $\Pi$, because of their NGB nature, only appear in combination with the 
decay constant $f$, through the dimensionless functions $F_{i\,n}$. Second, but this will not 
be discussed in detail until section~\ref{642}, the $G$ symmetry strongly constrains the 
possible contributions to the potential that arise at each given order in $\gsm$. 
This can be analyzed  simply using spurion's 
power counting performed by assigning $G$ transformation properties to $\gsm$.
% the functions 
%$F_{i\,n}$ can be determined and the potential, thanks to our assumption on the $G$-breaking 
%sources, is almost completely  under control.

The generic properties of the EW vacuum and of the scalar spectrum are readily derived from the above equation.
In the absence of tuning the generic minimum of the potential will be at $v=\langle \Pi \rangle \sim f$, 
and similarly the masses of the scalars scale as
\beq
m_\Pi^2\sim \frac{\grho^2}{16\pi^2}\gsm^2 f^2\left (1+O(\epsilon^2)+\dots\right )\,.
\label{heavyscalars}
\eeq
As quantitatively discussed in section~\ref{situation}, however, a certain amount of 
tuning in $v/f$ seems unavoidable for a realistic model. 
In order to perform such a tuning, one of the mass terms in the potential must be unnaturally 
reduced, while the quartic Higgs couplings must remain unaffected. This makes that the estimate 
of eq.~(\ref{heavyscalars}) is typically violated, and along the ``tuned  direction'' of the potential 
a lighter scalar $h$ emerges. Its  mass is given by
\beq
m_h^2\sim \frac{\grho^2}{16\pi^2}\gsm^2 v^2\, ,\label{higgsmass}
\eeq
that is parametrically smaller than  \eq{heavyscalars}.
Up to effects $v^2/f^2$, the scalar $h$  behaves as the  SM Higgs.

% Other scalars, charged or neutral, could obviously be made 
%light with extra tuning.

In the realistic cases the dominant source of the potential is given by the proto-Yukawas of the top quark, $y_L$ and $y_R$. These latter are indeed forced to be rather large because they have to reproduce 
the top Yukawa coupling $Y_t\simeq1$, which is given by the relation
\beq
Y_t\simeq\frac{y_Ly_R}{\grho}\,.\label{ytop}
\eeq
Because of eq.~(\ref{gsm}), one can deduce the lower bound
\beq
\min (y_L,y_R)\gsim Y_t \qquad\Rightarrow\qquad m_h^2\gsim N_c\, \frac{g_\rho^2\, Y_t^2}{16\pi^2} v^2\,,
\eeq
where an $N_c=3$ factor representing the number of QCD colors has been added to the estimate of 
eq.~(\ref{higgsmass}). Notice that the lower bound above is only reached in the limit of total 
$t_R$ compositeness, $y_L\simeq Y_t$ and $y_R\simeq g_\rho$, but in other situations
$h$ will be    heavier.
%while the lightest Higgs 
%particle could be significantly heavier in other situations, according to eq.~(\ref{higgsmass}). 
In realistic concrete cases (see sections~\ref{secC2model} and \ref{secC1Pmodel}) the 
estimate of eq.~(\ref{higgsmass}) might however be violated by an extra accidental cancellation 
of the quartic coupling, and the Higgs could remain light. This notably occurs in the 
minimal $\SO(5)/\SO(4)$ composite Higgs model (MCHM) \cite{Agashe:2004rs}.

%Moreover, in the specific models, the point $y_L\sim y_R\sim \sqrt {y_t\grho}$ is selected either as an upper bound motivated by EWPT or as the only point where $v$ can be tuned $\ll f$. In any case that implies the upper bound
%\beq 
%m_h^2\lsim  N_c\, \frac{g_\rho^3\,\lambda_t }{16\pi^2} v^2\, .
%\eeq
%With a coupling $g_\rho\sim 6$ (similar to the coupling of the $\rho$ in QCD) we obtain a Higgs mass around $300~\rm{GeV}$.
%As  we will argue in the next section, much smaller values of $\grho$ are disfavored by EWPT. 
%It is therefore reasonable to conclude that  pseudo-NGB Higgs models  favor the range $100\GeV\lsim m_h\lsim 300 \GeV$.

% 
% 
%From the above equation the scalar
%quartic couplings can be expanded as
%\beq
%\lambda=\frac{\grho^2}{16\pi^2}\left (a_0 \gsm^2+a_1\frac{\gsm^4}{\grho^2}+\dots\right )\, .\label{quartic}
%\eeq
%The mass of the which upon multiplication by $v^2$ yelds 

\subsection{An issue with $\hat T$}
\label{Tissue}

After the general considerations of the previous section, let us now focus on the case of 
two composite Higgs doublets. As we will now discuss, an extra and very large 
contribution to the $\widehat{T}$ parameter, which is structurally absent in the single-Higgs 
case, potentially emerges. This is however very easily avoided.

In the SM with an elementary Higgs doublet, the accidental $\SO(4)$ symmetry
of the Higgs sector ensures the survival, after EWSB, of an (approximate) custodial isospin 
$\SO(3)_c$. This symmetry is essential to  successfully reproduce electroweak precision data, in particular the relation $\rho \equiv m_W^2/m_Z^2 \cos^2 \theta_W \simeq 1$, or equivalently the bound on $\hat{T}$ (see \cite{Giudice:2007fh} for the conventions).
In the MCHM based on $\SO(5)/\SO(4)$, the $\SO(4)$ symmetry is a true symmetry of the strong dynamics, satisfied by all the non-linear $\sigma$-model interactions. The Higgs field, being a ${\bf 4}$ of $\SO(4)$, determines a generic vacuum that 
again respect a residual custodial $\SO(3)_c$.  An equivalent statement is that the  gauged $\SO(4)_g$ and the residual $\SO(4)_H$ in the coset, when embedded in $\SO(5)$, have at least a common $\SO(3)$ subgroup.
On the other hand, in non-minimal models with two Higgses  in the ${\bf 4}$ of $\SO(4)$ the generic residual symmetry of the vacuum will  only be $\SO(2)_c$. The equivalent statement is that $\SO(4)_g$ and $\SO(4)_H$, when embedded in $\SO(6)$ generically have {\it only} an $\SO(2)$ common subgroup.
 %This is  because  the scalar potential, generated by $SO(4)$ breaking interactions (for instance  the top Yukawa or the SM gauge couplings) will in general only respects the $SU(2)_L\times U(1)_Y$ subgroup of $SO(4)$.\footnote{The unbroken $SO(2)_c$ should of course coincide with $U(1)_Q$ in order to avoid a worse phenomenological problem.}
Thus even though the nonlinear interactions satisfy $\SO(4)$, an unacceptable contribution to $\hat T$ will arise for a generic vacuum structure.  
To discuss this problem in more detail, it is useful to use two parametrizations of  a
${\bf 4}$ of $\SO(4)$, the one as a 4-vector $\Phi=\{\phi_i\}$, $i=1,\dots,4$ and  the one as a $2\times 2$ matrix $\Phi\equiv \phi_4 +i\phi_k\sigma_k$ ($k=1,2,3$) transforming as  $\Phi\rightarrow L \Phi R^{\dagger}$ under $\SO(4) \cong \SU(2)_L \times \SU(2)_R$. We will use the same symbol $\Phi$ for both parametrizations, as it will be clear from the context which one we use~\footnote{In the matrix notation, the complex doublet is embedded as $\Phi = (\widetilde{H}, H)$ where $\widetilde{H} = i \sigma_2 H^*$.}.
%, while the interaction with the $\SU(2)_L$ gauge bosons comes from the covariant derivative(we neglect the $\U(1)_Y$ coupling, since is not relevant for our purpose) $D_{\mu} \Phi= \partial_{\mu} \Phi - i \, W_a \frac{\sigma_a}{2} \, \Phi$.}.

In a model with two Higgs fields $\Phi^\uno$ and $\Phi^\due$, up to $\SU(2)_L\times \U(1)_Y$ rotations, the generic charge preserving vacuum expectation value (VEV) is $\Phi^\uno=(0,0,0,v_4^\uno)$, $\Phi^\due=(0,0,v_3^\due,v_4^\due)$. In Higgs doublet notation this corresponds to,
\beq
H^\uno=\frac{1}{\sqrt{2}}\left (\begin{array}{c} 0\\v_4^\uno\\ \end{array}\right ) \qquad\qquad
H^\due=\frac{1}{\sqrt{2}}\left (\begin{array}{c} 0\\v_4^\due-iv_3^\due\\ \end{array}\right )\, ,
\eeq
where, up to effects $v^2/f^2$, we have  $v=\sqrt{(v_4^\uno)^2+(v_4^\due)^2+(v_3^\due)^2}\simeq 246\, \rm{GeV}$. 

It is easy to check that the operator
\beq
\label{Toperator}
%\frac{c_{T}}{4f^2} \, \Trs[{\Phi^\uno}^{\dagger} D_{\mu} \Phi^\due],
\frac{c_{T}}{f^2}  \left({\Phi^\uno}\cdot \overleftrightarrow D_{\mu} \Phi^\due\right)^2,
\eeq
(${\Phi^\uno}\cdot \overleftrightarrow D_{\mu} \Phi^\due = {\Phi^\uno}\cdot (D_{\mu} \Phi^\due) - (D_{\mu} {\Phi^\uno}) \cdot \Phi^\due$) which in general arises from the non-linearities of an $\SO(4)$-symmetric $\sigma$-model, generates a contribution
\beq
\widehat{T} = - 8 c_T \frac{{(v_4^\uno)}^2 {(v_3^\due)}^2}{f^2 [{(v_4^\uno)}^2+{(v_4^\due)}^2+{(v_3^\due)}^2]}\, ,
% \sim c_T \frac{v^2}{f^2}
% \simeq \frac{1}{8},
\label{Ttree}
\eeq
proportional to the square of the order parameter $v_4^\uno v_3^\due$ of  $\SO(4)\to \SO(2)_c$
breaking. Notice that a contribution to $\widehat T$ is associated to ${\rm Im\,} ({H^\uno}^\dagger H^\due)\not = 0$.
For $c_T\sim {\cal O}(1)$, as generically generated by $\sigma$-model interactions \footnote{In the particular case of the $\SO(6)/\SO(4) \times \SO(2)$ coset, one finds $c_T = -\frac{1}{4}$, that implies $\widehat{T} > 0$.},
and $v_4^\uno\sim  v_3^\due\sim v$, we would have  $\widehat T\sim v^2/f^2$. That would be
phenomenologically acceptable only at the price of significant tuning: $v^2/f^2\lesssim 0.002$. 

Two discrete symmetries, $C_1$ and $C_2$, control the order parameter $v_4^\uno v_3^\due$ and provide a useful organizing principle 
to describe vacuum dynamics:
\begin{itemize}
\item $C_1$ is the $Z_2$ subgroup of $\SO(4)$ acting on quadruplets as 
	\beq\label{eqC1definition}
\left (\phi_1,\,\phi_2,\,\phi_3,\,\phi_4\right )\,\to \,\left (-\phi_1,\,\phi_2,\,-\phi_3,\,\phi_4\right )\,,
\eeq
or simply $H\to H^*$ in doublet notation. $C_1$, being a subgroup of $\SO(4)$,
is respected by the strong sector in all models under consideration. It acts like charge conjugation on 
the Higgses, as we have seen, and on the 
$\SU(2)_L\times \U(1)_Y$ gauge bosons as well; 
it is thus broken when the SM fermions are taken into account. When fermions are included, $C_1$ may  become an approximate symmetry only when combined with parity $P$, and that is just $CP$. 
Throughout the paper $C_1P$ is defined to act as standard $CP$ on the SM states. In particular it acts like $\psi \to \bar \psi$ without extra phases on the SM Weyl fermions.

\item $C_2$ is a reflection in the $(\Phi^\uno,\Phi^\due)$ plane, which without loss of generality we can choose to be  $\Phi^\uno\to \Phi^\uno$, $\Phi^\due\to - \Phi^\due$. This second symmetry is external to $\SO(4)$, it commutes with it and it may well be exact even when fermions are included.
In  $\SO(6)/\SO(4)\times \SO(2)$ and $\SO(6)/\SO(4)$ the role of $C_2$ can be played by the six-dimensional parity $P_6$. In that case those cosets would respectively be lifted to O$(6)/\SO(4)\times$ O(2) and O(6)$/\SO(4)\times P_2$. In the case of $\SU(5)/ \SU(4)\times \U(1)$ the role of $C_2$ can be played by charge conjugation in $\SU(5)$. It should be stressed that at the two derivative level the $\sigma$-model Lagrangian for $\SO(6)/\SO(4)\times \SO(2)$ and $\SU(5)/ \SU(4)\times \U(1)$ are automatically  endowed with $C_2$: if the fundamental dynamics were to break $C_2$
that would only show up in the four- and higher-derivative Lagrangian, and in the interactions with the heavy composite states.
On the other hand, the generic $\SO(6)/\SO(4)$ Lagrangian breaks $C_2$ already at the two derivative level. In that case
$C_2$ can be imposed by suitably chosing the three independent coefficients
%  (or decay constant) $f^2_1$, $f^2_2$
that describe the two derivative $\sigma$-model action
(remember $\SO(6)/\SO(4)$ is a reducible coset).

\end{itemize}
Combining $C_1$, $C_2$ and $P$ we have thus the following possibilities:~\footnote{
In the discussion that follows it is implicitly assumed that the vacuum respects the  
discrete symmetry under consideration. This typically happens in a region of paremeter space with non-zero measure.}
\begin{enumerate}
\item $C_1P$  is an exact or approximate symmetry of the strong sector. If it is exact,  it can
also remain  exact  when only the third family fermions are included, but it will be definitely
broken  by the  Yukawa couplings of the light families.  Then the leading contribution to the Higgs potential will be $C_1P$ symmetric: $v_4^\uno v_3^\due$ will only arise from small effects and will be well under control.

% In this situation,
% depending on a continuum set of parameters in the Higgs potential (the sign of a pseudoscalar mass squared),  $C_2P$ may or may not be spontaneously broken. If it is not spontaneously broken, then $v_4^Av_3^B$ will only arise from the light flavor sector and be well under control.
\item $C_2$ is an exact or approximate symmetry. If it is exact, $H^\due$ acts like a composite inert Higgs \cite{Barbieri:2006dq}, and 
the contribution to $\widehat T$ from the $\sigma$-model vanishes.

\item $C_1P \cdot C_2$ is an approximate symmetry only broken by the light family Yukawas, and 
it plays the role of $CP$. 
In this situation $v_4^\uno v_3^\due \not =0$, while $v_4^\uno v_4^\due=0$ up to negligible effects, 
and  the Higgs VEVs are  anti-aligned. The custodial symmetry is maximally broken 
and the model is not viable. 
This is the situation encountered in the specific model discussed in ref.~\cite{Gripaios:2009pe} .

\item No combination of $C_1$, $C_2$ and $P$ is even an approximate symmetry.
In this situation $CP$ is  violated at $O(1)$ by the top-Higgs sector, and also the custodial symmetry is  broken at $O(1)$ by the VEV structure.
\end{enumerate}

The above list exhausts all possibilities. We conclude that, in composite two Higgs doublet models,  $\widehat T$ can be protected by either   (approximate)  $CP$ or  (approximate or exact) $C_2$ .
Moreover it seems to us that the conditions for this protection, case 1 and 2, are rather mild and generic. In the potentially realistic  models satisfying either condition 1 or 2,
the leading, and unavoidable,  new physics contribution to $\widehat T$ typically comes from the top sector
and its properties are the same as discussed in ref.~\cite{Giudice:2007fh}. 
%  In particular there is an important costraint from the correlation between $\widehat T$ and the correction to the $Zb\bar b$ vertex. 
After having discussed the structure of Yukawa couplings,
we shall review the issue of  electroweak  precision parameters  in section~\ref{situation}. There we will also make some novel remarks concerning the correlation between $\widehat T$ and the corrections to the $Zb\bar b$ vertex.

In the rest of the paper we shall mostly focus on the phenomenology of models of class 1 and 2.
There is however a third interesting possibility to control $\widehat T$, corresponding to a symmetry that allows to rotate $\Phi^\due$ parallel to $\Phi^\uno$, or, which is the same, to a symmetry that  constrains $c_T$ to vanish. Such a symmetry clearly cannot commute with $\SO(4)$ and should contain two $\SU(2)_R$'s under which the two doublets transform independently: \ie $\Phi^{\uno} \rightarrow L \Phi^{\uno} R_1^{\dagger}$ and $\Phi^{\due} \rightarrow L \Phi^{\due} R_2^{\dagger}$. The simplest coset where that occurs is $\Sp(6)/\SU(2)\times \Sp(4) $ in table~\ref{cosets}.
This third possibility  is indeed the one which is accidentally realized in the weakly coupled case, such as in Supersymmetry. In a renormalizable theory, the kinetic terms are the only operators that give a mass to the vector bosons, and these are invariant under $\SO(8)$, explicitly broken to $\SU(2)_L\times \Sp(4)$ by the gauging of $\SU(2)_L$. $\Sp(4)$ contains two $\SU(2)_R$ under which each doublet transforms as above so that a custodial diagonal combination of the three $\SU(2)^3$  is preserved after both Higgses have taken arbitrary VEVs, implying $\widehat{T} = 0$. Notice that for this to work only the kinetic terms must be invariant, not the entire Lagrangian.
We shall further discuss  the model building and phenomenology of this third class of models in section~\ref{extcust}.
%This $\SU(2)^3$ is precisely the custodial symmetry present in the coset $\Sp(6)/\Sp(4) \times \SU(2)$, but in this case at the full non-linear level, what guaranties $\widehat{T} = 0$.

\subsection{The Structure of Flavor }\label{flavorstruct}
One special feature of the renormalizable SM  is that there exists only one matrix of flavor breaking (Yukawa) interactions associated to the fermions of any given charge. This ensures the absence at tree level  of contributions to flavor changing neutral currents (FCNC) and is the zeroth order reason for the SM success in describing flavor breaking phenomena. This special feature, once called natural flavor conservation, and now dubbed Minimal Flavor Violation (MFV) \cite{mfv}, is ``structurally'' absent in virtually all extensions of the SM. That means that in the extensions of the SM to obtain the same simple structure additional symmetries or dynamical assumptions other than plain renormalizability~\footnote{Here, of course, we use the concept of renormalizability with its  modern effective field theory meaning: we perform an inverse mass expansion and keep only relevant or marginal couplings.}    must be invoked.

In the 2HDM, focussing just on quarks, the  most general  Yukawa  interaction is
\beq
\bar{q}_L \big( Y_1^u \widetilde H_\uno+ Y_2^u \widetilde H_\due\big) u_R + \bar{q}_L \big( Y_1^d H_\uno + Y_2^d H_\due \big) d_R + h.c. \, ,
\label{yuk2HDM}
\eeq
corresponding to four coupling matrices to generate the two mass matrices of the up and of the down quarks. The additional flavor breaking parameters give rise to dangerous flavor transitions via Higgs exchange, implying strong constraints on the parameters. 
A more plausible  model can be obtained by restoring MFV, which can be done either by 
symmetry or by an ansatz. Using the same notation of the previous section, we can consider the
Higgs parity symmetry $C_2$ under which $(H_\uno,H_\due)\to (H_\uno,-H_\due)$ and all fermions are even, and the isospin parity $C_I$ under which $(u_R, d_R)\to (u_R,-d_R)$ and all other fields are even. Then by imposing either $C_2$ or $C_I \cdot C_2$ the unwanted new sources of flavor violation are eliminated, and we go back to the minimal flavor violating structure of the SM. The two corresponding models are respectively  known as type I and type II 2HDM.
These and other options for the SM fermion parities,
corresponding to different types of models present in the literature, are given in the table below:
\begin{table}[h!]
	\begin{center}
\begin{tabular}{cccccc}
\hline
type & $u_R$ & $d_R$ & $e_R$ \\
\hline
I & $+$ & $+$ & $+$ \\
II & $+$ & $-$ & $-$ \\
X & $+$ & $+$ & $-$ \\
Y & $+$ & $-$ & $+$ \\
\hline
\end{tabular} \\
%\caption{Choices of $C_2$-parity for the SM right-handed fields reproducing the different types of models present in the literature.}
	\label{types}
	\end{center}
\end{table}

\noindent The third possibility, known as type III, amounts to making the ansatz $Y_1^u\propto Y_2^u$, $Y_1^d\propto Y_2^d$, effectively enforcing MFV without any extra symmetry. 
This ansatz is consistent with  selection rules
from the flavor symmetry $\SU(3)_{q_L}\times \SU(3)_{u_R}\times \SU(3)_{d_R}$ and could in principle be motivated in a suitable model for the origin of flavor.

In composite Higgs models there are, a priori,  extra sources of flavor violations in the Higgs sector 
\cite{Giudice:2007fh,Gripaios:2009pe,Agashe:2009di}.
For example, in the MCHM with only one Higgs doublet $H$ 
the most general structure of the Yukawa interactions (that is with zero derivatives)  is 
\footnote{Other sources of flavor violation are associated with generalized kinetic terms with multiple Higgs insertions:
these effects come at higher order in the Yukawa or proto-Yukawa couplings and are normally subdominant and not very problematic \cite{Agashe:2009di}. This is why we neglect them in our discussion.}
\beq
\bar{q}_L \big( Y_1^u \tilde{H} + Y_3^u \tilde{H}H^\dagger H/f^2+\dots \big) u_R + \bar{q}_L \big( Y_1^d H + Y_3^d H H^\dagger H/f^2+\dots \big) d_R + \textnormal{h.c.} \, .
\label{yukHcomp}
\eeq
%corresponding again to multiple sources of flavor violation. 
The matrices $Y_3^{u,d}$ generically give rise to flavor changing couplings to the neutral Higgs only suppressed, compared with the renormalizable ones in eq.~(\ref{yuk2HDM}), by $v^2/f^2$ which is typically not enough (see however the estimates that follow). 
The way out is again MFV, {\it{i.e.}} the conditions 
$Y_1^u\propto Y_3^u\propto\dots$ and similarly for the downs. 
Interestingly, this can be automatically enforced in PNGB composite Higgs models 
where selection rules of the global group $G$ 
can imply, at lowest order in the proto-Yukawa couplings, a factorized flavor structure 
\cite{Agashe:2009di}
\beq
\bar{q}_L \big( Y_1^u \tilde{H} F_u(H^\dagger H/f^2)\big ) u_R + \bar{q}_L \big( Y_1^d HF_d(H^\dagger H/f^2) \big) d_R + h.c. \, .
\label{yukHcont}
\eeq
This feature eliminates the leading contribution to Higgs-mediated FCNC.

Now, in the composite 2HDM the issues exemplified by eq.~(\ref{yuk2HDM}) and eq.~(\ref{yukHcomp}) will both be present, but at the same time one will be able to rely, as explained above,  on both, discrete symmetries or ans$\ddot{\textrm{a}}$tze and on $G$ selection rules.
\begin{figure}
\centering
%\hspace{-300pt}
\epsfig{file=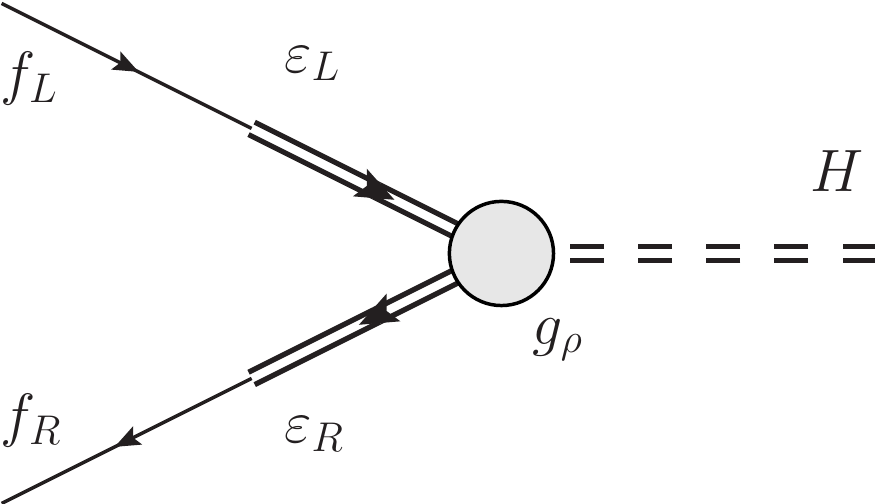,height=100pt}\ \hspace{20pt}
\ \epsfig{file=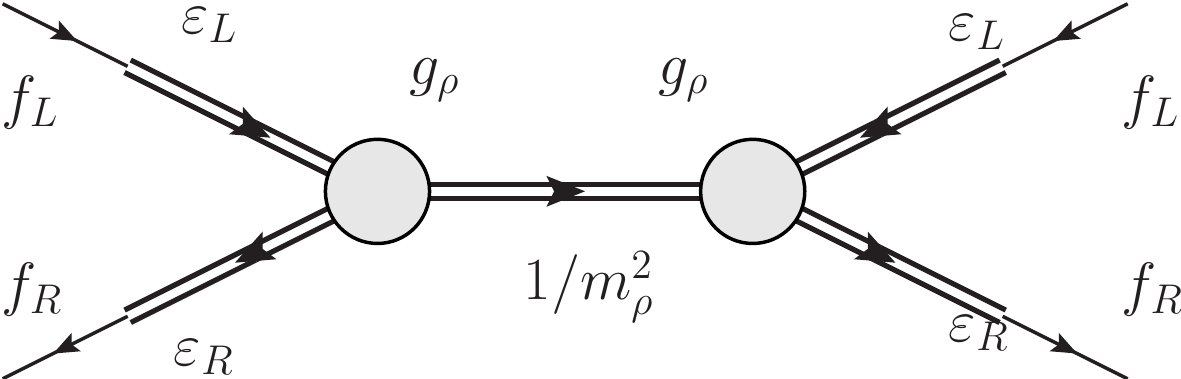,height=80pt}
\vspace{1cm}
\caption{The contribution from the exchange of heavy modes to the Yukawas and to the 
FCNC operators.}\label{fig1}
\end{figure}
Let us discuss in more detail how these mechanisms work and protect from Higgs-mediated 
flavor transitions. As previously explained, the SM fermions are coupled linearly to the 
strong sector through fermionic composite operators ${\cal O}_{f_L,f_R}$.
The latter describe couplings at microscopic scales, where  the  breaking 
$G\to H$ can be neglected, and therefore correspond to some representations of $G$
that we denote, respectively, as  ${\bf r}_L$ and ${\bf r}_R$. 
For one generation, eq.~(\ref{mix0}) can be rewritten more explicitly as
\beq
{\cal L}_{\rm mix}= 
		(\bar f_L)_{\overline{\alpha}} ({y_L}^{\overline{\alpha}})^{I_{f_L}} {\cal O}_{ I_{f_L}} 	+
		(\bar f_R) (y_R)^{I_{f_R}} {\cal O}_{I_{f_R}} 			+
		h.c.\, ,
\label{mixing}
\eeq
where the $I_{f_L}$ and $I_{f_R}$ indices of $y_{L,R}$ are in the conjugate representation 
of ${\bf r}_{L,R}$ while $\overline{\alpha}$ denotes the SM $\SU(2)_L$-doublet index.
As the notation suggests, in eq.~(\ref{mixing}) we have uplifted the $y_{L,R}$ couplings 
to  representations (spurions) of the $G\times SU(2)_W\times U(1)_Y$. This will allow us to exploit  
fully the constraints from $G$-invariance.

Adding flavor to eq.~(\ref{mixing}), amounts to adding an index $i$ to $f_L,\,y_L,\,y_R,\,{\cal O}_{ I_{f_L}},\,{\cal O}_{I_{f_R}} $. Notice that in general there is no notion of orthogonality for the composite operators, meaning that the correlator $\langle {\cal O}_{ I_{f_L}}^i {\cal O}_{ I_{f_L}}^j\rangle $ is in general non zero for any $i,j$ pair (similarly for 
${\cal O}_{I_{f_R}}^i $).
Effective Yukawa couplings, in principle of the general form of eqs.~(\ref{yuk2HDM}) and (\ref{yukHcomp}), arise at low energy via the exchange of  the heavy modes  excited by 
${\cal O}_{f_L,f_R}$  -- see fig.~\ref{fig1}. By applying  power counting as depicted in the figure, 
we expect for the $Y_1^{ij}$, $Y_2^{ij}$ and $Y_3^{ij}$ in eqs.~(\ref{yuk2HDM},\ref{yukHcomp}) the structure
\beq
Y_{1,2,3}^{ij} = \frac{y_L^i y_R^j}{g_{\rho}} \times a_{1,2,3}^{ij}= g_\rho \epsilon_L^i\,  
\epsilon_R^j\times a_{1,2,3}^{ij}\ , \qquad 
%(\textnormal{no sum over $i$, $j$}), \quad 
a_{1,2,3}^{ij} \sim O(1)\, ,
\label{genyuk}
\eeq
with $a_1^{ij}\not =a_2^{ij}\not =a_3^{ij}$ in general.
Notice that the size of the Yukawa of a given SM fermion is proportional to the degrees of mixing $\epsilon_L^i$ and $\epsilon_R^i$ of its chirality components to their
composite counterparts. Assuming the strong sector does not have any flavor structure 
($a^{ij}_{1,2,3} \sim O(1)$) these mixings have  to be hierarchical in order  to reproduce 
the observed Yukawas. It is then straightforward to estimate the typical size of flavor violating 
transitions. The transitions mediated by heavy modes, as again depicted in 
figure~\ref{fig1}, give, for instance, $LRLR$  4-fermi interactions
\beq
%{\cal L}_{4f}\sim
\epsilon_L^i\epsilon_R^j\epsilon_L^k\epsilon_R^\ell\frac{g_\rho^2}{m_\rho^2}\,\left (\bar f_L^i f_R^j \bar f_L^k f_R^\ell\right )\,.
\label{4fermion}
\eeq
For instance for the $( \bar d s)^2$,  $\Delta S=2$ transition, the coefficient is $\sim m_dm_s/v^2 m_\rho^2$ which is small enough for the real part, while it puts some pressure on the parameters for $\epsilon_K$ \cite{Csaki:2008zd}. Overall it is fair to say that this class of flavor violation can be under control with some, not totally implausible, mild tuning of parameters. On the other hand the FCNC mediated by the Higgses are usually larger.
Generically, the lightest scalar $h$, that behaves as the MCHM Higgs,
can mediate FCNC contributions from the flavor-changing couplings of
\eq{yukHcomp}, while  extra heavy scalars $S$  can mediate them  from
the couplings  of  eq.~(\ref{yuk2HDM}).
These two types of FCNC contributions are  respectively given by
\beq
%{\cal L}_{4f}^{h-med}\sim
\epsilon_L^i\epsilon_R^j\epsilon_L^k\epsilon_R^\ell\frac{g_\rho^2}{m_h^2}\frac{v^4}{f^4}\,\left (\bar f_L^i f_R^j \bar f_L^k f_R^\ell\right )\   ,
\qquad
\epsilon_L^i\epsilon_R^j\epsilon_L^k\epsilon_R^\ell\frac{g_\rho^2}{m_{ S}^2}\,
\left (\bar f_L^i f_R^j \bar f_L^k f_R^\ell\right )\, .
\label{4fermionHiggs}
\eeq
%The extra factor of $v^4/f^4$ arises, in the second term which is the one mediated by 
%the higher dimensional operators in eq.~(\ref{yukHcomp}),  because on-shell flavor 
%violating vertices with the Higgs are $O(v^2/f^2)$, while a $\Delta S=2$ transition
%requires two such vertices. 
Taking  the lightest Higgs mass to be $m_h^2\lsim Y_t^2 v^2$, 
as it happens, for example, in the MCHM, 
the contribution of the first term of eq.~(\ref{4fermionHiggs}) is
enhanced with respect to eq.~(\ref{4fermion}) by at least
$(m_\rho/m_h)^2 (v/f)^4\sim (g_\rho/Y_t)^2(v/f)^2\gg 1$. The second term of eq.~(\ref{4fermionHiggs})
%which instead originates from eq.~(\ref{yuk2HDM}), 
is potentially even more dangerous;
from  \eq{heavyscalars} we have  $m^2_S\lsim g_{SM}^2f^2$ that leads to a 
contribution  larger than
eq.~(\ref{4fermion}) by a factor $(m_\rho/m_S)^2 \sim (g_\rho/g_{SM})^2\gg 1$.

%This second effect is thus more problematic, and perhaps worth taking more seriously.

The group theoretical mechanism 
that can control
the above  Higgs-mediated FCNC
works as follows. 
At the leading order, at which loops of 
elementary states are neglected, the $f_{L,R}$ fields and the $y_{L,R}$ spurions always 
enter together in the combinations (again flavor indices $i$ not shown)
\beq
(f_L)_{\overline{\alpha}} ({y_L^*}^{\overline{\alpha}})^{I_{f_L}}/g_{\rho}\equiv
{\Psi_L}^{I_{f_L}}  \, ,
\qquad
(f_R) (y_R^*)^{I_{f_R}}/g_{\rho}\equiv {\Psi_R}^{I_{f_R}} \, .
\label{embed}
\eeq
In order to discuss what kind of terms will appear in eqs.~(\ref{yuk2HDM}) and (\ref{yukHcomp}) 
we have to classify all the possible operators compatible with the $G$ symmetry,  
with zero derivatives and any number of insertions of the NGB $\Pi$. 
This is best done by introducing the NGB  matrix $U(\Pi)$
\begin{equation}
\displaystyle
U(\Pi)= e^{i\frac1f{ \Pi^{\hat{a}} T^{\hat{a}}}}\,,
\end{equation}
where $T^{\hat{a}}$ denotes the broken generators of the coset. 
The NGB matrix transforms as
\beq
\displaystyle
U(\Pi)\to U(\Pi^{(g)})\,=\, g\, U(\Pi)\,h^\dagger(\Pi,g)\,,
\label{utr}
\eeq
where $h\in H$. As the previous equation makes 
manifest, via a  multiplication by $U^\dagger$, any representation of $G$ can be ``converted'' into a representation of 
$H$. As in the standard CCWZ construction \cite{CCWZ}, then, the $G$ invariants are provided 
by the $H$-invariants in the tensor product $ \textbf{r}_L\otimes \textbf{r}_R$. The most 
general such an invariant will read
\beq
{\cal L}_Y=m_\rho \sum_{A,i,j} a^A_{ij}\bar{\Psi}_{L}^i U(\Pi) P_A U^\dagger(\Pi) \Psi_R^j + \textnormal{h.c.}, 
\label{yukstruct}
\eeq
where $A$ indicates any $H$ invariant contained in $ \textbf{r}_L\otimes \textbf{r}_R$, while  $P_A$
represents the corresponding    projector. Since the couplings $y_{L,R}$ break  $G$,  the $\Psi_L$ and $\Psi_R$ in eq.~(\ref{embed}) are incomplete $G$ multiplets.  This explicit breaking 
of  $G$ leads, upon expansion of the above formula, to a set of Yukawa structures 
(\ref{yuk2HDM},\ref{genyuk}).

In the simplest situation, the proto-Yukawa matrices ${(y_L^i)}^{{\overline{\alpha}}{I_{f_L}}}$ for different flavors $i$ are proportional to one another, and similarly for $y^i_R$. That situation arises necessarily when, compatibly with the SM quantum numbers, there exists only one embedding of $f_L$ and $f_R$, in respectively $\textbf{r}_L$
and $\textbf{r}_R$.
In that case the number of independent Yukawa structures is clearly bounded by the number $N$ of invariants. Notice however that
for  the particular case $ \textbf{r}_L= \textbf{r}_R$, there exists one  trivial invariant
(corresponding to $P_A=1$ in eq.~(\ref{yukstruct}))
that does not depend on the NGB, and which will vanish when the 
$\Psi_{L,R}$ are put to their physical 
values in  eq.~(\ref{embed})
\footnote{This is simply because no gauge invariant bilinear $\bar f_L f_R$ can be written without the insertion of at least one Higgs field.}. In that case the number of invariants is $N-1$. Now, Higgs-mediated flavor violations are  absent if the number of non-trivial invariants is 1 for both the up and the down sector. This is
because in that case the flavor dependence will unavoidably factorize in eq.~(\ref{yukstruct})
leading to the structure of eq.~(\ref{yukHcont}). 

The one we have just described is the simplest situation. When there exists  
more than one inequivalent way to embed $f_L$ and $f_R$ into respectively  $ \textbf{r}_L$ and $\textbf{r}_R$,  the orientation of the matrices ${(y_L^i)}^{{\overline{\alpha}}{I_{f_L}}}$ (and similarly for $y_R^i$) can depend on $i$. In that case it is easy to conclude that the number of independent structures arising from eq.~(\ref{yukstruct}) is given by the number of non-trivial invariants times  the number of independent embeddings. For instance if there are two independent embeddings for $f_L$ but only one for $f_R$ we get twice as many structures, if there are two independent embeddings in both L and R we get 4 times as many Yukawa structures. 
In the minimal case 
studied in \cite{Agashe:2009di}   the doublet 
%with appropriate hypercharge 
and the singlet SM fermions are embedded in a unique way in the ${\mathbf{5}}$ of $\SO(5)$.  That model belongs thus to first simple class of models. On the other hand,
in the composite 2HDM  we will typically have multiple embeddings. We  then 
have to force the same embedding for all flavors  by either imposing a symmetry or by an ansatz, in the same spirit of MFV. Let us now see how all this works in explicit examples.

%Notice however that we have been working until now under the assumption that the 
%proto-Yukawas  
%$y_{L,R}$ in eq.~(\ref{mixing}) are flavor-diagonal; if this was not the case we could 
%generate Yukawa misalignment even in the presence of a unique invariant. 
%Flavor-diagonal proto-Yukawas are enforced if it exists a unique mixing, 
%compatibly with the SM group quantum numbers, of the 
%$f_{L,R}$ with respectively  ${\cal O}_{f_L,f_R}$. 
Consider first $\SO(6)/\SO(4)\times \SO(2)$ with $q_L\in \Psi_L=\textbf{20}'$ and $u_R, d_R\in \Psi_R=\textbf{1}$. In the tensor product $\textbf{20}'\otimes \textbf{1}$ there is obviously only one invariant, which seems already good. However $\textbf{20}'$ contains  two independent $({\bf 2,2})$, forming an $\SO(2)$ doublet,
% $Q_\alpha$ ($\alpha=5,6$), 
so that there exists two independent embeddings of $q_L$ into the $\textbf{20}'$. One of the embedding can be forbidden by demanding the coupling in eq.~(\ref{mixing}) to satisfy $C_2$,
that is just a reflection in the $5,6$ plane of $\SO(6)$. 
This leads to a composite 2HDM  of type I. One could also fold $C_2$ with isospin parity $C_I$ (under which $(u_R,d_R)\to (u_R,-d_R)$) and thus obtain the analogue of type II. Finally one could assume an ansatz according to which the embedding of $q_L$ into $\textbf{20}'$ is flavor independent. This would correspond to the composite version of type III. Of the three scenarios we outlined, the second and the third still requires an unbroken approximate $C_1P$ symmetry 
to control  $\widehat T$.

Consider now again $\SO(6)/\SO(4)\times \SO(2)$ but with matter embedded as $q_L\in \Psi_L=\textbf{6}$ and $u_R, d_R\in \Psi_R=\textbf{6}$. Decomposing ${\bf 6= 4+ 2}\equiv v_4\oplus v_2$ under $\SO(4)\times \SO(2)$, with an obvious notation, we find that  $\bf 6\otimes 6$ contains
3 invariants: $(v_4\cdot v_4)$, $v_2\cdot v_2$ and $v_2\wedge v_2$. One combination, 
$v_4\cdot v_4+v_2\cdot v_2$ is trivial and so we are left with two non-trivial invariants, that we can choose to be $v_4\cdot v_4$ and $v_2\wedge v_2$. In order to reduce the number of possible Yukawa structures we are forced to assume the strong sector respects $C_2$. Then depending on the overall $C_2$ parities of  ${\cal O}_{ I_{f_{L,R}}}$
either $v_4\cdot v_4$ or $v_2\wedge v_2$ will be eliminated. This is not yet enough
because there are two independent ways to embed $u_R, d_R\in \Psi_R=\textbf{6}$, either into the 5th or the 6th entry. This gives two possible Yukawa structures from the most general coupling in eq.~(\ref{mixing}). At this stage we can proceed like in the first model we discussed. If we assume that the mixing also respects $C_2$, the number of structures is just one, and obtain the analogue of type I. If we assume $C_I \cdot C_2$, we obtain the analogue of type II. And if we assume the embedding breaks $C_2$ while remaining flavor independent, we obtain the analogue of type III. But notice that even in this third case to eliminate one invariant we still need to assume  that the strong sector respects $C_2$.

\subsection{Electroweak  Precision Observables} \label{situation}
In this section we review the issue of  electroweak  precision tests and also take the opportunity to improve in a significant way the analysis of ref.~\cite{Giudice:2007fh}. 

The main advantage of PNGB Higgs models, compared to technicolor, is the possibility to tune $v$ to be somewhat smaller than the fundamental scale $f$. This permits to control dangerous corrections to  electroweak  observables. On the other hand, a model is  the more plausible the larger $v/f$ is. Because of that, electroweak precision tests (EWPT) still constrain significantly the structure of composite Higgs models. The first obvious constraint is given by the $S$-parameter
\beq
\widehat S\sim \frac{m_W^2}{m_\rho^2}\sim \frac{g^2}{\grho^2}\frac{v^2}{f^2}\label{Sparameter}\,.
\eeq
From the experimental constraint on $\widehat S$,    we obtain the lower bound $m_\rho\gsim 2 \TeV$, or equivalently \cite{Giudice:2007fh},
\beq
\xi\equiv\frac{v^2}{f^2}\lesssim  0.01\, g_\rho^2\simeq \,\frac{1.6}{N}\, ,
\label{boundxi}
\eeq
showing that the larger  $\grho$, {\it{i.e.}} the smaller $N$, the smaller the needed tuning on $\xi$. That gives one  sure reason for being interested in strongly coupled models.

The other relevant constraints are associated with the top couplings. Indeed, by eq.~(\ref{ytop}),  one, or both, $y_L$ and $y_R$ must be larger than $Y_t$,  giving potentially large effects. These can however be controlled by specific choices of the quantum numbers of the operators ${\cal O}_{f_L}$ and ${\cal O}_{f_R}$ in eq.~(\ref{mixing}). It is instructive to first just focus on the $\SO(4)\times \U(1)_X$ quantum numbers. Later we shall discuss the important changes due to the additional constraining power of $G/H$. For the choice ${\cal O}_L=\bf (2,1)_{1/6}$, ${\cal O}_R=\bf (1,2)_{1/6}$ the expected corrections to $Z\bar bb$ and $\widehat T$ are
\begin{equation}
\frac{\delta g_b}{g_b}\sim \frac{y_L^2}{g_\rho^2}\xi\ ,\qquad\qquad
\widehat T\sim
\frac{N_c y^4_{R}}{16\pi^2g_\rho^2}\xi\label{deltabdeltaT}\, .
\end{equation}
%These effects are computed by power counting the diagrams in Fig.?. 
Notice that for our choice of embedding, $y_L$ is an isospin singlet while $y_R$ is a spurion of custodial isospin $1/2$. Since $\widehat T$ corresponds to a violation of 2 units of isospin charge,  selection rules dictate the four powers of $y_R$ in eq.~(\ref{deltabdeltaT}).  
Now, the experimental  bounds, together with  eq.~(\ref{ytop})   imply $\xi<0.05$. This tight bound arises because  $\delta g_b/g_b$ demands a small $y_L$, $\widehat T$ demands a small $y_R$,  while the two couplings are constrained to have a sizable product to reproduce $Y_t$. A less constrained, and thus less tuned scenario, can arise in the case where
${\cal O}_L=\bf (2,2)_{2/3}$, ${\cal O}_R=\bf (1,1)_{2/3}$. We also should mention that in this case to generate the Yukawas of the down sector, assuming that the right chiralities couple to a 
$\bf (1,1)_{-1/3}$, we need to couple the quark doublet to a second operator in the $\bf (2,2)_{-1/3}$\footnote{We do not consider the possibility that down right-handed quarks couple to a $\bf (1,3)_{2/3}$ representation.}. This might in general give rise to flavor problems which can be avoided with appropriate UV assumptions, see \cite{Cacciapaglia:2007fw} . 
Now $y_R$ is an $\SO(4)$ singlet under  the custodial group and  drops out of~\eq{deltabdeltaT}.
However $y_L$  transforms as $\bf (1,2)$ under $\SO(4)$ and therefore one generically expects
\begin{equation}
\frac{\delta g_b}{g_b}\sim \frac{y_L^2}{g_\rho^2}\xi\ ,\qquad\qquad
\widehat T\sim  \frac{N_cy^4_{L}}{16\pi^2g_\rho^2}\xi\, .
\label{deltabdeltaT11}
\end{equation}
This result is more encouraging: for $y_L\sim Y_t$ and $y_R\sim g_\rho$ corresponding to a fully composite $t_R$,
the bound from $\delta g_b/g_b$ is comparable to the one from $\widehat S$, while the one from $\widehat T$ is much less severe. 

The situation might even be better though. It was pointed out in ref.~\cite{Agashe:2006at} that when the strong sector is  invariant under O(4)$=\SO(4)\times P_{LR}$ and not just $\SO(4)$, the contribution to $\delta g_b/g_b$ in eq.~(\ref{deltabdeltaT11}) vanishes. That result can be understood as follows. 
%As implied by Fig \textbf{???},  
Working at lowest order in $\gsm$, that amounts to  treating the SM fields as external sources, the strong sector is an exact O(4)$\times \U(1)_X/$ O(3)$\times \U(1)_X$ coset. Moreover, in the same limit we can neglect $m_W$ as compared to $m_\rho$. This amounts to computing the vertices of the vector bosons at zero momentum transfer, where they can be identified with the charges of the currents in O(4)$\times \U(1)_X$.  In particular the coupling to the neutral vectors is given by
\beq
g W^3_\mu J^{3\mu}_L+g'B_\mu( J^{3\mu}_R+J^\mu_X)\equiv g W^3_\mu( J^{3\mu}_V-J^{3\mu}_A)+g'B_\mu( J^{3\mu}_V+J^{3\mu}_A+J^\mu_X)\, .
\eeq
Now, the only correction to the current can come from the $J_A$ contribution, since $J_V$ and $J_X$ are conserved. However, on eigenstates of $P_{LR}$ the expectation value of the axial charge $Q^3_A$ clearly vanishes as $Q_A^3$ is odd. For these states $J_A^3$ does not contribute to the vector boson vertex, and in particular the coupling to the $Z$ is unaffected. Now in the fermion multiplet $\bf (2,2)_{2/3}$ the only eigenstate of $P_{LR}$
has electric charge $-1/3$, and plays the role of the bottom quark. 
This discussion can be complemented by an explicit effective Lagrangian analysis that makes full use of the $\SO(4)/\SO(3)$ CCWZ construction \cite{CCWZ}. 
Starting from a chiral fermion $Q_A$ transforming like $\bf (2,2)$, and using the 
NGB matrix $U_{A\bar A}$,  we can form the dressed fermions 
$\psi_i=Q_AU^*_{Ai}$ ($i=1,2,3$) transforming like a $\bf 3$ of $\SO(3)$ and $\eta=Q_AU^*_{A4}$ transforming like a singlet. Then it is straightforward to write all the possible interactions at lowest derivative order
\bea
{\cal O}_1 &=&\bar \psi \bar \sigma^\mu (\partial_\mu +{\cal E}_\mu) \psi\qquad\qquad \,{\cal O}_2=\bar \eta  \bar \sigma^\mu \partial_\mu\eta\\
{\cal O}_3 &=&\bar \psi_i\bar \sigma^\mu\eta {\cal D}_{i\mu}\qquad\qquad\qquad \quad{\cal O}_4=\bar \psi_i\bar\sigma^\mu\psi_j{\cal D}_{k\,\mu}\epsilon_{ijk}\label{operators}
\eea
where ${\cal E}_\mu$ and ${\cal D}_\mu$ are the $H$ connection and $G/H$ NGB respectively \cite{CCWZ}. ${\cal O}_{1,2,3}$ are manifestly $P_{LR}$ invariant, and  give no correction to $g_b$ upon weak gauging of the SM group. On the other hand ${\cal O}_4$ breaks $P_{LR}$ and does indeed renormalize $g_b$ \footnote{In the analysis of ref.~\cite{Agashe:2006at} only three operators are mentioned. The fourth operator left out is just the trivial kinetic term $\bar Q\bar \sigma_\mu\partial^\mu Q$ invariant under the {\it linearly realized} O(4) and corresponding to a linear combination of ${\cal O}_{1,2,3}$.}.

Now, what is remarkable, and was indeed missed in \cite{Giudice:2007fh}, is that when the Higgs scalar is itself a NGB residing into a bigger coset such as $\SO(5)/\SO(4)$ or $\SO(6)/\SO(4)\times \SO(2)$ the $P_{LR}$ arises as an accidental symmetry of the lowest derivative interactions. This is very similar to the case of $C_2$, an accidental symmetry of the 2-derivative $\sigma$-model. It is easy to prove that by extending the previous analysis  to $\SO(5)/\SO(4)$ and
assuming ${\cal O}_L={\bf 5}_{2/3}$. The corresponding fermion is $Q_A$, with $A=1,\dots ,5$. Dressing it with NGB, 
we obtain  $\psi_i=Q_AU^*_{Ai}$ ($i=1,2,3,4$) transforming like a $\bf 4$ of $\SO(4)$ and the singlet $\eta=Q_AU^*_{A5}$. Now we can still write the same $P_{LR}$ invariant contractions corresponding to  ${\cal O}_{1,2,3}$. However, at the one derivative level we cannot write the analogue of ${\cal O}_4$ since the Levi-Civita tensor of $\SO(4)$ has four indices! One can easily extend this analysis to $\SO(6)/\SO(4)\times \SO(2)$ with ${\cal O}_L$ either in the ${\bf 6}$ or ${\bf 20'}$. Again the main point is the impossibility of writing invariants that involve the Levi-Civita tensor.

In view of the latter result, in all the cases considered in previous literature $\SO(5)/\SO(4)$ or $\SO(6)/\SO(5)$ and in all the models studied in the present paper, experimental constraints allow a sizeable $y_L>Y_t$. Indeed the bound on $\widehat T$ (and also that on $B-\bar B$ mixing) can be met for $y_L$ as big as roughly $y_L\sim\sqrt{Y_t g_\rho}$. In that case a Higgs boson as heavy as $300\GeV$ could be obtained. 
%Of course, in order to make that acceptable  one would need some extra effect  bringing  $\epsilon_1$ and $\epsilon_3$ back into the experimental ellipse.
%(
%{\bf REMOVE? I decided not to put numbers for $m_h$ in the ''potential'' part, therefore there would 
%be no sense in putting the number here}
% \textbf{putting the numbers one gets even higher values})
%%Needless to say, in that case a positive but not too big  $\widehat T$ must compensate for the larger $m_h$ to satisfy EWPT.

\section{Explicit Models}
\label{secExplicitModels}

It is not difficult, making use of the general considerations outlined in the previous section, to construct potentially realistic scenarios with two composite PNGB Higgs doublets. The aim of the present section is to describe few examples that will be classified, as in section~\ref{Tissue}, in terms of the extra symmetry which will be assumed in order to deal with the  $\widehat T$ constraint. The case of discrete symmetries ($C_1P$ or $C_2$) will be considered below, 
restricting for definiteness to the $\SO(6)/\SO(4)\times \SO(2)$ coset, while the possibility of an extended global custodial group will be explored in section~\ref{extcust}. Each scenario will be defined by its $G/H$ coset, by extra discrete symmetries if needed, and by the SM fermion's embeddings into $G$ representations, {\it{i.e.}} the $G$ representations of the operators to which the SM fermions are assumed to mix. Within each model, the flavor structure will be described according to the general rules of section~\ref{flavorstruct}. Also, we will study the structure of the Higgs potential which, as we will see, is almost completely under control if extra assumptions are made on the $G$-breaking couplings external to the strong-sector. We will work under the rather strong assumption, dictated however by minimality, that the only sources of $G$-breaking are those unavoidably present, {\it{i.e.}} the SM gauge couplings and the fermion's couplings. This will allow us to parametrize the Higgs potential, at each given order in the gauge and fermion couplings, in terms of a limited number of coefficients and to check if they allow for EWSB and   the mild tuning \eq{boundxi}. We will also derive, in some specific model, interesting consequences on the spectrum of the physical Higgs scalars.

\subsection{${\bf \SO(6)/\SO(4)\times \SO(2)}$ Models}
\label{642}

To set the notation we will use the following basis for the generators (in the fundamental representation) of $\SO(6)$ algebra,
\begin{align}
%& (T^{a}_{R,L})_{ij} = -\frac{i}{2} \left[ \frac{1}{2} \epsilon^{abc} \left(  \delta^b_i \delta^c_j - \delta^b_j \delta^c_i \right) \pm \left(  \delta^a_i \delta^4_j - \delta^a_j \delta^4_i \right) \right] ,\nonumber \\
& (T^{a}_{R})_{IJ} = \frac{i}{2} \left[ \frac{1}{2} \epsilon^{abc} \left(  \delta^b_I \delta^c_J - \delta^b_J \delta^c_I \right) + \left(  \delta^a_I \delta^4_J - \delta^a_J \delta^4_I \right) \right] (-1)^{\delta^a_{2}},\nonumber \\
& (T^{a}_{L})_{IJ} = \frac{i}{2} \left[ \frac{1}{2} \epsilon^{abc} \left(  \delta^b_I \delta^c_J - \delta^b_J \delta^c_I \right) - \left(  \delta^a_I \delta^4_J - \delta^a_J \delta^4_I \right) \right] (-1)^{\delta^a_{1}} ,\nonumber \\
& (T_{S})_{IJ} = - \frac{i}{\sqrt{2}} \left(  \delta^5_I \delta^6_J - \delta^5_J \delta^6_I \right) ,\nonumber \\
& (T^{i}_\uno)_{IJ} = - \frac{i}{\sqrt{2}} \left(  \delta^{i}_I \delta^5_J - \delta^{i}_J \delta^5_I \right), \nonumber \\
& (T^{i}_\due)_{IJ} = - \frac{i}{\sqrt{2}} \left(  \delta^{i}_I \delta^6_J - \delta^{i}_J \delta^6_I \right),
\label{gen}
\end{align}
where $I, J = 1,\,\ldots,\, 6$, $i = 1,\,\ldots,\, 4$ and $a = 1,\,\ldots,\, 3$. The generators 
$T^{a}_{R,L}$ and $T_{S}$ represent, respectively, the $\SO(4) \cong \SU(2)_L \times \SU(2)_R$ 
and $\SO(2)$ subgroups while the $\SO(6)/\SO(4) \times \SO(2)$ coset is spanned by $T^{i}_{\alpha}$,  
with $\alpha=\widehat {1},\widehat {2}$. 
The broken generators $T^{i}_{\alpha}$ are associated with the NGB, transforming 
as a  $\bf (4,2)$ of  $\SO(4) \times \SO(2)$. 
Consistently with table~\ref{cosets} we therefore see that the coset  delivers two 
NGB $\SU(2)_L$-doublets $\Phi^{\alpha}=(\Phi^\uno,\Phi^\due)$. 

As in section~\ref{flavorstruct}, to derive the constraints from $\SO(6)$ symmetry we will introduce 
the NGB matrix $U(\Pi)$ transforming as in eq.~(\ref{utr}).
%It is of crucial importance, if wiling to explore the implications of the $SO(6)$ symmetry group, 
%to embed the Goldstone boson Higgses into the Goldstone matrix $U=U(\Phi)$. Under an 
%$SO(6)$ transformation $g$, $U$ transforms as
%\beq
%U(\Phi)\to U(\Phi^{(g)})\,=\,h(\Phi,g) U(\Phi)g^T\,,
%\label{utr}
%\eeq
%with $h\in SO(4)\times SO(2)$; this equation defines implicitly how $\Phi$ transforms. 
We will mostly use the  fundamental representation $U^{\mathbf 6}$. 
Using (\ref{gen}), the NGB matrix is given explicitly by
\beq
\displaystyle{
\left(
U^{\mathbf 6}
\right)^{I}_{\ \overline{I}}
= \left(e^{i \frac {\sqrt{2}\Pi} f}
\right)^{I}_{\ \overline{I}}
, \quad \Pi =  \, T^i_\alpha \Phi^\alpha_i   = 
\frac{i}{\sqrt{2}} \left( 
\begin{array}{c|c}
	0_{4\times4} 	& \Phi^\uno ~ \Phi^\due\\ \hline
	-\Phi^\uno 	& \multirow{2}{*}{$~0_{2\times2}$}	\\
	-\Phi^\due 	&
\end{array}
	 \right) \,,}
\eeq
where the index $I$ is in the fundamental of $\SO(6)$ while the index $\overline{I}$ transforms 
in a {\emph{reducible}} (non-linear)  representation of $\SO(6)$, that is  through  multiplication by the matrix $h$ in eq.~(\ref{utr}). In the case of the ${\mathbf 6}$ representation, 
$h$ is composed of two blocks, corresponding respectively to  an $\SO(4)$ and to  
an $\SO(2)$ rotation. The index $\overline{I}$ therefore runs over two 
components, $\overline{I}\equiv\{i,\,\alpha\}$,  such that $i$ labels the components of   an $\SO(4)$ 4-plet while $\alpha$ labels the components of an  $\SO(2)$ doublet.
Besides the global $\SO(6)$ group, we will also be interested in 
discrete symmetries and in particular in the $C_{1,2}$ parities defined 
in section~\ref{Tissue}.  The matrix $U^{\mathbf 6}$ transforms as 
$U^{\mathbf 6}\rightarrow {\mathcal C}_{1,2}^{\mathbf 6}\cdot U^{\mathbf 6}\cdot{\mathcal C}_{1,2}^{\mathbf 6}$ with
\beq
	{\mathcal C}_1^{\mathbf 6}=\textrm{diag}(-1,1,-1,1,1,1)\ ,
	\quad
	{\mathcal C}_2^{\mathbf 6}=\textrm{diag}(1,1,1,1,1,-1)\, .
	 \label{c12}
\eeq
%\begin{array}{c|c}
%	1_{4\times4} 	& 0~0	\\ \hline
%	0	& \multirow{2}{*}{$\sigma_3$}	\\
%	0	&
%\end{array}
%	 \right) \,.
%	 \label{c12}
%\end{equation}
%the ${\mathcal C}_{1,2}^{\mathbf 6}$ matrices  represent the action of $C_{1,2}$ in the fundamental of $SO(6)$. 
We see that $C_{1}$ is an element of the $\SO(4)$ unbroken subgroup and that $C_2$ acts as 
parity in $6$ dimensions, defined as the inversion of the last coordinate, on both the $I$ 
and $\overline{I}$ indices. Notice that an appropriate NGB matrix $U^{\bf r}$ might be 
defined for each $\SO(6)$ representation ${\bf r}$. For vectorial representations, such as the 
$\mathbf{20'}$ we will use below, $U^{\bf r}$ is trivially obtained in terms 
of products of $U^{\mathbf 6}$.

\subsubsection{Higgs Potential}

The Higgs potential originates from the $\SO(6)$ breaking effects, which we have assumed to be only 
due to the $\SU(2)_L\times \U(1)_Y$ gauge and fermion couplings. Among the latter, only 
those associated to the top quark mass will give a sizable contribution and will be considered in what 
follows. The structure 
will be determined by the $\SO(6)$ representations ${\bf r}_{Q,T}$ 
to which the $q_L=(t_L,\,b_L)$ and $t_R$ doublet and singlet are coupled to
\begin{equation}
	{\mathcal L}_\textrm{mix} = 
		(\bar q_L)_{\overline{\alpha}} ({y_L}^{\overline{\alpha}})^{I_{Q}} {\cal O}_{ I_{Q}} 	+
	(\bar t_R) (y_R)^{I_{T}} {\cal O}_{ I_{T}} 			+
		h.c.\, .
\label{lmix}
\end{equation}
%where the $y$'s have been uplifted to tensors with indices in the global group. To make 
%contact with the notation of the previous section we should identify $y$ with 
%$y\cdot \Gamma$, where $\Gamma$ is in eq.~(\ref{embed}).
As in the discussion below eq.~(\ref{mixing}), 
the implications of the symmetries can be worked out  regarding the $y$'s as
non-dynamical external spurionic fields. The $I_{Q,T}$ indices are, respectively, in the 
${\bf r}_{Q,T}$ representations of the $\SO(6)$ symmetry group of the strong sector, while 
${\overline{\alpha}}=1,2$ are indices of the ``elementary'' $\U(2)_L^{\rm el}$ group under which 
the $q_L^{\overline \alpha}$ rotate, the strong sector and in particular the Higgs fields being invariant \footnote{According to this formal viewpoint, the ``expectation" values of the external spurions $y_{L,R}$ and $g$ break the fictitious extended symmetry to a diagonal $\SU(2)\times \U(1)$ under which the Higgs multiplets have their usual quantum numbers.}. A second 
elementary group, under which $y_R$ is charged, is the $\U(1)_R^{\rm el}$ of $t_R$. 
Given that the Higgs is neutral, requiring the potential to be invariant under 
these additional elementary symmetries forces it  to depend on $y_{L,R}$ only via the combinations:
\begin{equation}
\begin{array}{rcl}
	(\Upsilon_L)^{I_{Q} J_{Q}} &=& (y^*_{L~\overline\alpha})^{I_{Q}}  (y_L^{\overline\alpha})^{ 
	J_{Q}}\,, \\ \\
	(\Upsilon_R)^{I_{T} J_{T}} &=& (y_R^*)^{I_{T}}  (y_R)^{J_{T}}\,.
\end{array}
\end{equation}
In the small-coupling expansion, making use of the power counting rule described in section~\ref{genS}, 
the Higgs potential at one loop order takes the form
\begin{equation}
\displaystyle
	V = \frac{m_\rho^4}{16\pi^2}\sum_{n_R,n_L}\frac1{({g_\rho^2})^{n_R+n_L}}\,
	\sum_{\delta} c_{\delta}^{(n_R,n_L)} ~\mathcal{I}_{(n_R,n_L)}^\delta\,,
	\label{pot1}
\end{equation}
where $\mathcal{I}_{(n_R,n_L)}^\delta$ denotes $\SO(6)$ invariant operators  constructed 
with the NGB and $n_{R,L}$ powers of $\Upsilon_{R,L}$, while 
$c_{\delta}^{(n_R,n_L)}$ are order one coefficients.

It is straightforward to classify these invariants at each given order proceeding similarly to section~\ref{flavorstruct}. 
The central objects are the dressed spurions $\overline\Upsilon_{L,R}$, 
\bea\displaystyle
\left(\overline\Upsilon_L\right)^{\overline{I}\,\overline{J}}\,\equiv\,
\left(U^{{\bf r}_Q\dagger}\right)^{\overline{I}}_{\  \ I}
\left(U^{{\bf r}_Q\dagger}\right)^{\overline{J}}_{\  \ J}\left(\Upsilon_L\right)^{IJ}\,,\nonumber\\
\left(\overline\Upsilon_R\right)^{\overline{I}\, \overline{J}}\,\equiv\,
\left(U^{{\bf r}_T\dagger}\right)^{\overline{I}}_{\  \ I}
\left(U^{{\bf r}_T\dagger}\right)^{\overline{J}}_{\  \ J}\left(\Upsilon_R\right)^{IJ}\,,
\label{upbar}
\eea
obtained by rotating $\Upsilon_{L,R}$ with the NGB matrix in the appropriate representation. 
Because of eq.~(\ref{utr}), and by the same argument we made in the previous section concerning the  $\overline I$ index in ${\mathbf 6}$, the $\overline\Upsilon_{L,R}$ form a
reducible non-linear representation of $\SO(6)$. More explicitly,  they transform as
\beq\displaystyle
\left( \overline\Upsilon_{L,R}\right)^{\overline{I}\overline{J}}\,\rightarrow\,
\left( h^{{\bf r}_{Q,T}}\left(\Phi,\,g\right) \right)_{\ \ \overline{K}}^{ \overline{I}}\,
\left( h^{{\bf r}_{Q,T}}\left(\Phi,\,g\right) \right)_{\ \ \overline{L}}^{ \overline{J}}
\left(\overline\Upsilon_{L,R}\right)^{\overline{K}\,\overline{L}}\,,
\eeq
where $h^{{\bf r}_{Q,T}}$ takes, as before, a block-diagonal form. 
To construct the $\SO(6)$ invariants we therefore simply have to 
classify all possible $\SO(4)\times \SO(2)$ invariants that can be built 
out of  $\overline\Upsilon_{L,R}$ at a given order. Notice  that among the  $\SO(4)\times \SO(2)$ invariants, the special ones that are also invariant under $\SO(6)$ are clearly not interesting. Indeed,  by the  definition~(\ref{upbar}),  $U^{\bf r}$ cancels out when forming $\SO(6)$ invariants,
which therefore give a  field independent constant contribution to the potential.

It will be relevant in our classification to establish the $C_2$ and $C_1P$ parities of each  
invariant. Given that the SM elementary fermions are $C_2$ even, 
invariance of ${\cal L}_{\textrm{mix}}$ in  eq.~(\ref{lmix}) is ensured by formally assigning the following $C_2$ transformation
\beq
\left(y_{L,R}\right)^I\,\rightarrow\,
\left(
{\mathcal C}_2^{\mathbf{r}_{Q,T}}\right)^I_{\ \ J}
\left(y_{L,R}\right)^{J}
\;\;\;\;\Rightarrow\;\;\;\;
\left(\overline\Upsilon_{L,R}\right)^{\overline{I}\overline{J}}\,\rightarrow\,
\left({\mathcal C}_2^{\mathbf{r}_{Q,T}}\right)_{\ \ \overline{K}}^{ \overline{I}}\,
\left({\mathcal C}_2^{\mathbf{r}_{Q,T}}\right)_{\ \ \overline{L}}^{ \overline{J}}
\left(\overline\Upsilon_{L,R}\right)^{\overline{K}\,\overline{L}}\,,
\label{c2}
\eeq
where ${\mathcal C}_2^{\mathbf{r}_{Q,T}}$ denotes the $C_2$ action in the appropriate 
representations. For vector-like representations this is again easily obtained from 
the one in the fundamental, ${\mathcal C}_{2}^{\mathbf 6}$, which is reported in 
eq.~(\ref{c12}). For what concerns the action of $C_1P$ (see section~\ref{Tissue}),  it
coincides with ``ordinary'' $CP$ on the elementary fermions,  on the SM gauge fields 
and on the Higgs. This last requirement fixes $C_1P$ 
to act on the NGB matrix as parity ($\vec{x}\rightarrow-\vec{x}$)  combined with 
the $C_1$ transformation defined in eq.~(\ref{c12}). On the fermionic operators of the 
strong sector, such as the ones that mix with the elementary fermions, 
we take $C_1P$ to be ordinary $CP$, ${\cal O}\rightarrow \overline{\cal O}$ in Weyl notation, 
combined with 
the $C_1$ transformation that we have introduced for the Higgs. 
Invariance of eq.~(\ref{lmix})  implies  for the couplings the following transformation
\beq
\left(y_{L,R}\right)^I\,\rightarrow\,
\left(
{\mathcal C}_1^{\mathbf{r}_{Q,T}}\right)^I_{\ \ J}
\left(y_{L,R}^*\right)^{J}\,.
\label{c1p}
\eeq
Remember however that ${\mathcal C}_1$ is an element of the symmetry 
group of the strong sector, and that the ${\mathcal I}^\delta$'s are automatically 
invariant under such transformations. For the purpose of establishing the $C_1P$ parities 
of the various invariants, the ${\mathcal C}_1$ part of the transformation can therefore be ignored 
and the action of $C_1P$ effectively reduces to  $y_{L,R}\rightarrow y_{L,R}^*$ or, even 
more simply
\beq
\left(\overline\Upsilon_{L,R}\right)^{\overline{I}\overline{J}}\,\rightarrow\,\left(\overline\Upsilon_{L,R}\right)^{\overline{J}\overline{I}}\,.
\label{c1tr}
\eeq

Let us now apply these general considerations to two specific choices of the $\mathbf{r}_{Q,T}$ 
representations that will be useful in the following: $\mathbf{r}_{Q,T}={\mathbf{6}}$ and $\{\mathbf{r}_{Q},\,\mathbf{r}_{T}\}=\{\mathbf{20'},\,\mathbf{1}\}$.

In our discussion we shall call  {\it spurionic} the symmetries that are formally satisfied by the
effective action when  the spurions are tranformed according to the rules we discussed. We shall instead call {\it residual} the symmetries that are truly unbroken, that is  when the spurions are not tranformed.

\subsubsection*{Fermion contributions with $\mathbf{r}_{Q,T}={\mathbf{6}}$:}

Both $q_L$ and $t_R$ couple, respecting $\SU(2)_L\times \U(1)_Y$, to 
fermionic operators in the ${\mathbf 6}$ with $X=2/3$ (as usual the 
hypercharge is given by $Y=T_R^3+X$). 
More precisely $q_L$ couples to the ${\bf 4_{2/3}}$  of $\SO(4)\times \U(1)_X$ which populates the first $4$ entries of the ${\bf 6_{2/3}}$. 
Given that there is a unique embedding of $q_L$ in the $\bf 4_{2/3}$, the physical value of 
the  $y_L$ spurion in eq.~(\ref{lmix}) which determines the $q_L$ coupling is uniquely fixed to be
\beq
\left(y_L^{{\overline\alpha}}\right)^I\,=\,\frac{y_L}{\sqrt{2}}\,
\left\{ \left( \vec v^{\,\bar1},0,0 \right),\left( \vec v^{\,\bar2},0,0 \right) \right\}\,,
%\left\{ \left( 0,0,i,1,0,0 \right),\left( -i,1,0,0,0,0 \right) \right\}\,,
\label{genL}
\eeq
where  $y_L$ has been made real by an $\U(1)$ rotation of the elementary $q_L$ and
we have defined the vectors,
\begin{eqnarray}
\vec v^{\,\bar1}=(0,0,i,-1)\, , \nonumber \\
\vec v^{\,\bar2}=(-i,1,0,0)\, .
\label{embeddingleft}
\end{eqnarray}
%$(v_{\alpha_R}^{\bar\alpha_L})_i$ and  $(v_{\alpha_R}^{\bar\alpha_L})^i$ is the projector on the $(T_3^R,T_3^L)=(\alpha_R,\alpha_L)$ component inside a $\mathbf{4}$ of $\SO(4)$ :
%\begin{equation}\label{eqProj22in4}\renewcommand\arraystretch{1.0}
%	\sqrt{2}(v_{\alpha_R}^{\bar\alpha_L})^i =
%	\begin{tabular}{|c|c|cc|}\cline{3-4}
%		\multicolumn{2}{c|}{} & \multicolumn{2}{c|}{$\alpha_R$} \\\cline{3-4}
%		\multicolumn{2}{c|}{} & - & + \\ \hline
%		\multirow{2}{*}{$\alpha_L$} 	
%		& 1 & $(0,0,i,1)$ & $(i,1,0,0)$\\
%		& 2 & $(-i,1,0,0)$& $(0,0,-i,1)$\\\hline
%	\end{tabular}
%\end{equation}
We see, comparing with eqs.~(\ref{c2},\ref{c1p}), that the $y_L$'s VEV is automatically 
invariant under both $C_1P$ and $C_2$. Provided  one of the two parities was a 
symmetry of the strong sector, $y_L$ will not induce new breaking effects. 
The situation is different for $y_R$. Given that the ${\mathbf 6}$ (with, again, $X=2/3$)
contains two $\SU(2)_L$ singlets with the hypercharge of the $t_R$, the most general 
form of its VEV is
\beq
%\left(y_R\right)^I\,=\,\left( 0,0,0,0, y_R^+,iy_R^- \right)\ ,%\qquad y_{R}^\pm\in\Re\, ,
\left(y_R\right)^I\,=\,\left( 0,0,0,0, \vec v_R + i \vec v_I \right)\ ,
\eeq
where $\vec v_{R,I}$ are two real $\SO(2)$ vectors.
%Looking again to eqs~(\ref{c2},\ref{c1p}), $C_1P$ and $C_2$ are broken if $\vec v_R\wedge \vec v_I\neq0$, 
%implying that the coupling anyway preserves $C_1P\cdot C_2$.
By combining an $\SO(2)$ strong sector's rotation with a $\U(1)_R^{\rm el}$ 
phase transformation, $\vec v_{R}$ and $\vec v_{I}$ can be aligned  respectively along 
$(1,0)$ and  $(0,1)$, allowing to parametrize the most general VEV of the $y_R$ spurion as
%with, a priori, complex $y_R^\pm$ coefficients. It is however not difficult to show that $y_R^\pm$ 
%can be made real by combining an $SO(2)$ strong sector's rotation with a $U(1)_R^e$ phase 
%transformation. The most general VEV of the $y_R$ spurion can therefore be written as 
\beq
\left(y_R\right)^I\,=\,y_R\,\left( 0,0,0,0,\cos\theta,i\sin\theta \right)\,.%\qquad y_{R}^\pm\in\Re\, ,
\label{genR}
\eeq
with $y_R$  real. %Both $C_1P$ and $C_2$ are preserved in the special case $\theta=0$.
This general VEV, looking again at eqs.~(\ref{c2},\ref{c1p}), breaks both $C_1P$ and $C_2$
while it preserves the product  $C_1P \cdot C_2$. Both $C_1P$ and $C_2$ are 
preserved in the special case $\theta=0$.

Let us now proceed, following the general method outlined before, to the classification of the 
possible contributions to the Higgs potential. The $\overline\Upsilon_{L,R}$ have two indices in 
the ${\mathbf 6}$, which decompose as $({\mathbf 4},{\mathbf 1})\oplus ({\mathbf 1},{\mathbf 2})$ 
under $\SO(4)\times \SO(2)$. At leading order ($n_{L,R}=1$) six $\SO(4)\times \SO(2)$ invariants 
can be formed, three  from $\overline\Upsilon_{L}$ and three from 
$\overline\Upsilon_{R}$. One linear combination of each of these three, corresponding to the $\SO(6)$ 
invariant $\left(\overline\Upsilon_{L,R}\right)^{\overline{I}\overline{J}}\delta_{\overline{I}\overline{J}}$, 
is independent of the Higgs field and must be removed from the counting. We are therefore left with four operators
%, one of which 
%($\epsilon_{\alpha\beta}(\overline\Upsilon_R)^{\alpha\beta}$) however vanishes because with the 
%choice (\ref{yrvev}) the $\Upsilon_L$ spurion is symmetric. Eventually, one has three invariants 
\bea
&&{\mathcal I}^1_{(1,0)}\,=\, \delta_{ij}(\overline\Upsilon_R)^{ij}\,,
\;\;\;\;\;\;\;\;\;\;\;\;
{\mathcal I}^1_{(0,1)}\,=\, \delta_{ij}(\overline\Upsilon_L)^{ij}
\,,\nonumber\\
&&{\mathcal I}^2_{(1,0)}\,=\, \epsilon_{\alpha\beta}(\overline\Upsilon_R)^{\alpha\beta}\,,
\;\;\;\;\;\;\;\;\;\;
{\mathcal I}^2_{(0,1)}\,=\, \epsilon_{\alpha\beta}(\overline\Upsilon_L)^{\alpha\beta}
\,,
\eea
the first two are even under both $C_1P$ and $C_2$ while the others are odd, as summarized 
in table~\ref{tabPot66}. The parities we refer to are the {\it spurionic} ones, obtained by transforming 
the spurions as in eqs.~(\ref{c2},\ref{c1p}) independently on whether their VEV preserves the 
symmetry or not. Under the assumption that the strong sector is invariant 
under either $C_1P$ or $C_2$, the spurionic parities determine whether a given operator can be 
generated or not. 

After the invariants are classified and written explicitly, the last 
step which is needed in order to compute their actual contribution to the Higgs potential is to 
substitute the spurions with their VEVs, which are given by eqs.~(\ref{genL},\ref{genR}). 
Given that the VEV of eq.~(\ref{genR}) breaks (for $\theta\neq0$) $C_1P$ and $C_2$, 
the parities of the corresponding contributions 
to the potential, which we denote as {\it residual parities}, do not coincide, in general, 
with the spurionic ones. 
These residual parities are shown in table~\ref{tabPot66}; notice that not all the operators have a definite 
residual parity because substituting a parity-breaking spurion VEV might make the operator acquire 
one parity-even and one parity-odd part.  From the table we can also read if each operator, again 
after the spurions have taken their VEV,  is invariant under the $\SO(4)$ custodial symmetry or not. 
Given that custodial is only broken by the $y_L$ spurions, 
it comes as no surprise that it is preserved by 
${\mathcal I}^{1}_{(1,0)}$ and ${\mathcal I}^{2}_{(1,0)}$ while it is broken by ${\mathcal I}^{2}_{(0,1)}$. 
The invariance of  
${\mathcal I}^{1}_{(0,1)}$ is more surprising and it can be understood as follows. 
Whenever the $q_L$ couples to a ${\mathbf 4}$ of $\SO(4)$ as in the present case 
(and in the one of $\{\mathbf{r}_{Q},\,\mathbf{r}_{T}\}=\{\mathbf{20'},\,\mathbf{1}\}$ discussed in the 
following paragraph),  the custodial-breaking part of the spurion multiplet is
\beq
%\left(y_L^{{\overline\alpha}}\right)^i\,=\,\frac{y_L}{\sqrt{2}}\,\left\{ \left( 0,0,i,1 \right),\left( -i,1,0,0 \right) \right\}\,,
\left(y_L^{{\overline\alpha}}\right)^i\,=\,\frac{y_L}{\sqrt{2}}\,\left\{ \vec v^{\,\bar 1}, \vec v^{\,\bar 2}\right\}\,,
\eeq
and, because of $\U(2)^{\rm el}_L$ invariance, it will only enter through the combination
\beq
\left(\Upsilon_L\right)^{i\, j} \,=\, \left(y^*_{L~\overline\alpha}\right)^{i}  
\left(y_L^{\overline\alpha}\right)^{ 
	j}\,=\,\frac{y_L^2}{2}
\left(\begin{array}{cccc} 1&i&0&0\\-i&1&0&0\\0&0&1&i\\0&0&-i&1\\ \end{array} \right)_{ij}
\,\equiv\, \left(\Upsilon_L^+\right)^{i\, j} \,+i\,\left(\Upsilon_L^-\right)^{i\, j} \,.
\label{iu}
\eeq
Here $\Upsilon_L^\pm$ denote, respectively, the symmetric and antisymmetric components 
of $\Upsilon_L$ that are even and odd under $C_1P$ because of eq.~(\ref{c1tr}).
The VEV of $\Upsilon_L^+$, as eq.~(\ref{iu}) shows, is proportional to the identity and therefore 
does not break custodial. We then understand why, at the leading order where a single power of  
$\Upsilon_L^\pm$ can be used to construct the potential, custodial breaking can only appear 
in a $C_1P$-odd term. That the latter is also $C_2$-odd is instead a peculiarity of the case under 
consideration and cannot be understood in general terms.

It is a simple exercise to continue the classification for the second order terms. The results are 
presented in table~\ref{tabPot66}. The only new subtlety, which is first encountered at 
this order, concerns the coefficient of the operators, whose estimate does not always coincide with 
eq.~(\ref{pot1}). This is because eq.~(\ref{pot1}) assumes all the operators to be generated by one 
single loop of the elementary fermions (from which the $1/16\pi^2$ pre-factor), while many operators  
in table~\ref{tabPot66} start being generated at the two-loop level. 
% given that they are not of the ``single-trace'' type. 
The latter operators are formally subleading  even though in practice the suppression 
might be small, being $O(g_\rho^2/16\pi^2)$.

Let us summarize the main results of this classification, that will be further discussed in the following. 
At leading order, $\propto y^2$, all the operators are either even or odd under both the discrete symmetries
$C_1P$ and $C_2$, so that, assuming the strong sector to respect one discrete symmetry, automatically implies the other. 
Moreover, all the even operators will respect $\SO(4)$, $C_1P$ and $C_2$ after the spurions will acquire VEVs,
in spite of the fact that all these symmetries were broken by the spurion's VEVs. This leaves 
many accidental symmetries in the Higgs potential. Notice also that, even when it is not preserved by the strong sector, $C_1P \cdot C_2$ arises as an accidental spurionic  symmetry of the potential 
at  leading order. Given that $C_1P\cdot C_2$ is also preserved by  the VEV of the spurions, it will therefore remain as an accidental residual symmetry of the  leading order potential. 
These features are lost at  order $y^4$. 
Indeed, two operators, $\mathcal{I}^5_{(2,0)}$ \& $\mathcal{I}^2_{(1,1)}$, break $C_1P\cdot C_2$ and $\SO(4)$ is broken by even operators ($\mathcal{I}^1_{(1,1)}$, $\mathcal{I}^1_{(0,2)}$, \dots).
However, in the particular situation we will consider below, where the spurion VEV respects both $C_1P$ and $C_2$ ($\theta=0$), at order $y^4$  it remains true that 
$C_2$ invariance of  the strong sector implies an accidental $C_1P$ in the potential.

\begin{table}[!p]
\centering
\renewcommand{\Re}{\mathfrak{Re}}
\renewcommand{\Im}{\mathfrak{Im}}
\renewcommand{\arraystretch}{1.5}
\scalebox{0.8}{
\begin{tabular}{|c|rclp{0cm}||c|c||c|c||c|c||c||cccc|}\cline{2-16}
\multicolumn{1}{c|}{}					&
\multicolumn{4}{c||}{\multirow{3}{*}{Operator}} 	&
\multicolumn{2}{c||}{Spurionic} 			&
\multicolumn{4}{c||}{Residual} 			&&
\multicolumn{4}{c|}{\multirow{2}{*}{Belongs to}} 
\\ \cline{8-11}
\multicolumn{1}{c|}{}&
&&&&
\multicolumn{2}{c||}{Parity} 			&
\multicolumn{2}{c||}{$\SO(4)$} 			&
\multicolumn{2}{c||}{Parity} 			&
loops 
&&&&
\\ \cline{6-11}\cline{13-16}
\multicolumn{1}{c|}{}&
&&&&
$C_2$ 			&
$C_1P$ 			&
$\theta = 0$		&
$\theta \neq 0$		&
$C_2$ 			&
$C_1P$ 			&&
A&B&C&D
\\ \hline
%%%%%%%%%%%%%%%%%%%%%%%
% R %%%%%%%%%%%%%%%%%%%
%%%%%%%%%%%%%%%%%%%%%%%
\multicolumn{1}{|c|}{\multirow{2}{*}{$y_R^2$}}
% (10)1 
&
$\mathcal{I}_{(1,0)}^1$ &=& $\delta_{ij} (\overline\Upsilon_R)^{ij}$ & 
&
$+$ & $+$ &
$\checkmark$ & $\checkmark$ & $+$ & $+$ & 1 
&$\checkmark$ & $\checkmark$ &$\checkmark$ & $\checkmark$ 
\\ \cline{2-16}
% (10)2 
&
$\mathcal{I}_{(1,0)}^2$ &=& $\epsilon_{\alpha\beta} (\overline\Upsilon_R)^{\alpha\beta}$ & &
$-$ & $-$ &
$0$ & $\checkmark$ & $+$ & $+$ & 1 
&$\checkmark$ & $\times$ &$\times$ & $\times$ 
\\ \hline \hline
%%%%%%%%%%%%%%%%%%%%%%%
% L %%%%%%%%%%%%%%%%%%%
%%%%%%%%%%%%%%%%%%%%%%%
\multicolumn{1}{|c|}{\multirow{2}{*}{$y_L^2$}}
% (01)1 
&
$\mathcal{I}_{(0,1)}^1$ &=& $\delta_{ij} (\overline\Upsilon_L)^{ij}$ & 
&
$+$ & $+$ &
$\checkmark$ & $\checkmark$ & $+$ & $+$ & 1 
&$\checkmark$ & $\checkmark$ &$\checkmark$ & $\checkmark$ 
\\ \cline{2-16}
% (01)2 
&
$\mathcal{I}_{(0,1)}^2$ &=& $\epsilon_{\alpha\beta} (\overline\Upsilon_L)^{\alpha\beta}$ & 
&
$-$ & $-$ &
$\times$ & $\times$ & $-$ & $-$ & 1 
&$\checkmark$ & $\times$ &$\checkmark$ & $\times$ 
\\ \hline \hline
%%%%%%%%%%%%%%%%%%%%%%%
% RR %%%%%%%%%%%%%%%%%%%
%%%%%%%%%%%%%%%%%%%%%%%
\multicolumn{1}{|c|}{\multirow{5}{*}{$y_R^4$}}
% (20)1 
&
$\mathcal{I}_{(2,0)}^1$ &=& $\Re\left[\delta_{il}\delta_{jk} (\overline\Upsilon_R)^{ij}(\overline\Upsilon_R)^{kl}  \right]$ &&
$+$ & $+$ &
$\checkmark$ & $\checkmark$ & $+$ & $+$ & 1 &
$\checkmark$ & $\checkmark$ & $\checkmark$ & $\checkmark$
\\ \cline{2-16}
% (20)2 
&
$\mathcal{I}_{(2,0)}^2$ &=& $\Re\left[\epsilon_{\alpha\delta}\epsilon_{\beta\gamma} (\overline\Upsilon_R)^{\alpha\beta}(\overline\Upsilon_R)^{\gamma\delta}  \right]$ &&
$\phantom{^{(a)}}+^{(a)}$ & $+$ &
$0$ & $\checkmark$ & $+$ & $+$ & 1 & 
$\checkmark$ & $\times$ & $\times$ & $\times$
\\ \cline{2-16}
% (20)3
&
$\mathcal{I}_{(2,0)}^3$ &=& $\Im\left[\delta_{ij}\epsilon_{\alpha\beta} (\overline\Upsilon_R)^{i\alpha}(\overline\Upsilon_R)^{\beta j}  \right]$ &&
$-$ & $-$ &
$0$ & $\checkmark$ & $+$ & $+$ & 1 & 
$\checkmark$ & $\times$ & $\times$ & $\times$
\\ \cline{2-16}
% (20)4 
&
$\mathcal{I}_{(2,0)}^4$ &=& $\Re\left[\delta_{ik}\delta_{jl} (\overline\Upsilon_R)^{ij}(\overline\Upsilon_R)^{kl}  \right]$ &&
$+$ & $+$ &
$\checkmark$ & $\checkmark$ & $+$ & $+$ & 2 & 
$\times$ & $\checkmark$ & $\times$ & $\times$
\\ \cline{2-16}
% (20)5 
&
$\mathcal{I}_{(2,0)}^5$ &=& $\Im\left[ \mathbb{I}_{IK}\delta_{jl} (\overline\Upsilon_R)^{Ij}(\overline\Upsilon_R)^{Kl} \right]$ &&
$+$ & $-$ &
$0$ & $\checkmark$ & $-$ & $+$ & 2 & 
$\checkmark$ & $\checkmark$ & $\times$ & $\times$
\\ \hline\hline
%%%%%%%%%%%%%%%%%%%%%%%
% RL %%%%%%%%%%%%%%%%%%%
%%%%%%%%%%%%%%%%%%%%%%%
\multicolumn{1}{|c|}{\multirow{6}{*}{$y_R^2y_L^2$}}
% (11)1 
&
$\mathcal{I}_{(1,1)}^1$ &=& $\Re\left[ \delta_{il}\delta_{jk} (\overline\Upsilon_R)^{ij}(\overline\Upsilon_L)^{kl} \right]$ &&
$+$ & $+$ &
$\checkmark$ & $\times$ & $$ & $$ & 1 & 
$\checkmark$ & $\checkmark$ & $\checkmark$ & $\checkmark$
\\ \cline{2-16}
% (11)2 
&
$\mathcal{I}_{(1,1)}^2$ &=& $\Re\left[\epsilon_{\alpha\beta}\delta_{ij} (\overline\Upsilon_R)^{\alpha i}(\overline\Upsilon_L)^{j\beta}  \right]$ &&
$-$ & $+$ &
$\checkmark$ & $\checkmark$ & $-$ & $+$ & 1 & 
$\checkmark$ & $\times$ & $\checkmark$ & $\times$
\\ \cline{2-16}
% (11)3 
&
$\mathcal{I}_{(1,1)}^3$ &=& $\Re\left[ \epsilon_{\alpha\delta}\epsilon_{\beta\gamma} (\overline\Upsilon_R)^{\alpha\beta}(\overline\Upsilon_L)^{\gamma\delta} \right]$ &&
$\phantom{^{(a)}}+^{(a)}$ & $+$ &
$\checkmark$ & $\times$ & $$ & $$ & 1 & 
$\times$ & $\times$ & $\checkmark$ & $\times$
\\ \cline{2-16}
% (11)4 
&
$\mathcal{I}_{(1,1)}^4$ &=& $\Im\left[ \epsilon_{\alpha\beta}\delta_{ij} (\overline\Upsilon_R)^{\alpha i}(\overline\Upsilon_L)^{j\beta} \right]$ &&
$-$ & $-$ &
$\times$ & $\times$ & $$ & $$ & 1 & 
$\checkmark$ & $\times$ & $\checkmark$ & $\times$
\\ \cline{2-16}
% (11)5 
&
$\mathcal{I}_{(1,1)}^5$ &=& $\mathcal{I}_{(10)}^1\mathcal{I}_{(01)}^1$ &&
$+$ & $+$ &
$\checkmark$ & $\checkmark$ & $\checkmark$ & $\checkmark$ & 2 & 
$\times$ & $\checkmark$ & $\times$ & $\checkmark$
\\ \cline{2-16}
% (11)6 
&
$\mathcal{I}_{(1,1)}^6$ &=& $\Re\left[\epsilon_{ijkl} (\overline\Upsilon_R)^{ij}(\overline\Upsilon_L)^{kl}  \right]$ &&
$+$ & $+$ &
$0$ & $\checkmark$ & $\times$ & $\times$ & 2 & 
$\times$ & $\checkmark$ & $\times$ & $\times$
\\ \hline\hline
%%%%%%%%%%%%%%%%%%%%%%%
% LL %%%%%%%%%%%%%%%%%%%
%%%%%%%%%%%%%%%%%%%%%%%
\multicolumn{1}{|c|}{\multirow{4}{*}{$y_L^4$}}
% (02)1
&
$\mathcal{I}_{(0,2)}^1$ &=& $\Re\left[ \delta_{il}\delta_{jk} (\overline\Upsilon_L)^{ij}(\overline\Upsilon_L)^{kl} \right]$ &&
$+$ & $+$ &
$\times$ & $\times$ & $+$ & $+$ & 1 & 
$\checkmark$ & $\checkmark$ & $\checkmark$ & $\checkmark$
\\ \cline{2-16}
% (02)2 
&
$\mathcal{I}_{(0,2)}^2$ &=& $\Re\left[ \epsilon_{\alpha\delta}\epsilon_{\beta\gamma} (\overline\Upsilon_L)^{\alpha \beta}(\overline\Upsilon_L)^{\gamma\delta} \right]$ &&
$\phantom{^{(a)}}+^{(a)}$ & $+$ &
$\times$ & $\times$ & $+$ & $+$ & 1 & 
$\times$ & $\times$ & $\checkmark$ & $\times$
\\ \cline{2-16}
% (02)3 
&
$\mathcal{I}_{(0,2)}^3$ &=& $\Im\left[ \delta_{ij}\epsilon_{\alpha\beta} (\overline\Upsilon_L)^{i\alpha}(\overline\Upsilon_L)^{\beta j} \right]$ &&
$-$ & $-$ &
$\times$ & $\times$ & $-$ & $-$ & 1 & 
$\times$ & $\times$ & $\checkmark$ & $\times$
\\ \cline{2-16}
% (02)4 
&
$\mathcal{I}_{(0,2)}^4$ &=& $(\mathcal{I}_{0,1}^1)^2$ &&
$+$ & $+$ &
$\checkmark$ & $\checkmark$ & $+$ & $+$ & 2 & 
$\times$ & $\checkmark$ & $\times$ & $\checkmark$
\\ \hline
%\multicolumn{12}{c|}{} &
%\rotatebox{90}{$11+1$}&
%\rotatebox{90}{$5+5$}&
%\rotatebox{90}{$11+0$}&
%\rotatebox{90}{$5+2$}\\\cline{13-16}
\end{tabular}}\\
\caption{
The independent invariants that contribute to the Higgs potential, up to order $y_{L,R}^4$, in the case $\mathbf{r}_{Q,T} = \mathbf{6}$. For each operator, the first two columns contain the spurionic $C_2$ and $C_1P$ parities, the third and fourth ones indicate whether it will respect the $\SO(4)$ symmetry after the spurions acquire VEV, while the following two show the $C_2$ and $C_1P$ parities of the generated potential.
Whether the operator can be generated at one or two loops is written in the seventh column.
The last columns indicate which operators should be used in a given setup;
A: no constraints, B: $C_2$ in the strong sector, C: $C_2$ in the fermion coupling, and D: $C_2$ both in the strong sector and the fermion coupling.
To order the operators we have given priority to 1 loop against 2 loops, and further assumed $g_\rho>y_R>y_L$.
The shape of the potential is not affected by this choice, only the Naive Dimensional Analysis (NDA) associated to the various coefficients would be modified.
$(a)$: the intrisic $C_2$ is positive because it is the product of two $C_2$ odd contributions.
In case of a $C_2$ symmetric strong sector, these operators would come at two loops.
}
\label{tabPot66}

%\vspace{1cm}
%
%\begin{tabular}{|c|c||c|c|}\hline%\cline{3-4}
%	\multicolumn{2}{|c||}{}&\multicolumn{2}{c|}{Strong sector} \\\cline{3-4}
%	\multicolumn{2}{|c||}{}&$\slashed{C}_2$&$C_2$\\\hline\hline
%		\multirow{2}{*}{$y_R$}&
%		$\slashed{C}_2$ & A : 11+1	& B : 5+5	\\\cline{2-4}
%	&	$C_2$		& C : 11+0	& D : 5+2	\\\hline
%\end{tabular}\\
%	\caption{{\bf This Table is going to be suppressed if no one puts any decent comment 
%	about it in the text}
%	Number of 1 loop + 2 loop independent invariants contributing to the Higgs potential, up to order $y_{L,R}^4$, in the case $\mathbf{r}_{Q,T}={\mathbf{6}}$ depending on the $C_2$ symmetry of the strong sector and of the right coupling ($\theta=0$ or $\neq0$).}
%	\label{tabNumContrib66}
\end{table}

\subsubsection*{Fermion contributions with $\{\mathbf{r}_{Q},\,\mathbf{r}_{T}\}=\{\mathbf{20'},\,\mathbf{1}\}$:}

The $\mathbf{20'}$ representation is the symmetric and traceless product of 
two $\mathbf{6}$, and it decomposes under $\SO(4)\times \SO(2)$ as
\beq
\mathbf{20'}\,=\,({\mathbf{9}},{\mathbf{1}})\oplus ({\mathbf{4}},{\mathbf{2}}) \oplus ({\mathbf{1}},{\mathbf{2}})
\oplus ({\mathbf{1}},{\mathbf{1}})\,.
\label{dec20}
\eeq
Operators in this representation and $X=2/3$ can be coupled to $q_L$ as in the case of the $\mathbf{6}$. 
Unlike that case, however, we now have two four-plets of 
$\SO(4)$ to which the doublet could mix. The $y_L$-spurion's VEV is therefore not uniquely determined in general. 
Assuming the VEV to be either $C_1P$ or  $C_2$ invariant uniquely fixes the embedding,
\begin{equation}
\label{ylvev}
	\renewcommand\arraystretch{1.2}
\left(y_L^{{\overline\alpha}}\right)^{IJ} =y_L
\scalebox{0.8}{$
\left\{ 
\left( 
\begin{tabular}{c|cc}
	$0_{4\times4}$ & $(\vec v^{\,\bar1})^T$ & $0_{4\times1}$ \\ \hline
	$\vec v^{\,\bar1}$	& \multicolumn{2}{c}{\multirow{2}{*}{$0_{2\times2}$}} 
	\\
	$0_{1\times4}$	&&
\end{tabular}
\right)
~,~
\left( 
\begin{tabular}{c|cc}
	$0_{4\times4}$ & $(\vec v^{\,\bar2})^T$ & $0_{4\times1}$ \\ \hline
	$\vec v^{\,\bar2}$	& \multicolumn{2}{c}{\multirow{2}{*}{$0_{2\times2}$}} 
	\\
	$0_{1\times4}$	&&
\end{tabular}
\right)
\right\}\,,
$}
\end{equation}
so that, as for the $y_R$ spurion in the previous $\bf \{6,6\}$ case, imposing the VEV to respect 
one of the symmetries automatically implies the other. Unlike for the $\bf \{6,6\}$, we will only consider the  $C_2$, $C_1P$ symmetric  yukawa of eq.~(\ref{ylvev}) .
%For simplicity, and because it doesn't affect the following discussion, we have aligned the VEV along the direction $5$.
%We might have easily considered the 
%general VEV of $y_R$, which would have made  an angle appear in the above equation 
%similarly to eq.~(\ref{genR}), but this will not be needed for the applications of the 
%following section {\red DON'T UNDERSTAND!}.
Out of $y_L$ we build 
$\left(\overline\Upsilon_L\right)^{ IJKL}$, which has now four indices, and classify the 
$\SO(4)\times \SO(2)$-invariants. Fortunately, as discussed in the following section, the leading 
order terms will be sufficient for our purposes, and are shown in table \ref{tabPot201}. 
The number of independent invariants is again obtained by counting, given the decomposition 
in eq.~(\ref{dec20}), the $\SO(4)\times \SO(2)$ singlets one can form with two $\mathbf{20'}$s. 
There are $6$ of them, one of which however should be removed given that it corresponds 
to the trivial $\SO(6)$ invariant which does not contribute to the potential. 
Finally, the $y_R$ spurion will not contribute to 
the potential because the coupling of $t_R$ 
with an $\SO(6)$ singlet does not break the NGB symmetry. As already mentioned, 
$t_R$ could even be a completely composite state,  corresponding to $y_R\rightarrow g_\rho$.
%and no $y_R$ coupling would be introduced at all. 
%Considering this case corresponds, in the present and in the following section, to the limit 
%of maximal compositeness $y_R\rightarrow g_\rho$.

The results are similar to the ones obtained in the case of the $\mathbf{6}$: at the $y^2$ order 
imposing any one of the discrete symmetries automatically implies the other and also  $\SO(4)$ 
invariance. Moreover, $C_1P\cdot C_2$ is an accidental symmetry of the potential. Unlike the case of two ${\bf{6}}$, spurionic and residual 
symmetries coincide because we chose to restrict to a spurion $y_L$ that preserves $C_1P$ and $C_2$.
\begin{table}[!hbp]
\centering
\renewcommand{\arraystretch}{1.5}
\scalebox{1.0}{
\begin{tabular}{|c|ll||c|c||c|}\cline{2-6}
	\multicolumn{1}{c|}{}& \multicolumn{2}{|c||}{\multirow{3}{*}{Operator}} & \multicolumn{2}{c||}{Spurionic} &	Residual	\\
\multicolumn{1}{c|}{}& & &
\multicolumn{2}{|c||}{Parity}& $SO(4)$ \\\cline{4-5}
\multicolumn{1}{c|}{}& & &
$C_2$& $C_1P$&  \\\hline
% I(1,0) 1
\multirow{5}{*}{$y_L^2$} &
$\mathcal{I}_{(0,1)}^1$ & $=\delta_{ij}\delta_{kl}(\overline\Upsilon_L^{~\mathbf{20'}})^{ijkl}$ & 
$+$ & $+$& $\checkmark$\\ \cdashline{2-6}[1pt/1pt]
% I(1,0) 2
&$\mathcal{I}_{(0,1)}^2$ & $=\delta_{ik}\delta_{jl}(\overline\Upsilon_L^{~\mathbf{20'}})^{ijkl}$ & 
$+$ & $+$& $\checkmark$\\ \cdashline{2-6}[1pt/1pt]
% I(1,0) 3
&$\mathcal{I}_{(0,1)}^3$ & $=\delta_{\alpha\gamma}\delta_{\beta\delta}(\overline\Upsilon_L^{~\mathbf{20'}})^{\alpha\beta\gamma\delta}$ & 
$+$ & $+$& $\checkmark$\\ \cdashline{2-6}[1pt/1pt]
% I(1,0) 4
&$\mathcal{I}_{(0,1)}^4$ & $=\epsilon_{\alpha\gamma}\delta_{\beta\delta}(\overline\Upsilon_L^{~\mathbf{20'}})^{\alpha\beta\gamma\delta}$ & 
$-$ & $-$& $\times$\\ \cdashline{2-6}[1pt/1pt]
% I(1,0) 5
&$\mathcal{I}_{(0,1)}^5$ & $=\epsilon_{\alpha\gamma}\delta_{ij}(\overline\Upsilon_L^{~\mathbf{20'}})^{i\alpha j\beta}$ & 
$-$ & $-$& $\times$\\ \hline
\end{tabular}
}
\caption{The independent invariants that contribute to the Higgs potential, up to order $y_{L,R}^2$ for $\{\mathbf{r}_{Q},\mathbf{r}_T\}=\{{\mathbf{20'}},\mathbf{1}\}$.
For each operator, the first two columns contain its spurionic $C_2$ and $C_1P$ parities, the third one indicates whether it will respect the $\SO(4)$ symmetry 
after the spurions will have taken VEV.}
\label{tabPot201}
\end{table}

\subsubsection*{Gauge Contributions:}
\begin{table}[!hbp]
\centering
\renewcommand{\arraystretch}{1.5}
\scalebox{1.0}{
\begin{tabular}{|c|ll||c|c||c|}\cline{2-6}
	\multicolumn{1}{c|}{}& \multicolumn{2}{|c||}{\multirow{3}{*}{Operator}} & \multicolumn{2}{c||}{Spurionic} &	Residual	\\
\multicolumn{1}{c|}{}& & &
\multicolumn{2}{|c||}{Parity}& $SO(4)$ \\\cline{4-5}
\multicolumn{1}{c|}{}& & &
$C_2$& $C_1P$&  \\\hline
% Ig 1
\multirow{3}{*}{$g^2$} &
$\mathcal{I}_g^1$ & $=\delta_{ik}\delta_{jl}(\overline\Gamma_{g})^{ijkl}$ &
$+$ & $+$& $\checkmark$\\ \cdashline{2-6}[1pt/1pt]
% Ig 2
&$\mathcal{I}_g^2$ &
$=\delta_{\alpha\gamma}\delta_{\beta\delta}(\overline\Gamma_{g})^{\alpha\beta\gamma\delta}$
&
$+$ & $+$& $\checkmark$\\ \cdashline{2-6}[1pt/1pt]
% Ig 3
&$\mathcal{I}_g^3$ & $=\epsilon_{ijkl}(\overline\Gamma_{g})^{ijkl}$ &
$+$ & $+$& $\checkmark$\\ \hline\hline
% Ig' 1
\multirow{4}{*}{${g'}^2$} &
$\mathcal{I}_{g'}^1$ &
$=\delta_{ik}\delta_{jl}(\overline\Gamma^+_{g'})^{ijkl}$ &
$+$ & $+$& $\times$\\ \cdashline{2-6}[1pt/1pt]
% Ig' 2
&$\mathcal{I}_{g'}^2$ &
$=\delta_{\alpha\gamma}\delta_{\beta\delta}(\overline\Gamma^+_{g'})^{\alpha\beta\gamma\delta}$
&
$+$ & $+$& $\times$\\ \cdashline{2-6}[1pt/1pt]
&% Ig' 3
$\mathcal{I}_{g'}^3$ & $=\epsilon_{ijkl}(\overline\Gamma^+_{g'})^{ijkl}$ &
$+$ & $+$& $\checkmark$\\ \cdashline{2-6}[1pt/1pt]
&% Ig' 4
$\mathcal{I}_{g'}^2$ &
$=\epsilon_{\alpha\beta}(\overline\Gamma^-_{g'})^{\alpha\beta}$ &
$-$ & $-$& $\times$\\ \hline
\end{tabular}
}
\caption{Gauge contributions to the potential, constructed with the dressed spurions $\overline\Gamma_{g}$, $\overline\Gamma^+_{g'}$, and $\overline\Gamma^-_{g'}$. For each operator, the first two columns contain the intrinsic 
$C_2$ and $C_1P$ parities, and the third one indicates whether it will respect the $\SO(4)$ symmetry 
after the spurions have acquires VEVs.}
\label{tabPotGauge}
\end{table}

Let us now discuss the gauge contributions to the potential, where  few modifications of the above  procedure will be needed. The starting point are now the couplings of the elementary 
$\SU(2)_L\times \U(1)_Y$ gauge fields ($W$ and $B$) to the strong sector,
given by
\begin{equation}
\displaystyle
	{\cal L}_\textrm{gauge} =
	         -\,W_{\mu\,{\overline a}} \left(g^{{\overline a}}\right)^{JI}\, J^\mu_{IJ}\,-\,
	         B_\mu\left({g'}\right)^{JI}\, J^\mu_{IJ}\,-\,
	         B_\mu \, g'_X\, J^\mu_{X}\,,
	         \label{gmix}
\end{equation}
where $J^\mu_{X}$ denotes the $\U(1)_X$ current while $J^\mu_{IJ}$ is defined, in terms of 
the $\SO(6)$ currents $J^\mu_{A}$, by 
$$
J^\mu_{IJ}\,\equiv\, J^\mu_{A}\,T^A_{IJ}\,.
$$
The Lagrangian in eq.~(\ref{gmix}) has precisely the same structure of eq.~(\ref{lmix});
it describes the coupling, due to the partial gauging of the strong sector's global group, 
of the elementary gauge fields to the global currents.  These couplings, {\it{i.e.}} VEVs of the 
spurions $g$ and $g'$ in eq.~(\ref{gmix}), are determined by identifying the 
$\SU(2)_L$ SM 
group factor  with the $\SU(2)_L$ (in the notation of eq.~(\ref{gen})) subgroup of $\SO(6)$ and 
hypercharge with $T_R^3+X$. They are given by
\beq
\displaystyle
\left(g^{{\overline a}}\right)^{IJ}\,=\,g\,\left(T_L^{\overline a}\right)^{IJ}\,,
\;\;\;\;\;
\left(g'\right)^{IJ}\,=\,g'\,\left(T_R^{3}\right)^{IJ}\,,
\;\;\;\;\;
g'_X\,=\,g'\,.
\label{sgvev}
\eeq
where $g$ and $g'$ are the $\SU(2)_L$ and $\U(1)_Y$ gauge couplings.

The $g$-spurion $\left (g^{{\overline a}}\right)^{IJ}$ has, on top of the antisymmetric $[I,\,J]$ pair  of $\SO(6)$ indices, 
an extra  ${\overline a}=1,2,3$ index in the adjoint of the elementary 
$\SU(2)^{\rm el}_L$. One can easily see that invariants can be formed by either contracting the $g$-spurion with itself or  with at least two powers of the $y_L$-spurion. The
second possibility, corresponds, however, to higher-order terms.  We thus conclude  that, at leading order, the $g$-spurion can only 
enter  the potential through the combination
\beq
\left(\Gamma_g\right)^{IJKL}\,\equiv\,\left(g_{{\overline a}}\right)^{IJ}\left(g^{{\overline a}}\right)^{KL}\,,
\label{g}
\eeq
which is obviously antisymmetric in the $[I,\,J]$ and $[K,\,L]$ indices and symmetric under the simultaneous 
exchange of the $[I,\,J]$ and $[K,\,L]$ pairs.
For what concerns  $\left(g'\right)^{IJ}$ and $g'_X$, because of the symmetry $B_\mu\rightarrow - B_\mu$, 
$g'\rightarrow -g'$, they can only enter via two combinations
\beq
\left(\Gamma_{g'}^+\right)^{IJKL}\,\equiv\,\left(g'\right)^{IJ}\left(g'\right)^{KL}\,,\;\;\;\;\;
\left(\Gamma_{g'}^-\right)^{IJ}\,\equiv\,\left(g'\right)^{IJ}g'_X\,,
\label{gp}
\eeq
having ignored the $(g'_X)^2$ term that, being an $\SO(6)$ singlet will not contribute to the potential. 

Starting from the building blocks in eqs.~(\ref{g},\ref{gp}), we can classify   
the possible contributions to the potential in terms of the $\SO(4)\times \SO(2)$ invariants that 
can be built out of $ \Gamma_{g}$ and $ \Gamma_{g'}^\pm$. At the leading order, it is very simple 
to count the invariants, if one remembers that each $[I,\,J]$ and $[K,\,L]$ pair actually forms 
a single index in the adjoint, and that the adjoint decomposes as
\begin{equation}
\mathbf{15} =(\mathbf{6},\mathbf{1})\oplus
(\mathbf{4},\mathbf{2})\oplus
(\mathbf{1},\mathbf{1})\, , 
\end{equation}
under  $\SO(4)\times \SO(2)$. Out of $\Gamma_{g}$, which is the product of two 
$\mathbf{15}$, $4$ invariants can be formed \footnote{One should not forget, to perform the correct counting, that the
$\mathbf{6}$ is reducible
because it coincides with the adjoint $(\mathbf{3},\mathbf{1})\oplus
(\mathbf{1},\mathbf{3})$ 
in the $\SU(2)_L\times \SU(2)_R$ notation. Also, one of the two invariants one could form with 
two $(\mathbf{4},\mathbf{2})$ actually vanishes because it is antisymmetric.} 
one of which is however trivial, being associated 
to the $\SO(6)$ invariant. The same applies to $\Gamma_{g'}^+$, while $\Gamma_{g'}^-$ only 
leads to one invariant, associated to the unique singlet in the decomposition of the $\mathbf{15}$. 
We are therefore left with $7$ invariants, that are listed in table~\ref{tabPotGauge} together 
with their  $C_1P$ and $C_2$ parities.

Concerning $C_1P$, few more comments are needed. We defined $C_1P$ to act as the 
standard $CP$ conjugation in the elementary sector; it therefore acts on the gauge fields as
\beq
\displaystyle
W_{\overline a}\,\rightarrow\,-\,\left(-\right)^{\delta^2_{{\overline a}}} W_{\overline a}^{(P)}\,,
\;\;\;\;\;\;\;\;\;\;
B\,\rightarrow\,-B^{(P)}\,,
\label{elcp}
\eeq
where the ``${}^{(P)}$'' superscript denote the action of ordinary parity. On the strong sector, $C_1P$
acts again as ordinary $CP$, but convoluted with the ${\mathcal C}_1$ $\SO(6)$ 
rotation defined in eq.~(\ref{c12}). Under ordinary $CP$ each representation goes into its conjugate, so that 
the strong sector's currents transform as \footnote{The resulting signs simply follow from the fact that the generators of $\SO(6)$ are purely imaginary, while that of $\U(1)_X$ is real.}
\beq
\displaystyle
J_{IJ}\,\rightarrow\,-\,\left(J_{IJ}^{(P)}\right)^*\,=\,J_{IJ}^{(P)}\,,
\;\;\;\;\;\;\;\;\;\;
J_{X}\,\rightarrow\,-\,\left(J_{X}^{(P)}\right)^*\,=\,-\,J_{X}^{(P)}\,.
\label{stcp}
\eeq
Convoluting the above equation with ${\mathcal C}_1$ and making use of eq.~(\ref{elcp}), we 
immediately find that the spurions transform as
\bea
\displaystyle
&&  \left(g^{{\overline a}}\right)^{IJ}\,\rightarrow\,-\,\left(-\right)^{\delta^2_{{\overline a}}}
\left({\mathcal C}_1^{\mathbf 6}\right)^{I}_{\ \ K}
\left({\mathcal C}_1^{\mathbf 6}\right)^{J}_{\ \ L}
\left(g^{{\overline a}}\right)^{{K}\,{L}}\,,\nn\\
&&\left(g'\right)^{IJ}\,\rightarrow\,-\,
\left({\mathcal C}_1^{\mathbf 6}\right)^{I}_{\ \ K}
\left({\mathcal C}_1^{\mathbf 6}\right)^{J}_{\ \ L}
\left(g'\right)^{{K}\,{L}}\,,\nn\\
&&g'_X\,\rightarrow\,g'_X\,.
\label{eqC1Pgauge}
\eea
It is straightforward at this point to check that the spurion's VEVs in eq.~(\ref{sgvev}) are invariant, meaning 
that the gauging of the SM group does not break $C_1P$ (provided that it was present 
as a symmetry of the strong sector). 
From the above equation one can also derive the $C_1P$ action on $\Gamma_{g}$ and on 
$\Gamma_{g'}^\pm$. Up to the ${\mathcal C}_1$ rotation, we find that $\Gamma_{g}$ and  
$\Gamma_{g'}^+$ are even while $\Gamma_{g'}^-$ is odd; this explains the results of 
table~\ref{tabPotGauge}. 
%As a final comment, we remark that not all the operators of table~\ref{tabPotGauge} can 
%be generated at the one-loop level, in spite of being of the single-trace form. This is a 
%difference with the case of fermions, in which instead all the single-trace operators are 
%generically generated at one loop, and it is due to gauge invariance. Gauge invariance  
%indeed fixes uniquely, at the leading order in the derivative expansion, the Higgs 
%interactions with the gauge fields and allows few invariant structures to be 
%generated at one loop. Of all the operators of table~\ref{tabPotGauge}, only 
%$2$ {\bf CHECK!!!} combinations are generated and these are 
%$\mathcal{I}_g^1+\mathcal{I}_g^2$ and $\mathcal{I}_{g'}^1+\mathcal{I}_{g'}^2$. 
%The others are suppressed by the factor $g_\rho^2/16\pi^2\simeq 1/N_c$ with respect 
%to the estimate of eq.~(\ref{pot1}).

\subsubsection{$C_2$ Invariant Models}\label{secC2model}

Armed with the technical tools of the previous section, we now describe some 
specific composite-Higgs scenarios, based on the $\SO(6)\to \SO(4)\times \SO(2)$ 
symmetry breaking pattern. In this section we will focus on the case in which the 
strongly interacting sector also possesses the additional $C_2$ symmetry, while 
the case of $C_1P$ invariance will be described in the following one. Also, we 
restrict to the case of $\mathbf{r}_{Q,T}={\mathbf{6}}$, in which left- and right-handed
elementary fermions couple to composite operators in the ${\mathbf{6}}$; 
however the choice of $\{\mathbf{r}_{Q},\,\mathbf{r}_{T}\}=\{\mathbf{20'},\,\mathbf{1}\}$ 
might equally well be considered. Notice that $C_2$-invariance of the strong sector 
does not imply that this is an exact symmetry, given that it might be broken explicitly 
by the coupling of elementary fermions or spontaneously by the VEV of the 
second Higgs $\Phi^\due$; both possibilities will be considered in what follows. 

%For the up-type sector, the structure of elementary-composite fermion couplings is given by eq.~(\ref{lmix}), or better 
%to its obvious generalization to the different families $q_L^f$ and $u_R^f$. The most general form of the couplings 
%which is allowed by the SM group reads
Introducing flavor indices in eq. (\ref{genL}), $f=u,c,t$, we have for the up sector,  
\begin{equation}\renewcommand\arraystretch{1.3}\begin{array}{rcl}
\left({\left(y_L\right)^\textrm{u}_f}^{\overline{\alpha}}\right)^{I}&=&
\dfrac{\left(y_L\right)^\textrm{u}_f}{\sqrt{2}}\,\left\{ \left( \vec v^{\,\bar1},0,0 \right),\left( \vec v^{\,\bar2},0,0 \right) \right\}\,,\\
\left(\left(y_R\right)^\textrm{u}_f\right)^I&=&\left(y_R\right)^\textrm{u}_f\,\left( 0,0,0,0,\cos{\theta^{\textrm{u}}_f},i\,e^{i\phi^{\textrm{u}}_f} \sin{\theta^{\textrm{u}}_f} \right)\,,
\label{emu6}
\end{array}\end{equation}
where $\left(y_L\right)^\textrm{u}_f$ and $\left(y_R\right)^\textrm{u}_f$ have been made real by, respectively, a $\U(1)_L^{\rm el}$ and  $\U(1)_R^{\rm el}$ flavor-dependent rotation of the elementary fields. The vectors $\vec v^{\,\bar1,2}$ are  defined in eq. \eqref{embeddingleft}. By an $\SO(2)$ rotation in the strong sector group one can also eliminate, as we did in eq.~(\ref{genR}), the phase $\phi^{\textrm{u}}_t$ associated to the
top quark, while the others remain physical. With more than one family, therefore, $C_1P\cdot C_2$ is not any longer automatically preserved (provided it
was a symmetry of the strong sector to start with) by the up-type couplings. 

The discussion is easily extended to the down-type Yukawa coupling. 
If, as we assume,  the right-handed down quarks are coupled to a singlet of custodial symmetry, 
the Yukawas can be generated by coupling the left-handed  SM doublet to a second operator  in the $\bf\mathbf (2,2)_{-1/3}$ representation of $\SU(2)_L\times \SU(2)_R\times \U(1)_X$.
This can be realized with fermionic operators, ${\cal O'}_{ I_{Q'} }$, again in the ${\mathbf 6}$ representation, but with $X=-1/3$  charge:
\begin{equation}
	{\cal L}_\textrm{mix}= 
		(\bar q_L)_{\overline{\alpha}} \left({y_L}^{\overline{\alpha}}\right)^{I_{Q'}} {\cal O'}_{ I_{Q'} } 	+
		(\bar b_R) \left(y_R\right)^{I_B} {\cal O}_{I_B} 			+
		\textrm{h.c.}\,,
		\label{dwn}
\end{equation}
where the flavor indices are understood. The embedding of the SM quarks in the representations above is given by
\begin{equation}\renewcommand\arraystretch{1.3}\begin{array}{rcl}
\left({\left(y_L\right)^\textrm{d}_f}^{\overline{\alpha}}\right)^{I}&=&
\dfrac{\left(y_L\right)^\textrm{d}_f}{\sqrt{2}}\,\left\{ \left( (\vec v^{\,\bar2})^*,0,0 \right),\left( (\vec v^{\,\bar1})^*,0,0 \right) \right\}\,,\\
\left(\left(y_R\right)^\textrm{d}_f\right)^I&=&\left(y_R\right)^\textrm{d}_f\,\left( 0,0,0,0,\cos{\theta^{\textrm{d}}_f},i\,e^{i\phi^{\textrm{d}}_f} \sin{\theta^{\textrm{d}}_f} \right)\,,
\end{array}
\label{emd6}
\end{equation}
where $f=d,s,b$, and 
$\left(y_{R}\right)^\textrm{d}_f$ has been made real by a $\U(1)^{\rm el}_R$ elementary rotation of the $d_R^f$ quarks. Of the remaining phases, 
one could be eliminated by a $\U(1)_X$ elementary rotation while all the others are physical.
We stress that, as discussed in section~\ref{flavorstruct}, a generic choice with flavor dependent ${\theta^{\textrm{d}}_f},{\theta^{\textrm{u}}_f},
{\phi^{\textrm{d}}_f},{\phi^{\textrm{u}}_f}$ would lead to Higgs-mediated FCNC.
%In the above equation the couplings are assumed to be diagonal in flavor space in the same basis where the couplings 
%of the up quarks are. This cannot be achieved in general through a rotation of the elementary fields but is necessary in order to
%avoid larger FCNC.

The only $C_2$-invariant choice of the couplings is $\theta^{\textrm{u},\textrm{d}}_f=0$, 
while, by taking $\theta^{\textrm{u}}_f=0$ and 
$\theta^{\textrm{d}}_f=\pi/2$, we preserve $C_I\cdot C_2$, where  $C_I$ is the isospin parity defined in section~\ref{flavorstruct} under which the $d_R^f$
elementary quarks change sign~\footnote{Together with $\theta^{\textrm{d}}_f=\pi/2$ one can also take, by field redefinitions, 
$\phi^{\textrm{d}}_{f}=0$.}. Similarly, one could also introduce the lepton couplings and show that all the different scenarios of table~\ref{types} can be implemented by suitable choices of the mixing angles. In each of these scenarios, as discussed in section~\ref{flavorstruct}, large Higgs-mediated FCNC are avoided. In the present framework, however, the type-I scenario results more natural than the others because it is the only one that does not require the second Higgs to take a VEV in order for the masses to be generated after EWSB. As we will see, the second Higgs acquiring (a not too large) VEV requires a certain additional fine-tuning.

Having introduced the general framework, let us now discuss its possible vacuum structures, which are determined  by the 
form of the Higgs potential. The latter is insensitive to the light fermion couplings, so that the discussion which follows is independent of the choices of the mixing angles and phases, apart of course from those of the top quark in eqs.~(\ref{genL},\ref{genR}). 
We denote, as in the previous section, $\theta^{\rm u}_t$ as $\theta$.
Having assumed $C_2$-invariance of the strong sector, the allowed contributions to the potential are the intrinsic $C_2$-even operators listed in tables~\ref{tabPot66} and \ref{tabPotGauge}. 
As already explained, no accidental symmetry is expected 
in the general case in which $\theta\neq0$ in eq.~(\ref{genR}) 
as both $C_1P$ and $\SO(4)$ are broken by the top quark proto-Yukawas.
But, by choosing the top-quark coupling to respect $C_2$ ({\it{i.e.}}, $\theta=0$), the potential becomes separately invariant under $C_1 P$ and $C_2$. Moreover, at the leading $y^2$ order one 
obtains accidental $\SO(4)$.
%As already mentioned, these operators accidentally respect $C_1P$, even though this might not at all be a symmetry of the strong sector. 
%Given that the top mixings in eq.s~(\ref{genL},\ref{genR}) unavoidably respect $C_1P\cdot C_2$, the latter symmetry is accidentally present in the Higgs potential to the order we are working at. 
%Moreover, if choosing the top-quark mixing to respect $C_2$ ({\it{i.e.}}, $\theta=0$), the potential becomes separately invariant under both  $C_1 P$ and $C_2$. 
Besides these important symmetry considerations, we can make a more concrete use of the results in tables~\ref{tabPot66} and \ref{tabPotGauge}. 
Expanding in powers of $\Phi_{\uno,\,\due}/f$, each invariant will give a specific contribution to the parameters of the general renormalizable 2HDM potential
\begin{equation}\renewcommand\arraystretch{2.0}\begin{array}{rl}
	V(\Phi_\uno, \Phi_\due) ~=& \phantom{+}
\dfrac{1}{2} m_{11}^2 \Tr[\Phi_\uno^{\dagger} \Phi_\uno] + \dfrac{1}{2} m_{22}^2 \Tr[\Phi_\due^{\dagger} \Phi_\due] 
{+}
\dfrac{1}{2} \Tr[\Phi_\uno^{\dagger} \Phi_\due (m_{12}^2 + i \, \tilde{m}_{12}^2 \sigma_3)] \\
& + \dfrac{1}{4} \lambda_1 \Trs[\Phi_\uno^{\dagger} \Phi_\uno] + \dfrac{1}{4} \lambda_2 \Trs[\Phi_\due^{\dagger} \Phi_\due] + \dfrac{1}{4} \lambda_3 \Tr[\Phi_\uno^{\dagger} \Phi_\uno] \Tr[\Phi_\due^{\dagger} \Phi_\due]  \\
&+ \dfrac{1}{4} \lambda_4 \Trs[\Phi_\uno^{\dagger} \Phi_\due] + \dfrac{1}{4} \tilde{\lambda}_4 \Trs[\Phi_\uno^{\dagger} \Phi_\due \sigma_3] + i \, \dfrac{1}{4} \lambda_5 \Tr[\Phi_\uno^{\dagger} \Phi_\due] \Tr[\Phi_\uno^{\dagger} \Phi_\due \sigma_3]  \\
& + \dfrac{1}{4} \Tr[\Phi_\uno^{\dagger} \Phi_\uno] \Tr[\Phi_\uno^{\dagger} \Phi_\due (\lambda_6 + i \tilde{\lambda}_6 \sigma_3) ] + \dfrac{1}{4} \Tr[\Phi_\due^{\dagger} \Phi_\due] \Tr[\Phi_\uno^{\dagger} \Phi_\due (\lambda_7 + i \tilde{\lambda}_7 \sigma_3)]\, .
\label{2HDMV}
\end{array}\end{equation}
These contributions are summarized in tables~\ref{contf} and \ref{contg}; it is understood that each term is proportional to the corresponding coefficient $c_{\delta}^{(n_R,n_L)}$ appearing in eq.~(\ref{pot1}). 
From the tables we see that the coefficient $\lambda_5$ is not generated at the order we are working, since it breaks $\SO(4)$ while being $C_1P$ odd and $C_2$ even. The first contribution to $\lambda_5$  comes at order $y_L^4 y_R^2$.
For $\theta=0$, $m_{12}^2$, ${\tilde m}_{12}^2$, $\lambda_{6,7}$ and ${\tilde\lambda}_{6,7}$ also vanish due to the separate $C_2$ and accidental $C_1P$ symmetries.

\begin{table}[!tb]
	\newcommand{\jloop}{\left( \dfrac{g_\rho}{4\pi} \right)^2}
	\newcommand{\tY}{\theta}
	\begin{center}
		\renewcommand\arraystretch{2.1}
		%\rotatebox{90}
		{\scalebox{0.7}{
		\begin{tabular}{|c||c|c||c|c|c||c|c|c||c|c|} \hline
operator & 
$\mathcal{I}^{1}_{(0,1)}$ & $\mathcal{I}^{1}_{(1,0)}$ & 
$\mathcal{I}^{1}_{(2,0)}$ & $\mathcal{I}^{4}_{(2,0)}$ & $\mathcal{I}^{5}_{(2,0)}$ &
$\mathcal{I}^{1}_{(1,1)}$ & $\mathcal{I}^{5}_{(1,1)}$ & $\mathcal{I}^{6}_{(1,1)}$ &
$\mathcal{I}^{1}_{(0,2)}$ & $\mathcal{I}^{4}_{(0,2)}$ \\

\hline
$\dfrac{1}{16\pi^2}\times$ &
$-\dfrac{y_L^2g_\rho^2}{2}$ & $y_R^2g_\rho^2$ &
$\dfrac{y_R^4}{4}$ & $\dfrac{y_R^4}{4}\jloop$ & $\dfrac{y_R^4}{4}\jloop$ &
$\dfrac{y_R^2y_L^2}{4}$ & $y_R^2y_L^2\jloop$ & $-y_R^2y_L^2\jloop$ &
$-\dfrac{y_L^4}{2}$ & $-y_L^4\jloop$
\\\cdashline{1-11}[1pt/1pt] %%%%%%%%%%%%%%%%%%%%%%%%%%%%%%%%%
$m_{11}^2/f^2$ &
$1$ & $\cos^2\tY$ &
$0$ & $0$ & $0$ &
$\cos^2\tY$ & $\cos^2\tY$ & $0$ &
$1$ & $1$
\\
$m_{22}^2/f^2$ &
$1$ & $\sin^2\tY$ &
$0$ & $0$ & $0$ &
$\sin^2\tY$ & $\sin^2\tY$ & $0$ &
$1$ & $1$
\\
$m_{12}^2/f^2$ &
$0$ & $0$ &
$0$ & $0$ & $\sin4\tY$ &
$0$ & $0$ & $0$ &
$0$ & $0$
\\
$\tilde{m}_{12}^2/f^2$ &
$0$ & $0$ &
$0$ & $0$ & $0$ &
$-\sin2\tY$ & $0$ & $\dfrac{1}{2}\sin2\tY$ &
$0$ & $0$

\\ \cdashline{1-11}[1pt/1pt] %%%%%%%%%%%%%%%%%%%%%%%%%%%%%%
$\lambda_1$ &
$-\dfrac{1}{3}$ & $-\dfrac{1}{3}\cos^2\tY$ &
$2 \cos^4\tY$ & $2 \cos^4\tY$ & $0$ &
$-\dfrac{4}{3}\cos^2\tY$ & $-\dfrac{7}{12}\cos^2\tY$ & $0$ &
$-\dfrac{7}{12}$ & $-\dfrac{11}{24}$
\\
$\lambda_2$ &
$-\dfrac{1}{3}$ & $-\dfrac{1}{3}\sin^2\tY$ &
$2 \sin^4\tY$ & $2 \sin^4\tY$ & $0$ &
$-\dfrac{4}{3}\sin^2\tY$ & $-\dfrac{7}{12}\sin^2\tY$ & $0$ &
$-\dfrac{7}{12}$ & $-\dfrac{11}{24}$

\\
$\lambda_3$ &
$0$ & $0$ &
$\sin^2\tY$ & $-\sin^2\tY$ & $0$ &
$0$ & $-\dfrac{1}{4}$ & $0$ &
$0$ & $-\dfrac{1}{4}$

\\
$\lambda_4$ &
$-\dfrac{2}{3}$ & $-\dfrac{1}{3}$ &
$0$ & $2 \sin^22\tY$ & $0$ &
$-\dfrac{4}{3}$ & $-\dfrac{1}{3}$ & $0$ &
$-\dfrac{7}{6}$ & $-\dfrac{2}{3}$

\\
$\tilde{\lambda}_4$ &
$0$ & $0$ &
$0$ & $0$ & $0$ &
$0$ & $0$ & $0$ &
$\dfrac{1}{2}$ & $0$

\\
$\lambda_5$ &
$0$ & $0$ &
$0$ & $0$ & $0$ &
$0$ & $0$ & $0$ &
$0$ & $0$

\\
$\lambda_6$ &
$0$ & $0$ &
$0$ & $0$ & $-\dfrac{1}{3}\sin4\tY$ &
$0$ & $0$ & $0$ &
$0$ & $0$

\\
$\tilde{\lambda}_6$ &
$0$ & $0$ &
$0$ & $0$ & $0$ &
$\dfrac{2}{3}\sin2\tY$ & $0$ & $-\dfrac{1}{12}\sin2\tY$ &
$0$ & $0$

\\
$\lambda_7$ &
$0$ & $0$ &
$0$ & $0$ & $-\dfrac{1}{3}\sin4\tY$ &
$0$ & $0$ & $0$ &
$0$ & $0$

\\
$\tilde{\lambda}_7$ &
$0$ & $0$ &
$0$ & $0$ & $0$ &
$\dfrac{2}{3}\sin2\tY$ & $0$ & $-\dfrac{1}{12}\sin2\tY$ &
$0$ & $0$

\\

\hline

\end{tabular}}}

	\end{center}	
	\caption{Contribution to the parameters of the general 2HDM potential \eq{2HDMV} from fermions in the \textbf{6}. 
	%We parametrize the couplings as $y_R^+ = y_R \cos \theta$ and $y_R^- = y_R \sin \theta$. 
The individual contributions of the $\SO(6)/\SO(4) \times \SO(2)$ operators of table~\ref{tabPot66} are shown. The first line indicates the power-counting estimate of the pre-factor.}
\label{contf}
\end{table}

\begin{table}[h!]
\begin{center}
	   \renewcommand{\arraystretch}{2.1}
	   \scalebox{1.0}{
		\begin{tabular}{|c||c|c|c||c|c|c|} \hline
operator & 
$\mathcal{I}^{1}_{g}$ & $\mathcal{I}^{2}_{g}$ & $\mathcal{I}^{3}_{g}$ &
$\mathcal{I}^{1}_{g'}$ & $\mathcal{I}^{2}_{g'}$ & $\mathcal{I}^{3}_{g'}$ \\

\hline
$\dfrac{g_{\rho}^2}{16\pi^2}\times$ &
$\dfrac{3}{4}g^2$ & $\dfrac{3}{2}g^2$ & $-\dfrac{1}{8}g^2$ &
$\dfrac{1}{4}{g'}^2$ & $\dfrac{1}{8}{g'}^2$ & $-\dfrac{1}{2}{g'}^2$
\\\cdashline{1-7}[1pt/1pt] %%%%%%%%%%%%%%%%%%%%%%%%%%%%%%%%%
$m_{11}^2/f^2$ &
$1$ & $0$ & $1$ &
$1$ & $0$ & $1$
\\
$m_{22}^2/f^2$ &
$1$ & $0$ & $1$ &
$1$ & $0$ & $1$
\\ \cdashline{1-7}[1pt/1pt] %%%%%%%%%%%%%%%%%%%%%%%%%%%%%%
$\lambda_1$ &
$-\dfrac{1}{3}$ & $0$ & $-\dfrac{1}{12}$ &
$-\dfrac{1}{3}$ & $0$ & $-\dfrac{1}{12}$
\\
$\lambda_2$ &
$-\dfrac{1}{3}$ & $0$ & $-\dfrac{1}{12}$ &
$-\dfrac{1}{3}$ & $0$ & $-\dfrac{1}{12}$
\\
$\lambda_3$ &
$-\dfrac{1}{3}$ & $2$ & $-\dfrac{1}{2}$ &
$0$ & $0$ & $-\dfrac{1}{2}$
\\
$\lambda_4$ &
$-\dfrac{1}{3}$ & $-2$ & $\dfrac{1}{3}$ &
$-\dfrac{2}{3}$ & $0$ & $\dfrac{1}{3}$
\\
$\tilde{\lambda}_4$ &
$0$ & $0$ & $0$ &
$1$ & $2$ & $0$
\\
\hline
\end{tabular}}
\end{center}	
\caption{Contribution to the parameters of the general 2HDM potential \eq{2HDMV} from $\SU(2)_L$ and $\U(1)_Y$ gauge bosons. The individual contributions of the $\SO(6)/\SO(4) \times \SO(2)$ operators of table~\ref{tabPotGauge} are shown. %Only $C_2$ invariant operators have been considered.
The first line indicates the NDA pre-factor.
%We notice that the contributions to the operators $\mathcal{I}^{1}_{g}-\mathcal{I}^{2}_{g}$ and $\mathcal{I}^{3}_{g}$ are suppressed by $g_{\rho}^2/16 \pi^2$ with respect to $\mathcal{I}^{1}_{g}+\mathcal{I}^{2}_{g}$. The same applies to the corresponding $g'$ operators.
}
\label{contg}
\end{table}

\subsubsection*{Composite Inert Higgs}

Let us now consider the case in which, by choosing $\theta^{\textrm{u},\textrm{d}}_f=0$, $C_2$ is  preserved by all the fermion couplings. 
Provided the potential allows $\Phi^\due$ to have zero VEV, we obtain a composite realization of the inert Higgs scenario \cite{Barbieri:2006dq}
in which the $C_2$ symmetry is completely unbroken. In this case, the lightest component of the $C_2$-odd Higgs doublet  becomes absolutely stable, 
providing a potential dark matter candidate. 
%We will see in the following that this lightest state is preferentially the neutral one, we will discuss in section~4 to what 
%extent and in which conditions it can provide a viable explanation of the DM abundance in the Universe.

Treating the coefficients $c^{(n_R,n_L)}_\delta$ as $O(1)$ free parameters, and assuming that they 
can take both signs, 
a ``large'' region of the parameter space is easily identified where 
EWSB  occurs with  the second Higgs not  taking a  VEV, so that $C_2$ is unbroken.
However, we need something more for a potentially realistic composite inert-Higgs model. 
As we explained before,  satisfying the EWPT requires the VEV of the  first Higgs to satisfy \eq{boundxi}, that, unless $g_\rho$ is maximally large, implies $\xi\ll 1$.
This can only be achieved by advocating a 
cancellation among the different  contributions to  the mass-parameter $m_{11}^2$ in eq.~(\ref{2HDMV}), which must become negative 
and smaller than what was expected by the NDA counting of eq.~(\ref{pot1}). We see from table~\ref{contf} that at the leading $y^2$-order the two separated contributions to $m_{11}^2$ can be canceled one with each other. This cancellation, however,   also reduces the quartic $\lambda_1$, making it  useless for reducing the Higgs VEV, which remains  of order $f$. This accident, which also occurs in the models of refs.~\cite{Agashe:2004rs,Contino:2006qr}, 
renders the $O(y^2)$ potential not tunable, and is the very same
reason why we have been obliged to retain the higher order ($y^4$, $g^2$ and ${g'}^2$) contributions. 

If the higher-order terms are taken into account, the tuning becomes possible;
% and it is achieved in the following way. 
we must demand
the leading $y^2$ contributions to $m_{11}^2$ to be a factor $\xi$ smaller than the
subleading  one, those   of order  $y^4$ and $g^2$.
% Second, this tuned $y^2$ potential is subsequently tuned with the higher order one in order to obtain a small  VEV.  
In this case the quartic $\lambda_1$ is  dominated by  the 
higher-order  contributions:
\beq
\displaystyle
\lambda_1\,\sim\,\frac1{16\pi^2}\text{Max} \{N_c\,y^4_L, N_c\, y^4_R, N_c\, y^2_Ly^2_R,g^2g_\rho^2,{g'}^2g_\rho^2,\}\, ,
\label{l1}
\eeq
where the $N_c=3$ color factor has been included in the estimates of the fermion's contributions. The coefficients controlling the contributions of $y_L^2$ and $y_R^2$ to $m_{11}^2$ are plausibly expected to be  comparable. Then  a cancellation in $m_{11}^2$ would require  $y_L\sim y_R$ which implies, given eq.~(\ref{ytop}), $y_L\sim y_R\sim \sqrt{Y_t g_\rho}$.  We will assume those values for our estimates below. We stress that this choice does not create a tension with EWPT thanks to the accidental $P_{LR}$ discussed  in section~\ref{situation}. Because  $\lambda_1$ is given by \eq{l1},  we have that the lightest scalar in the spectrum  is expected to be the $C_2$-even  neutral scalar, $h$, which is 
contained in the first Higgs doublet $\Phi_\uno$. The mass of $h$ can be estimated as
\begin{equation}
\label{mhiggs1}
m_{h}^2 \sim \lambda_1 v_1^2 \sim  (250\textrm{ GeV})^2\left(\frac{3}{N}\right) 
%\left( \frac{\sqrt{Y_tg_\rho}}{y_L} \right)^4
\, ,
\end{equation}
where $N$ is defined in \eq{grhon}.
The potential of the second (inert) Higgs doublet $\Phi_\due$ is dominated by the leading order $y^2$ contribution,
which is basically fixed up to an $O(1)$ overall coefficient once a cancellation in $m_{11}^2$ is assumed
\begin{equation}
	V\simeq N_c\,\frac{ g_\rho Y_t }{16\pi^2}\,m_\rho^2 \left( 
	 	\Tr[\Phi_\due\cdot \Phi_\due] 
		- \frac{1}{6\,f^2}\Tr{}^2[\Phi_\due\cdot \Phi_\due]
		- \frac{1}{6\,f^2}\Tr{}^2[\Phi_\uno\cdot \Phi_\due] 
		\right) \,.
\label{pottt}
\end{equation}
Decomposing $\Phi_\due $ in its $\SO(3)_c$ triplet and singlet components, $H^a$ $(a=1,2,3)$ and  $H$ respectively, we see that 
the first term in the above equation gives a common contribution to the masses of all the components of order
\begin{equation}
	m_{H_2}^2 \sim  
	N_c\,\frac{g_\rho^3 Y_t}{16\pi^2} \,f^2 \simeq (1.3\textrm{ TeV})^2 % 544 GeV
	\left( \frac{3}{N} \right)^{\frac{3}{2}} \left( \frac{0.25}{\xi} \right)\, ,
	\label{massh2}
%	N_c\,\frac{g_\rho Y_t}{16\pi^2} \,m_\rho^2 \simeq (700\textrm{ GeV})^2 % 740 GeV 
%	\sqrt{\frac{3}{N}}
%	\left( \dfrac{m_\rho}{2\ {\rm TeV}} \right)^2\, ,
\end{equation}
where we have used the relation $m_{\rho} \simeq g_{\rho} f$.
The third term of \eq{pottt} induces a singlet-triplet splitting. Given that the overall sign of eq.~(\ref{pottt}) must 
be positive in order for $m_{H_2}^2$ to be positive, the sign of the splitting is fixed and the singlet $H$ 
is always lighter than the triplet $H^a$: 
\beq
m_{H}^2\simeq \left(1-\frac{\xi}{3}\right) m_{H^a}^2\, .
\label{splitt}
\eeq
As discussed in the previous section and explicitly shown in eq.~(\ref{pottt}), the $y^2$ potential is $\SO(4)$-invariant 
so that the $\SO(3)_c$ breaking splittings among the charged and neutral triplet components, 
defined respectively as $H^\pm=(H^2\pm i H^1)/\sqrt{2}$ and $A=H^3$, only come at order ${g'}^2$ and $y^4$.
These splittings can be respectively estimated as 
\begin{equation}\begin{array}{rcl}
	\left|\dfrac{m_{H^{\pm}}-m_{A}}{m_{H^{\pm}}}\right|_{g'}	
		&\sim& 	\dfrac{{g'}^2\xi}{g_\rho Y_t }
		\simeq		0.004 \, 
%				\left( \dfrac{\sqrt{Y_tg_\rho}}{y_L} \right)^2
				\sqrt{\dfrac{N}{3}}
				\left( \dfrac{\xi}{0.25} \right)\, ,	\\ \\
	\left|\dfrac{m_{H^{\pm}}-m_{A}}{m_{H^{\pm}}}\right|_{y^4}
		&\sim&  
		\dfrac{Y_t\xi}{g_\rho}
		\simeq		0.03 \, 
		%\left( \dfrac{y_L}{\sqrt{Y_tg_\rho}} \right)^2
				\sqrt{\dfrac{N}{3}}
				\left( \dfrac{\xi}{0.25} \right)\,.
\end{array}\end{equation}

\subsubsection*{Spontaneous  $C_2$ Breaking}
%Both Higgses taking a VEV
\label{spontaneousC2}

Still assuming that $C_2$  is preserved by the couplings, we now consider the 
possibility that the second Higgs $\Phi^\due$ also acquires a VEV. In this case $C_2$ is 
spontaneously broken, and the second Higgs multiplet is no longer  inert. Also, 
a VEV of $\Phi^\due$ is compulsory in order for the alternative scenarios 
(type-I, II, X and Y defined in  section~\ref{flavorstruct}) to become viable. The discussion which follows
applies to these scenarios as well. 

If the VEV of $\Phi_\due$ is non-zero, so breaking the  discrete symmetry ($C_2$ or $C_I\cdot C_2$, depending on the flavor 
embedding), large corrections to $\widehat T$  could be generated from 
the misalignment of the two VEVs, see section.~\ref{Tissue}. Avoiding these corrections was 
the very reason to advocate $C_2$, that was crucially assumed to be unbroken.  Since
we are now interested in choosing the parameters such that the vacuum is $C_2$-breaking, 
in this case the $C_2$ symmetry of the Lagrangian 
does not protect us  anymore from large corrections to $\widehat T$. Fortunately this  does not happen due to the accidental $C_1P$-invariance of the potential which 
is automatically present at the order we are working at. Because of $C_1P$, the two VEVs are aligned and 
large contributions to $\widehat T$ are avoided. We stress that it is only because of 
this accidental symmetry of the potential that the scenario of spontaneous $C_2$ breaking become 
phenomenologically viable in the present framework. The sub-leading effects that induce
$\widehat T$  come from the breaking of 
$C_1P$ in the potential. The leading contribution comes   at order $y_L^4y_R^2 $ and gives rise to  $\lambda_5\sim y_L^4y_R^2/(4\pi g_\rho)^2 \sim Y_t^3 g_{\rho}/16\pi^2$.
This generates a nonzero VEV $v_3^\due$ which can be estimated as
\begin{equation}
	v_3^\due \sim  \frac{\lambda_5(v_4^\uno)^2}{m_{22}^2}v_4^\due \sim
	0.05 ~ \sqrt{\frac{N}{3}} ~v\, ,
%	5\cdot10^{-3}% true value : 4.7 E-3
%	\left( \frac{y_L}{\sqrt{Y_tg_\rho}} \right)^2 \left( \frac{y_R}{\sqrt{Y_tg_\rho}} \right)^2 
%\left( \frac{3}{N} \right)\left( \frac{\xi}{0.25} \right) ~v\, ,
\end{equation}
where we  assumed a sizable spontaneous breaking of $C_2$, $v_4^\uno \sim v_4^\due \sim v/\sqrt{2}$.
The contributions to $\widehat T$ from this effect are under control
\begin{equation}
\widehat{T} = \frac{(v_4^\uno)^2 (v_3^\due)^2}{f^2 v^2}
\sim
	6\cdot10^{-4}
	\left( \frac{N}{3} \right) \left( \frac{\xi}{0.25} \right)\, .
%	5\cdot10^{-6}% true value : 5.6 E-3
%	\left( \frac{y_L}{\sqrt{Y_tg_\rho}} \right)^4 \left( \frac{y_R}{\sqrt{Y_tg_\rho}} \right)^4 
%\left( \frac{3}{N} \right)^2\left( \frac{\xi}{0.25} \right)^3\, .
\end{equation}

We have explicitly checked that the free parameters of our potential naturally allow for the two 
Higgses to take (aligned, as we have seen) VEVs, but the request that both VEVs are smaller than  $f$
%, $v_{\uno,\due}\ll f$,
clearly requires  fine-tuning. 
The amount of fine-tuning is the square of that in 
the single-VEV case because both Higgs mass terms $m_{11}^2$  and $m_{22}^2$ need 
now to be reduced independently, such that  $O(y^2)$ terms in the potential  are comparable to $O(y^4)$ terms. 
Looking at tables~\ref{contf} and \ref{contg}, we see that this requires that the coefficients of both operators arising at $O(y^2)$ be small, due to some (perhaps unappealing) peculiarity of the strong sector.  In that case, 
the patterns in the Higgs spectrum  described in the previous section for the inert Higgs scenario are not anymore present, and  
no sharp predictions  can be made. We can estimate 
that all the masses will now be reduced by these  tunings, and therefore all of them will be 
comparable and of the order  of $m_h$, given in eq.~(\ref{mhiggs1}).

\subsubsection*{{Explicit  $C_2$ Breaking}}

For ${\theta^{\textrm{u,d}}_f}\not= 0$ 
the $C_2$ and $C_1P$ symmetries are broken  by the couplings of the SM  fermions to the strong sector
%although the symmetry $C_1PC_2$ is respected. In this case, we are, 
and, in principle, we are no longer  protected 
against  sizable contributions to  FCNC processes or the $\widehat{T}$ parameter.

Nevertheless, as we mentioned above, flavor problems  can be avoided if for some reason
the proto-Yukawa matrices are aligned:
%of different flavors are proportional: 
${\theta^{\textrm{u,d}}_f}\equiv {\theta^{\textrm{u,d}}}$ and ${\phi^{\textrm{u,d}}_f}\equiv {\phi^{\textrm{u,d}}}$. In this situation we have the composite version of type III 2HDM. The remarkable propery of this model is that the $O(y^2)$ potential is invariant under $C_2$, $C_1P$ and $\SO(4)$ custodial. As long as we are interested in vacua with a hierarchy $v\ll f$, the leading sources of breaking of those symmetries are given by the mass terms $\tilde m_{12}^2$ and $m_{12}^2$. The former respects only $C_1P\cdot C_2$ and arises at order $y_L^2 y_R^2\sim Y_t^2g^2_\rho$  (see table~\ref{contf}). The latter breaks $C_2$ but preserves custodial and $C_1P$ and arises at order $g_\rho^2 y_R^4/16\pi^2 \sim Y_t^2g^4_\rho/16 \pi^2$, and is thus normally further suppressed with respect to $\tilde m_{12}^2$. Assuming electroweak symmetry breaking is primarily triggered by  the expectation value of $\langle \Phi_\uno\rangle =(0,0,0,v_4^\uno)$ we then have the following estimates for the entries in $\langle \Phi_\due\rangle $
%\footnote{This is,  however, not required for leptons in which the induced family transitions for different values of ${\theta^{\textrm{e}}_f}$ are small \cite{}.}. 
%In this case the hierarchical structure of Yukawa couplings
%cannot arise from $y_R^i$ but from the strong dynamical parameters $a^{ij}$
%of Eq.~(\ref{genyuk}).
%The value of   $y_R$ cannot be very small if it must reproduce the 3rd family mass;
%then, we have, for example, $y_R\gtrsim  y_t$ for the up-type quarks.
%This can be naturally achieved if we impose a global flavor symmetry for the SM fields,
% that it is only broken  by operators in the  new  strong sector 
% that become important in the IR. By the AdS/CFT correspondence this implies
% to have a warped 5D model where the flavor symmetry is broken by terms in the IR-boundary.
%The contributions to $\widehat{T}$  are generated from a nonzero value of  $v^\due_3$,
%that will be of order $\tilde m^2_{12}v_4^\uno /m_{22}^2$.  
%To generate $\tilde m^2_{12}$, however,  not only $C_2$ and $C_1P$  but also
%the   custodial symmetry must be broken.  That  only happens at order  $y_L^2 y_R^2\sim Y_t^2g^2_\rho$ 
%(see table~\ref{contf}), thus leading to the estimate   
\bea
v_3^\due & \simeq& \frac{\tilde m^2_{12}}{m_{22}^2}\, v_4^\uno\sim
\frac{Y_t \sin 2\theta}{g_{\rho}}  v_4^\uno\ll v\, 
\label{v2tadpole1}\\
v_4^\due  &\simeq& \frac{m^2_{12}}{m_{22}^2}\, v_4^\uno\sim
\frac{Y_t g_\rho   \sin 4\theta}{16\pi^2}  v_4^\uno\ll v
\label{v2tadpole2}
\eea
corresponding to a small breaking of $C_2$, $C_1P$ and $\SO(4)$. The resulting contribution to $\widehat{T}$ is given by
\beq
\widehat{T} = \frac{(v_4^\uno)^2 (v_3^\due)^2}{f^2 v^2} \sim \frac{(v_4^\uno)^4}{f^2 v^2} \frac{Y^2_t}{g_{\rho}^2} \sin^2 2\theta 
\simeq  10^{-3} \left(\frac{\sin 2\theta }{1/2}\right)^2  \left(\frac{N}{3}\right) \left(\frac{\xi}{0.25}\right)\, .
%\simeq  4\times 10^{-3} \left(\frac{\sin 2\theta }{1/2}\right)^2    \left(\frac{2\ {\rm TeV}}{m_\rho}\right)^2\, .
\label{Tcpv}
\eeq
As one can see, the experimental constraint $\widehat T \lesssim 2\times 10^{-3}$ can reasonably be satisfied.
% for a moderate departure from maximal breaking of $C_2$ and $C_1P$.
Interestingly, this contribution to $\widehat T$ is positive, which might even help to fit the 
EWPT.
%If the breaking of $C_2$ comes from only the down or lepton right-handed  sector, we
%have that $y_R y_L$, proportional to the fermion mass, is much smaller,
%and the contribution to 
%$\widehat T$ can be easily accommodated.
%On the other hand, the real part of $\langle \Phi_\due\rangle $, $v^\due_4$,  is  protected by the 
%$C_1P\cdot C_2$, which is only broken (see table~\ref{contf})  
%by terms of order   $g_\rho^2 y_R^4/(16\pi^2)$.   We then have
%\beq
%v_4^\due  \sim \frac{m^2_{12}}{m_{22}^2}\, v_4^\uno\sim
%\left( \frac{Y_t g_\rho   \sin 4\theta}{16\pi^2} \right) v_4^\uno\ll v.
%\label{v2tadpole2}
%\eeq
%We conclude that, although the $C_2$ is broken,  $\langle \Phi_\due\rangle $ is small.

In models with explicit breaking of $C_2$ the Higgs spectrum follows the estimates of the inert Higgs. 
The mixing of $h$  with $H$ ($A$) is small due to the    approximate $C_1P\cdot C_2$ (custodial) symmetry.
The only relevant phenomenological implication, as we will explore later, is that   $\Phi_\due$ couples now to fermions:
\begin{equation}
i\tan \theta^{\textrm{u}}\, \bar q_L\frac{M_{\textrm{u}}}{v/\sqrt{2}} u_R\, \widetilde H_\due+
ie^{i\phi^{\textrm{d}}} \tan \theta^{\textrm{d}}\, \bar q_L
\frac{M_{\textrm{d}}}{v/\sqrt{2}} d_R\,  H_\due+h.c.\, ,
\label{yukawasc2}
\end{equation}
where $M_{\textrm{u,d}}$ are the fermion mass matrices.

\subsubsection{$C_1P$ Invariant Models}
\label{secC1Pmodel}

We now turn to models based on the $C_1P$ invariance  of the strong sector. 
$C_1P$ and $C_2$ might also be imposed simultaneously with results similar to section~\ref{spontaneousC2}, but we now want 
to consider the case where $C_2$ is maximally broken. To construct 
a model of this sort we cannot employ the setup of the previous section in which the quarks mix to composite operators  with  $\{{\bf r}_Q,{\bf r}_T\}=\{{\bf6},{\bf6}\}$, because four Yukawa structures (two for the up-type and two for the down-type) 
would be present and, as discussed in section~\ref{flavorstruct}, this would lead to large Higgs-mediated FCNC. 
In order to avoid the second set of Yukawas, forbidden in the previous section by the $C_2$ symmetry of the strong sector, 
we will use the $\{{\bf r}_Q,{\bf r}_T\}=\{{\bf 20'},{\bf 1}\}$ representations.  A single invariant is now allowed 
by the $\SO(6)$ symmetry (see section~\ref{flavorstruct}). Then Higgs-mediated FCNC are avoided  if one assumes the proto-Yukawa matrices have a flavor independent orientation (corresponding to a flavor independent embedding of the left-handed doublets into the ${\bf 20'}$).

%Since the couplings respect $C_1P\cdot C_2$ and $C_1P$ is unbroken the couplings will
%accidentally be $C_2$ invariant also.
Assuming  the up-quark proto-Yukawas  are $C_1P$ invariant, and  using $\SO(2)$ rotations, we can in general write them in the form (see eq.~(\ref{ylvev})) 
\begin{equation}
	\renewcommand\arraystretch{1.2}
\left({(y_L)_f^{\textrm u}}^{{\overline\alpha}}\right)^{IJ} ={(y_L)_f^{\textrm u}}
\scalebox{0.8}{$
\left\{ 
\left( 
\begin{tabular}{c|cc}
	$0_{4\times4}$ & $(\vec v^{\,\bar1})^T$ & $0_{4\times1}$ \\ \hline
	$\vec v^{\,\bar1}$	& \multicolumn{2}{c}{\multirow{2}{*}{$0_{2\times2}$}} 
	\\
	$0_{1\times4}$	&&
\end{tabular}
\right)
~,~
\left( 
\begin{tabular}{c|cc}
	$0_{4\times4}$ & $(\vec v^{\,\bar2})^T$ & $0_{4\times1}$ \\ \hline
	$\vec v^{\,\bar2}$	& \multicolumn{2}{c}{\multirow{2}{*}{$0_{2\times2}$}} 
	\\
	$0_{1\times4}$	&&
\end{tabular}
\right)
\right\}\,,
$}
\label{upyukawa}
\end{equation}
which is found to accidentally respect  $C_2$ as well.

%Obviously, that $C_2$ is preserved by the coupling should a priori lead to any interesting consequence, 
%because $C_2$ was not a symmetry of the strong sector to start with. 
In principle, since $C_2$ is not a symmetry of the strong sector, the fact that it is preserved 
by the coupling of elementary fermions should not imply any relevant consequence. 
Nevertheless, we find that some important terms generated by the strong sector will be accidentally  $C_2$
invariant, and the model will, for this reason, resemble the Inert Higgs in several aspects, as we now discuss. 
One accidentally $C_2$-invariant term is the (unique, as remarked above) up-type generalized Yukawa term 
constructed with the  ${\bf 20'}$ and the singlet.
Provided the second Higgs does not have a VEV, then  all the interactions mediated by this 
term ({\it{i.e.}}, remarkably, the ones involving the $t_R$ quark) will respect 
$C_2$. 
The down-type Yukawas are also generated by the coupling of the $q_L$ to another
${\mathbf{20}}'$, with $X=-1/3$, while the $d_R$ couples to a $\bf 1_{2/3}$. The right-handed proto-Yukawa  $y_R$ is a singlet,
while the most general $C_1P$ invariant $y_L$ proto-Yukawa for the downs reads
%structure of these couplings is the one of eq.~\ref{dwn}, but now $y_L$
%is in the ${\mathbf{20}}'$ and reads
\begin{equation}
	\renewcommand\arraystretch{1.2}
\left({(y_L)_f^{\textrm d}}^{{\overline\alpha}}\right)^{IJ} ={(y_L)_f^{\textrm d}}
\scalebox{0.8}{$
\left\{ 
\left( 
\begin{tabular}{c|cc}
	$0_{4\times4}$ & $\cos\downangle(\vec v^{\,\bar2})^{\dagger}$ & $\sin\downangle(\vec v^{\,\bar2})^\dagger$ \\ \hline
	$\cos\downangle(\vec v^{\,\bar2})^*$	& \multicolumn{2}{c}{\multirow{2}{*}{$0_{2\times2}$}} \\
	$\sin\downangle(\vec v^{\,\bar2})^*$	&&
\end{tabular}
\right)
~,~
\left( 
\begin{tabular}{c|cc}
	$0_{4\times4}$ & $\cos\downangle(\vec v^{\,\bar1})^\dagger$ & $\sin\downangle(\vec v^{\,\bar1})^\dagger$ \\ \hline
	$\cos\downangle(\vec v^{\,\bar1})^*$	& \multicolumn{2}{c}{\multirow{2}{*}{$0_{2\times2}$}} \\
	$\sin\downangle(\vec v^{\,\bar1})^*$	&&
\end{tabular}
\right)
\right\}\, .
$}
\label{tildetheta}
\end{equation}
Notice that  $\tilde \theta$ cannot  be rotated away since  $\SO(2)$ invariance has already been used to put the up proto-Yukawa into the form of eq.~(\ref{upyukawa}).
Therefore the down-type proto-Yukawas do not  respect the accidental $C_2$.

% in general, because 
%compatibly with the condition of flavor-universality these can always point to another $\SO(2)$ direction 
%$\downangle\not=0$. The down-type interactions will therefore not respect the accidental $C_2$.
%, 
%leading  to a quite peculiar phenomenology of the second Higgs that we will 
%explore in section~4.

\begin{table}[h!]
\begin{center}
		\renewcommand\arraystretch{2.1}
\begin{tabular}{|c||c|c|c|}\hline
Operator & $\mathcal{I}^{1}_{(0,1)}$ & $\mathcal{I}^{2}_{(0,1)}$ & $\mathcal{I}^{3}_{(0,1)}$ \\ \hline
$\dfrac{1}{16\pi^2}\times$ 	& 
$y_L^2g_{\rho}^2$	& $-\dfrac{5}{2}y_L^2g_{\rho}^2$ 	& $y_L^2g_{\rho}^2$		
\\ \cdashline{1-4}[1pt/1pt] %%%%%%%%%%%%%%%%%%%%%%%%%%%%%%
$m_{11}^2/f^2$	& 
$1$ & $1$ & $1$				\\
$m_{22}^2/f^2$	& $0$			& $0$				& $\dfrac{1}{2}$		
\\ \cdashline{1-4}[1pt/1pt] %%%%%%%%%%%%%%%%%%%%%%%%%%%%%%
$\lambda_{1}$	& $-\dfrac{4}{3}$	& $-\dfrac{11}{15}$		& $-\dfrac{4}{3}$		\\ 
$\lambda_{2}$	& $0$			& $0$				& $-\dfrac{1}{6}$		\\ 
$\lambda_{3}$	& $0$			& $-\dfrac{1}{5}$		& $-\dfrac{1}{2}$		\\ 
$\lambda_{4}$	& $-\dfrac{4}{3}$	& $-\dfrac{8}{15}$		& $-1$		\\ 
$\tilde{\lambda}_{4}$ & $0$		& $0$				& $0$				\\
%${\lambda}_{5}$ & $0$			& $0$				& $0$				\\

\hline

\end{tabular}\\
\end{center}
\caption{Contribution to the parameters of the general 2HDM potential \eq{2HDMV} from fermions $\{\mathbf{r}_Q,\mathbf{r}_T\}=\{\mathbf{20}',\mathbf{1}\}$.  The individual contributions of the $\SO(6)/\SO(4) \times \SO(2)$ operators of table~\ref{tabPot201} are shown.
The first line indicates the NDA pre-factor.}
\label{tab201}
\end{table}

\subsubsection*{Almost Inert  Higgs}

Even if some terms in the Lagrangian, such as the up Yukawas previously discussed, are accidentally $C_2$-invariant, 
one would naively expect maximal breaking of $C_2$ elsewhere, in particular in the potential, given that the strong sector by assumption breaks $C_2$. That would  force both $\Phi_\uno$ and $\Phi_\due$ to acquire a VEV, and this compatibly with unbroken $C_1P$.
%resulting interactions 
%will not respect the symmetry if the second Higgs has a VEV and $C_2$ is spontaneously broken. To show the similarity of the 
%present scenario with the inert Higgs case, therefore, it is crucial to discuss the vacuum structure, which in turn is determined 
%by the Higgs potential. A priori, there is no reason why the second Higgs should not be ``obliged'' to take a VEV by a tadpole term 
%in the potential: this term is not forbidden by the $C_1P$ symmetry we have imposed. 
At leading order ($\propto y_L^2$), however, 
the only three $C_1P$-even contributions to the potential are also $C_2$-even, as  table~\ref{tabPot201} shows. The potential is thus accidentally approximately $C_2$ invariant. Furthermore, we have found that the  leading $C_2$-odd contribution  ({\it{i.e.}}, $C_1P$-even) comes at order $y_L^4$.  The $C_2$-odd mass $m_{12}^2$ that controls the $\langle \Phi_\due\rangle$ is therefore
\begin{equation}
	m_{12}^2\,\sim\, \frac{N_c}{16\pi^2}y_L^4\, f^2 \simeq (70\textrm{ GeV})^2 \left( \frac{y_L}{1} \right)^4 \left( \frac{0.25}{\xi} \right)\, .
\end{equation}
The associated phenomenological implications will be discussed in the section~\ref{almost}.

The contributions to the renormalizable potential in eq.~(\ref{2HDMV}) arising from each of the three allowed operators are shown in table \ref{tab201}, 
from which several interesting consequences can be drawn. First, we see that the leading order potential is tunable, without need of including 
sub-leading corrections. This is clearly an advantage as compared with the $\{{\bf 6},{\bf 6}\}$ model of the previous section, 
and also the MCHM \cite{Agashe:2004rs}: in the model at hand less fine-tuning is required to reach the same value of $\xi$. Second, we see that there is a unique 
contribution to both $m_{22}^2$ and $\lambda_2$;
their ratio being fixed, makes not possible to tune the VEV of  $\Phi^\due$.
This means that the VEV of $\Phi^\due$ cannot be fine-tuned to be smaller than $f$, so that even if a $C_2$-breaking vacuum existed, it would be difficult to make it phenomenologically viable.
Third, one can check, by studying explicitly the leading order potential, that a stable vacuum with both $\Phi^\uno$ and $\Phi^\due$ taking a VEV does not exist. Therefore not only we know that a $C_2$-preserving vacuum exists (because of the accidental $C_2$ symmetry), but we also 
see that spontaneous $C_2$ breaking cannot be achieved by the leading order potential. The model is therefore ``forced'' to resemble the Inert Higgs.

In this setup, differently from the one of the previous section, the quartic $\lambda_1$ is not reduced by the tuning 
and the Higgs mass therefore reads 
\begin{equation}
	m_{h}^2 	\sim \frac{N_c}{16\pi^2}y_L^2 g_\rho^2\,v^2
	\simeq (250 \textrm{ GeV})^2 \left( \frac{y_L}{1} \right)^2\left( \frac{3}{N} \right)\, .
\end{equation}
The masses of the other scalars, the $\SO(3)_c$ triplet, $H^a$, and  singlet, $H$, are dominated by a common $\SO(4)$-symmetric contribution 
\begin{equation}
	m_{H_2}^2 	\sim \frac{N_c}{16\pi^2}y_L^2 g_\rho^2\,f^2
	\simeq (500\textrm{ GeV})^2   \left( \frac{y_L}{1} \right)^2\left( \frac{3}{N} \right) \left( \frac{0.25}{\xi} \right)\, .
\end{equation}
After EWSB, $H$ gets an additional contribution through the $\lambda_4$ coefficient.
We can calculate this triplet-singlet mass splitting using table~\ref{tab201}.
At order $y^2$ we have only three operators and therefore the potential depends 
only on their  three unknown coefficients. %pre-factors.
 The tuning $v \ll f$ gives an approximate relation
between the three coefficients, which can be used to eliminate one.
The other two can be traded  for $m_{h}$ and  $m_{H^a}$.
We can then  obtain a prediction for $m_{H}$ as a function of these masses:
% it turns out that $\lambda_4>0$ under the constraints $m_{11}^2<0$, $m_{22}^2>0$ ,   $\lambda_1>0$, leading to the splitting 
\begin{equation}
\frac{m_{H}^2- m_{H^a}^2}{m_{H}^2}\simeq  \frac{1}{3}\left(\frac{m_{h}^2}{m_{H}^2}  + \xi\right)\sim \xi \, .
\label{prediction201}
\end{equation}
Custodial-breaking splitting comes from gauge contributions ($\propto{g'}^2$), and higher orders in $y_L$ ($\propto y_L^4$).
These splittings can be estimated as
\begin{equation}\begin{array}{rcl}
	\left|\dfrac{m_{H^{\pm}}-m_{A}}{m_{H^{\pm}}}\right|_{{g'}}	
		&\sim& 	 \left(\dfrac{g'}{y_L}\right)^2\xi
		\simeq		0.03 
				\left( \dfrac{1}{y_L} \right)^2
				\left( \dfrac{\xi}{0.25} \right)\, ,	\\ \\
	\left|\dfrac{m_{H^{\pm}}-m_{A}}{m_{H^{\pm}}}\right|_{ y_L^4}
		&\sim&  \left(\dfrac{y_L}{g_\rho}\right)^2\xi
		\simeq		0.005 \left( \dfrac{y_L}{1} \right)^2
				\left( \dfrac{N}{3} \right)
				\left( \dfrac{\xi}{0.25} \right)\, .
\end{array}\end{equation}

\subsection{Extended Custodial Symmetry}
\label{extcust}

In this section we wish to briefly discuss a last possibility, where  
large corrections to $\widehat T$ are avoided thanks to an $\SU(2)^3$ custodial symmetry in the Higgs sector. 
That symmetry allows arbitrary Higgs VEVs to preserve a diagonal $\SU(2)=\SO(3)_c$ 
which guarantees that $\widehat T=0$ at leading order. The simplest realization of the idea is provided by the coset
\begin{equation}
\frac {\Sp(6)}{\SU(2)\times \Sp(4)}\, ,
\end{equation}
which  delivers 8 NGB in the $\mathbf{(2,4)}$ representation of the unbroken 
group, corresponding to two Higgs
doublets. The unbroken symmetry coincides with the one of the renormalizable 2HDM 
after gauging $\SU(2)_L$.
This is easily seen by embedding the two Higgses $\Phi^{\uno,\,\due}$, in the $2\times2$ 
matrix notation, into a $2\times4$ matrix
\beq
\displaystyle
M\,=\,\left(\Phi^\uno,\,\Phi^\due\right)\,.
\label{emb2h}
\eeq
If $M$ was a generic matrix, we could act on it with an $\SU(2)$ rotation on the left (which 
correspond to the $\SU(2)_L$ SM group) and with an element of $\SU(4)$ on the right. 
The renormalizable 2HDM Lagrangian, once rewritten in terms of the matrix $M$, 
is immediately seen to be invariant under this $\SU(2)\times\SU(4)$ group. The 
pseudo-reality condition of $\Phi^{\uno,\,\due}$, however, implies
\beq
\displaystyle
M^*\,=\,\sigma_2\,M\,\Sigma_2\,,
\eeq
where \mbox{$\Sigma_2=$ diag$(\sigma_2,\,\sigma_2)$}. The above condition breaks the 
group of allowed transformations to $\SU(2)\times\Sp(4)$, as was to be shown. 

This mechanism could be also be extended to  $N$ Higgses. Here the relevant coset is
\begin{equation}
\frac {\Sp(2 N+2)}{\SU(2) \times \Sp(2 N)}\, .
\end{equation}
which produces $N$ doublets. The group $H$ is the symmetry group of the renormalizable 
model, and contains the subgroup $\SU(2)^{N+1}$ which protects the $\rho$ parameter 
in the case of $N$ Higgs doublets. 

We will focus on the $N=2$ coset in what follows. 
Under the extended custodial subgroup  $\SU(2)_L\times \SU(2)_{R1}\times 
\SU(2)_{R2}$  
of $H=\SU(2)\times \Sp(4)$ the NGB decompose as follows,
\begin{eqnarray}
\Phi^\uno&=&\mathbf{(2,2,1)}\,,\nonumber \\
\Phi^\due&=&\mathbf{(2,1,2)}.
\end{eqnarray}
We identify the hypercharge with the linear combination
\begin{equation}
Y=T^3_{R1}+T^3_{R2}+X\,,
\label{hypercharge}
\end{equation}
where the extra $\U(1)_X$ charge $X$ is needed to obtain the correct fermionic hypercharges, 
while the Higgs is $X$-neutral. By the above choice, the two Higgs doublets 
have the same  hypercharge. 

Let us now turn to the fermions. The smallest representations of $\Sp(6)$ decompose under 
$H$ as
\begin{eqnarray}
&\mathbf{6}= \mathbf{(2,1)}\oplus \mathbf{(1,4)}\,,\nonumber \\
&\mathbf{14}= \mathbf{(1,1)}\oplus\mathbf{(1,5)}\oplus \mathbf{(2,4)}\,,\nonumber \\
&\mathbf{14'}= \mathbf{(1,4)}\oplus \mathbf{(2,5)}\,,\nonumber \\
&\mathbf{21}=\mathbf{(3,1)}\oplus\mathbf{(1,10)}\oplus\mathbf{(2,4)}\,.
\label{deco}
\end{eqnarray}
We see that several possibilities exist for embedding the SM doublets and singlets. 
The safer option for EWPT is, as discussed in section~\ref{situation}, to embed the 
$t_R$ in a singlet of the custodial group, so that the $y_R$ proto-Yukawa does not 
contribute to $\widehat{T}$, and to take the limit of total $t_R$ compositeness 
$y_R\rightarrow g_\rho$. This is because in that case we can take a small
$y_L\simeq Y_t$ thus controlling 
$\widehat{T}$ and $\delta g_b/g_b$. 
Those considerations  easily generalize 
to the present case, though we should remember that the custodial group is 
not $\SO(4)$, but the larger $\SU(2)_L\times \SU(2)_{R1}\times \SU(2)_{R2}$ group. In particular 
$t_R$ should be a complete singlet of the latter group. On the other hand, it is not immediately evident  to see whether $P_{LR}$ plays a role in further suppressing $\delta g_b/g_b$, as discussed in section~\ref{situation}.
%and based on the accidental $P_{LR}$ symmetry, can also be generalized to the present case or 
%not. 
In the affirmative, one could depart from the limit of total $t_R$ compositeness, 
but we will ignore this possibility in the following.

Compatibly with the previous discussion, $t_R$ could either be embedded in a $\mathbf{14}$, or $\mathbf{21}$ or in 
a total singlet, with $X=2/3$ charge. Let us choose for definiteness the case of the 
$\bf 1_{2/3}$, while for the $q_L$ doublet we pick a $\bf {14}_{2/3}$.
Within the $\bf {14}_{2/3}$, 
$q_L$ can be embedded in
either the $\bf {(2,2,1)}_{2/3}$ or the $\bf {(2,1,2)}_{2/3}$ of 
$\SU(2)_L\times \SU(2)_{R1}\times \SU(2)_{R2}$ (which are in turn contained in the  the $\bf {(2,4)}_{2/3}$ of $\SU(2)_L\times \Sp(4)$).
Concerning flavor, applying the general consideration of 
section~\ref{flavorstruct}, this model behaves similarly to the one with the 
${\bf20'}$ and the singlet discussed in section~\ref{secC1Pmodel}. 
There is a unique Yukawa structure, because a unique $H$ singlet can be 
formed among the $\mathbf{14}$ and the singlet (see \eq{deco}), but 
two embeddings are present for the $q_L$ in the $\mathbf{14}$, as previously 
discussed. As in the model of section~\ref{secC1Pmodel}, in order to avoid Higgs-mediated FCNC, 
the embeddings must be taken flavor independent.

The above discussion shows that no obstructions seems to arise when trying to 
construct explicit models with an extended custodial group. 
A detailed phenomenological 
study of the model we have just described, and of the other possibilities that 
may be envisaged in the context of the $\Sp(6)/\SU(2)\times \Sp(4)$ coset, 
lies however outside the scope of the present paper.

\section{Phenomenology of Composite 2HDM}
\label{pheno}
In this section we will study the  phenomenological implications of  the composite 2HDM,
starting with the generic properties, and focussing later on the explicit examples we constructed. 
For this purpose we will derive the effective Lagrangian describing  Higgs physics 
at energies below $m_\rho$ . That can be written in terms of an expansion in
powers of $H_{i}/f$ and   $\partial_\mu^2/m_\rho^2$.
The leading dimension-4 operators  describe the Lagrangian of an elementary 
({\it i.e.} renormalizable) 2HDM. It is essential, for the purpose of the following discussion, 
to recall the properties of that limiting case.

\subsection{Elementary 2HDM}
\label{elementary}
The Lagrangian of the elementary 2HDM, focussing just on the top Yukawa terms,
can be written as
\begin{equation}
{\cal L}=|D_\mu H_1|^2+|D_\mu H_2|^2+ 
Y_t\, \bar q_L (\tilde H_1+a_t \tilde H_2) t_R
%\bar q_L (Y_1^u \tilde H_1 + Y_2^u \tilde H_2) u_R
+h.c.+V(H_1,H_2)\, .
\label{lare}
\end{equation}
It will be  convenient in this section  to work in the basis in which only one doublet  acquires a VEV. In  the unitary gauge  this is given by
\begin{equation}
H_1=\left(\begin{array}{c} 0  \\ (v+h^0)/\sqrt{2} \end{array}\right)
\ , \ \  H_2=
\left(\begin{array}{c} H^+  \\ (H^0+iA^0)/\sqrt{2} \end{array}\right)\, .
\label{smartbasis}
\end{equation}
In this basis the mass eigenstates, in the limit where $CP$ is not violated, are
\begin{equation}
h^0_1=h^0\cos\theta_h+H^0\sin\theta_h\ ,\ \ \  h^0_2=-h^0\sin\theta_h+H^0\cos\theta_h\ ,\ \ \  A^0\ \ \   {\rm and}\ \ \  H^\pm\, ,
\end{equation}
where $\theta_h$ is determined by the potential $V(H_1,H_2)$.
The angle $\theta_h$ is obviously proportional to the source  of  $C_2$ breaking,
but also to $v^2$. That is because in the limit of unbroken $\SU(2)_L$, {\it i.e.}  $v\rightarrow 0$,  $H^0$ should not mix with $h^0$ and  become degenerate with $A^0$ and $H^+$.
Therefore  in composite 2HDM one expects  $\theta_h \propto (v/f)^2$, possibly times an extra reduction factor 
associated to the small breaking of $C_2$.
% Of course in the presence of  $C_2$
%we would have $\theta_h=0$, as well as $a_t=0$. The only allowed term leading to bilinear mixing is 
%$\lambda_\textrm{mix}(|H_1|^2-v^2) H_1^\dag H_2 + h.c.$
% which leads to $\theta_h\propto \lambda_\textrm{mix}(v/m_{h_2^0})^2$.
%In the case of a composite Higgs it leads to $\theta_h \sim (v/f)^2$.

The specific form of the kinetic terms in eq.~(\ref{lare}), when expressed in terms of the mass eigenstate fields, leads to a set of relations  
between the masses of vector bosons and their coupling to scalars. Assuming $CP$ invariance, those can be written as the sum rules
\begin{subequations}\label{sumrules}
\begin{align}
	S_{\phi V V} &\equiv
	\dfrac{\sum_{i} g_{h_i^0 W^+W^-}^2}{g^2 m_W^2} =
	%{(g_{hW^+W^-}^\textrm{SM})^2} =
	\dfrac{\sum_{i} g_{h_i^0 ZZ}^2}{g^2 m_Z^2/\cW^2} = 
	%{(g_{hZZ}^\textrm{SM})^2} =
	\dfrac{\sum_{i} g_{h_i^0 W^+W^-} g_{h_i^0 ZZ}}{g^2 m_Z^2}
	%{g_{hW^+W^-}^\textrm{SM}g_{hZZ}^\textrm{SM}} 
	= 1 ~, \\ \nonumber \\
	S_{\phi \phi V} &\equiv
	\dfrac{\sum_{i} g_{h_i^0 H^+W^-}^2}{g^2} =
	\dfrac{\sum_{i} g_{h_i^0 A^0Z}^2}{g^2\cW^2} =
	\dfrac{\sum_{i} g_{h_i^0 H^+W^-}g_{h_i^0 A^0Z}}{g^2/\cW} 
	= 1 ~, \\ \nonumber \\
	S_{\phi V} &\equiv
	\dfrac{\sum_{i} g_{h_i^0 W^+W^-} g_{h_i^0 A^0Z}}{g^2 m_W / \cW} =
	\dfrac{\sum_{i} g_{h_i^0 W^+W^-} g_{h_i^0 H^+W^-}}{g^2 m_W} = 
	\dfrac{\sum_{i} g_{h_i^0 Z Z} g_{h_i^0 A^0Z}}{g^2 m_Z / \cW^2} =
	\dfrac{\sum_{i} g_{h_i^0 Z Z} g_{h_i^0 H^+W^-}}{g^2 m_Z / \cW} 
	= 0 ~.
\end{align}
\end{subequations}
and as the relations
\begin{subequations}
\label{relations}
\begin{align}
	\dfrac{g_{A^0H^+W^-}^2}{g^2/4} =
	\dfrac{g_{H^+H^-Z}^2}{g^2/4 \cW^2} 
	&= 1 ~, \\ \nonumber \\
	g_{H^+W^-Z} &=0 ~,%, \\ \nonumber \\
%	g_{A^0ZZ}=g_{A^0W^+W^-} &=0.
\end{align}
\end{subequations}
where the sums run over the two neutral $CP$-even scalars $h_1^0$ and $h_2^0$, while, with an obvious notation, the  interaction vertices are defined by
\begin{eqnarray} {\cal L}_{\rm int}^{\phi V_a V_b}&=&g_{\phi V^aV^b}\, \phi V^a_\mu V_b^\mu\, ,\\{\cal L}_{\rm int}^{\phi \phi'V_a }&=& g_{\phi \phi' V^a}\,  \phi \overleftrightarrow \partial_\mu\phi' V_a^\mu\, .\end{eqnarray}
It is noteworthy the vanishing of $g_{H^+W^-Z}$ at the renormalizable level.
A further set of useful relations following from the sum rules is
\begin{align}
\frac{g_{h_1^0 Z Z}}{g_{h_2^0 A^0Z}} &= - \frac{g_{h_2^0 Z Z}}{g_{h_1^0 A^0Z}} = m_Z\, ,
\label{ratios}
\end{align}
and similarly for ratios involving $g_{h_i^0 W^+ W^-}$ and $g_{h_j^0 H^+W^-}$, with $i \not= j$.

In the very same way the scalar couplings to the top multiplet satisfy specific relations. In the mass eigenstate basis the Yukawa interactions read (in the $CP$ conserving limit)
\begin{equation}
{\cal L}_{top}=Y_t\left \{ \frac{1}{\sqrt{2}} \bar t t \left [(\cos\theta_h +a_t\sin\theta_h)h_1^0+(a_t\cos\theta_h -\sin\theta_h)h_2^0\right]+\frac{ia_t}{\sqrt{2}} \bar t\gamma_5t A^0+ a_t(\bar b_Lt_RH^-+{h.c.})\right \}\, ,
\end{equation}
where only one parameter, $a_t$,    accounts for  four couplings. 
This results in  three relations that can be written as 
\begin{eqnarray}
y_{h_1^0}\cos\theta_h -y_{h_2^0}\sin\theta_h&=&\frac{m_t}{v}\, ,\\
y_{h_1^0}^2+y_{h_2^0}^2-\left(\frac{m_t}{v}\right)^2&=&y_{A^0}^2=\frac{y_{H^+}^2}{2}\, .
\label{fermionrules}
\end{eqnarray}

The above relations follow from a renormalizable, weakly coupled Lagrangian. Their violation  leads
to a growth with energy for the scattering amplitudes involving scalars and/or longitudinally polarized vectors. 
In the case of our PNGB,  the growth of the scattering amplitudes is dictated by the 
$\sigma$-model derivative interactions.

%Indicating the interaction vertices by any two scalar Parametrizing the couplings as 
% We have chosen to group couplings which are related by custodial symmetry. 
%If only dimension-4 operators are present in the theory, perturbative unitarity 
%for the  scattering amplitudes of type , $V^a \phi \rightarrow V^b \phi$, $V^aV^b \rightarrow V^c \phi$, where $V^a = W^+,W^-, Z$  and $\phi=h^0_i, H^a$, with $H^a = H^+,H^-,A^0$,
%leads to a set of sum rules. 

%\begin{align}
%%VV \rightarrow VV: 
%\label{sumr}
%& \sum_{i} g_{h_i^0 WW}^2 = g^2 m_W^2, \\
%%& \quad \sum_{i} g_{h_i^0 ZZ}^2 = \frac{g^2}{4 \cos^2 \theta_W} m_Z^2, \\
%%V \phi \rightarrow V \phi:
%& \sum_{i} g_{h_i^0 ZZ}^2 = \frac{g^2 m_Z^2}{\cos^2\theta_W}, \\
%& \sum_{i} g_{h_i^0 WW}\,  g_{h_i^0 ZZ} = g^2 m_Z^2, \\
%& \sum_{i} g_{h_i^0 H^+W}^2 = \frac{g^2}{4} \\
%& \sum_{i} g_{h_i^0 A^0 Z}^2 = \frac{g^2}{4 \cos^2\theta_W} \\
%& \sum_{i} g_{h_i^0 H^+W}\,  g_{h_i^0 A^0 Z} = \frac{g^2}{4 \cos\theta_W} \\
%& g_{A H^+W}^2 = \frac{g^2}{4}\, ,
%\end{align}

\subsection{Composite 2HDM}

We want now to discuss how things change in the composite 2HDM when including higher-dimensional operators. 
% restricting to two derivative terms. 
Among these operators, those  involving the Higgs fields without any derivative, for example $|H_1|^2|H_2|^4/f^2$,
will  not be relevant for us since they only modify the potential $V(H_1,H_2)$, but obviously not the relations derived in the previous section.
Higher derivative terms are suppressed by $O(\partial_{\mu}^2/m_\rho^2)$ and thus normally subleading, at sufficiently low energies, with respect to the  $O(v^2/f^2)$ effects coming from non-linear 
$2$-derivatives scalar interactions \cite{Giudice:2007fh}.  Let us first comment on the general case, i.e. working to all order in $v/f$ and without assuming a specific $\sigma$-model structure. 

In the general case,  unlike in the renormalizable one, 
we cannot find an operator basis that diagonalizes the kinetic terms, upon expansion, 
at any point in the field space.
Because of that, in any parametrization where $\langle H_2\rangle = 0$ there will still be kinetic mixings between the components of $H_2$ and the NGB living inside $H_1$. In this situation the standard unitary gauge of eq.~(\ref{smartbasis}) is no longer eliminating the bilinear mixings between vectors and scalars. Assuming custodial invariance the ``canonical" parametrization is of the form
\begin{equation}
H_1=\left(\begin{array}{c} \epsilon_T  H^+  \\ (v+h^0+\epsilon_H H^0 + i\epsilon_T A^0)/\sqrt{2} \end{array}\right)
\ , \ \  H_2=
\left(\begin{array}{c} H^+  \\ (H^0+iA^0)/\sqrt{2} \end{array}\right)\, ,
\label{smartbasis2}
\end{equation}
where  $\epsilon_T$ is fixed by the  gauge choice so as to eliminate the mixing of  $H^+$ and $A^0$ with vectors, while $\epsilon_H$ is chosen by requiring vanishing kinetic $h^0-H^0$ mixing (though these, in general, are not yet mass eigenstates). The above represents the general case. However in the specific case where the  kinetic terms respect an $\SO(2)$ symmetry under which $(H_1,H_2)$ form a doublet, it is easy to see that $\epsilon_T=\epsilon_H=0$. 
Moreover,  assuming $\SO(4)$ invariance, but not $\SO(2)$, and limiting the analysis to dimension 6 operators, one is not forced to use the general parametrization of eq.~(\ref{smartbasis2}). Indeed, as
shown in Appendix A, by performing the most general field redefinition of the form $H\to H+H^3/f^2$
one can reduce the set of dimension $6$ terms to the eight operators listed in eq.~(\ref{indops}). In principle after the non-linear field redefinition both $H_1$ and $H_2$ will have a non vanishing VEV,
but it is easy to see that the basis (not the individual operators)  in eq.~(\ref{indops}) is invariant under simple rotations in the $(H_1,H_2)$  plane. By one such rotation we can therefore always choose $\langle H_2\rangle =0 $.
By inspecting  the operators in  eq.~(\ref{indops}) one finds $\epsilon_T=0$ and 
$\epsilon_H = -(c_{H_1 H_{12}}/2)  (v^2/f^2)$.

Substituting eq.~(\ref{smartbasis2}) in the operators of eq.~(\ref{indops})
we  obtain  corrections of order  $v^2/f^2$  to  the Higgs  couplings.
The relevant ones for the trilinear 
%$\phi VV$, $\phi \phi' V$ and $\phi_1 \phi_2 \phi_3$
couplings are
% {\red [THIS Lagrangian IS NOT IN THE PHYSICAL BASIS FOR THE NEUTRAL SCALARS]}
\begin{equation}
\renewcommand\arraystretch{2.5}
\begin{array}{rl}
{\cal L}_{3-int}=
%\Big\{ 
&
\dfrac{g m_W}{2} \left[ \left(1+\dfrac{v^2}{f^2}\ca  \right)  h^0 +\dfrac{v^2}{f^2}\cb\,  H^0  \right] V^\mu_a V_\mu^a 		%\\ &
+ig   %\left[\dfrac{v^2}{f^2}\cc   h^0+
\left( 1+\dfrac{v^2}{f^2} \cc \right) H^0 %\right]
V^\mu_a \partial_\mu H^a 	\\ &
+ \dfrac{g}{2} \left( 1+\dfrac{v^2}{f^2} \cd \right)  \epsilon_{abc} H^a \partial^\mu H^b A_\mu^c 	%\\ &
+\dfrac{v}{f^2} \ce\, h^{0\, 2}\partial^2_\mu H^0\, ,	%	\\ &
%+\dfrac{v^2}{f^2}\dfrac{c_8}{f}(H^0)^2\partial^2_\mu h^0
%~ ~ \Big\}	
\end{array}
\label{extrac}
\end{equation}
%\begin{align}
%{\cal L}_{3-int} &=
%\frac{g m_W}{2} \left[ \left(1+\dfrac{v^2}{f^2}c_{1}  \right)  h^0 +\dfrac{v^2}{f^2}c_3  H^0  \right] A^\mu_a A_\mu^a \nonumber \\
%&+ ig   \left[\dfrac{v^2}{f^2}c_4 h^0+\left( 1+\dfrac{v^2}{f^2}c_5 H^0 \right)  \right]A^\mu_a \partial_\mu H^a \nonumber \\
%&
%\label{extrac}
%\end{align}
where 
%$A^a_\mu$  are the SM massive gauge bosons, $H^a\ni H^+,H^-,A^0$  and 
\begin{equation}\renewcommand\arraystretch{1.8}\begin{array}{rcl}
V^\mu_a V_\mu^a &\equiv& 2 (W^+)_\mu (W^-)^\mu+\dfrac{1}{\cos^2\theta_W}Z_\mu Z^\mu\, ~,\\
iV^\mu_a \partial_\mu H^a &\equiv& iW^{-\mu}  \partial_\mu H^+ +h.c. + \dfrac{1}{\cos\theta_W}Z^\mu \partial_\mu A^0\, ~, \\
\epsilon_{abc} H^a \partial^\mu H^b V_\mu^c &\equiv& 
\dfrac{i}{\cos\theta_W} H^- \partial^\mu H^+Z_\mu
%+i A \partial^\mu H^- W_\mu^+
+ A^0 \partial^\mu H^- W_\mu^+
%-i \partial^\mu A H^-W_\mu^+
- H^+ \partial^\mu A^0 W_\mu^- ~
+h.c.\, ,
%\dfrac{i}{\cos\theta_W}( H^-\partial^\mu H^+ - H^+\partial^\mu H^-)Z_\mu
%+i A (\partial^\mu H^- W_\mu^+ - \partial^\mu H^+ W_\mu^-)
%-i \partial^\mu A (H^-W_\mu^+-H^+W_\mu^-)
\end{array}\end{equation}
and
\begin{equation}\begin{array}{rclcrcl}
	\ca	=	-c_{H_1}
	\ ,\
	\cb	=	- \frac{1}{2} c_{H_1 H_{12}}
	\ ,\
	\cc	=	- \frac{1}{4} c_{H_{12}} + c_T
	\ ,\
	\cd	=	c_{\epsilon}
	\ ,\
	\ce	=	- \frac{1}{2} c_{H_1 H_{12}}\, ,
\end{array}\end{equation}
in terms of the coefficients of the dimension 6 operators in \eq{indops}. 
In eq.~(\ref{extrac}),
terms proportional to $\partial_{\mu} V^{\mu}$ have been omitted since they do not play any role in the production or decay of the Higgs bosons. 
%Besides, we have used a unitary rotation of $h^0$ and $H^0$ (of order $v^2/f^2$), to eliminate a coupling $h^0 V^\mu_a \partial_\mu H^a$. 
Notice also the absence of $H^+W^-Z$ and $A^0 V^aV^a$ couplings that, 
as will be discussed in more details below,  is a  consequence of custodial symmetry.
%{\red since the only operator at leading derivative expansion that could generate the vertex also generates kinetic mixing. [ALSO GENERATES MIXING H-NGBS, RIGHT?]} The same for $A^0 VV$ interactions.  
%Still, once the physical basis $h_i^0$ is adopted, both neutral scalars couple to $A^{\mu}_a \partial_{\mu} H^a$.
We recall that $h^0$ and $H^0$ are not in general physical states, since they mix at order $\theta_h \sim v^2/f^2$ in the potential. 

The corrections to the vertices of eq.~(\ref{extrac}) modify the relations (\ref{sumrules}) and (\ref{relations}). 
We now obtain
\begin{subequations}
\begin{align}
%	\dfrac{\sum_{i} g_{h_i^0 W^+W^-}^2}{(g_{hW^+W^-}^\textrm{SM})^2} =
%	\dfrac{\sum_{i} g_{h_i^0 W^+W^-}^2}{(g_{hZZ}^\textrm{SM})^2} =
%	\dfrac{\sum_{i} g_{h_i^0 W^+W^-}g_{h_i^0 ZZ}}{g_{hW^+W^-}^\textrm{SM}g_{hZZ}^\textrm{SM}} 
S_{\phi V V}
	&= \left(1+2 \ca \dfrac{v^2}{f^2}\right)\, , \\ \nonumber \\
%	\dfrac{\sum_{i} g_{h_i^0 H^+W^-}^2}{g^2/4} =
%	\dfrac{\sum_{i} g_{h_i^0 A^0Z}^2}{g^2/4\cos^2\theta_W} =
%	\dfrac{\sum_{i} g_{h_i^0 H^+W^-}g_{h_i^0 A^0Z}}{g^2/4\cos\theta_W} 
S_{\phi \phi V}
	&= \left(1+2 \cc \dfrac{v^2}{f^2}\right)\, , \\ \nonumber \\
S_{\phi V}
	&= \cb \dfrac{v^2}{f^2}\, ,  \\ \nonumber \\
	\dfrac{g_{A^0H^+W^-}^2}{g^2/4} &=
	\dfrac{g_{H^+H^-Z}^2}{g^2/4\cW^2} 
	= \left(1+2 \cd \dfrac{v^2}{f^2}\right)\, .
	\label{sumrulesmod}
\end{align}
\end{subequations}
%where $c_3$ and $c_4$ only enter at higher order, since the coupling they parametrise is absent at leading order.
The ratios \eq{ratios} get also modified:
\begin{align}
& \frac{g_{h_1^0 ZZ}}{g_{h_2^0 AZ}}=
%\frac{c_3-(f^2/v^2+c_1)\tan\theta_h}{c_4+(f^2/v^2+c_5)\tan\theta_h}...
m_Z \left[ 1+ \left( \ca - \cc \right) \frac{v^2}{f^2} %+ \cb \tan \theta_h \frac{v^2}{f^2}
\right]\ ,
& \frac{g_{h_2^0 ZZ}}{g_{h_1^0 AZ}}= - m_Z \left[ 1- \cb \frac{1}{\tan \theta_h}\frac{v^2}{f^2} \right]\, .
%-\frac{c_3+(f^2/v^2+c_1)\cot\theta_h}{c_4-(f^2/v^2+c_5)\cot\theta_h}...
\label{tworatios}
\end{align}
%Notice that 
%corrections to the second ratio are of order one since both $c_2$ and $\theta_h$ are proportional to the breaking of the $C_2$ symmetry, so we expect $\tan\theta_h\sim c_2 (v/f)^2$.

% and, if measured, will be a genuine signal of composite Higgs.

Similarly to the genuine $\sigma$-model corrections to the purely bosonic interactions, we can study the implications of compositeness on the interactions between scalars and fermions. As already emphasized in ref.~\cite{Giudice:2007fh}, for approximately elementary fermions the leading effects come from ``higher order" Yukawa interactions, obtained by sprinkling powers of $H_1$ on the leading order result. Sticking to the parametrization $\langle H_2\rangle = 0$ and focussing on terms that affect trilinears, we have  three new dimension 6
operators
\begin{equation}
Y_t  (\bar q_L \tilde{H}_1 t_R ) \left[\frac{c_{t}^{(1)}}{f^2} H_1^\dagger H_1 +\frac{ c_{t}^{(2)}}{f^2} H_1^\dagger H_2+\frac{ c_{t}^{(3)}}{f^2} H_2^\dagger H_1\right]+
Y_t  a_t\frac{c_{t}^{(4)}}{f^2} (\bar q_L \tilde{H}_2 t_R )\,  H_1^\dagger H_1 
+h.c.\, ,
\label{fermsigma}
\end{equation}
where $c_{t}^{(i)}$ are $O(1)$ coefficients.
%The first term was already discussed in ref.~\cite{Giudice:2007fh}. 
Because of these four new coefficients all  three relations in eq.~(\ref{fermionrules}) are modified at order $v^2/f^2$.  

Another potentially interesting implication of  higher-derivative  terms involving PNGB and
fermions arises  in the  class of models where  accidental symmetries
appear at lowest order in the action, as those discussed in section~\ref{secExplicitModels}.
For example, in the particular model of section~\ref{secC1Pmodel} 
where $C_2$ is badly broken in the strong sector, but 
accidentally  preserved in the zero derivative terms of the scalar potential and Yukawa interactions,
it is  interesting to investigate whether  this breakdown shows up unsuppressed in the derivative interactions between PNGB and composite fermions. 
To be definite let us focus on the case in which $t_R$ is a fully composite object. In principle we could expect terms involving the current $\bar t_R\gamma^\mu t_R$ and NGB currents. In the case $\SO(6)/\SO(4)\times \SO(2)$ one can quickly see that no such term can be written. Since the ${\cal D}^{i\alpha}_\mu$ is a $\bf (4,2)$ of $\SO(4)\times \SO(2)$, an invariant term must involve a top current with the same quantum numbers, which is not the case in any model we constructed, and probably of any sensible model in general.
The simplest seemingly sensible possibility is to have $t_R$ charged under $\SO(2)$ in which case the covariant derivative
\begin{equation}
\bar t_R\gamma^\mu (\partial_\mu +{\cal E}_\mu)t_R\, ,
\label{extrafer}
\end{equation}
would break $C_2$ and give rise to a term $\bar t_R\gamma^\mu t_R \Phi_\uno\overleftrightarrow \partial_\mu\Phi_\due$ upon expanding in the NGB fields. 
Unfortunately, in the model with $\{\bf 20',1\}$ we have that  $t_R$ can be fully composite but is a singlet, while  in the model with $\bf \{6,6\}$ we have that  $t_R$ is not a singlet
but  the symmetry $C_2$ is preserved by the strong interactions and therefore the term in eq.~(\ref{extrafer}) cannot be generated.
As a last remark we notice that in $\SO(6)/\SO(4)$ there is an additional NGB transforming as a singlet of $\SO(4)$. The corresponding ${\cal D}_\mu$ can be coupled to the right-handed top current and generates  a sizable effect. But again this theory has an additional 
neutral scalar and it is not just a 2HDM.

\subsection{$H^\pm W^\mp Z$ and the Role of Symmetries}
One might have hoped that the richer kinetic structure of these models would allow for the presence of interactions which, in a 
renormalizable theory, are absent at tree-level.
This is for example the case of $H^\pm W^\mp Z$, which would be a golden channel for $H^+$ detection.
In  the renormalizable 2HDM, 
the $H^\pm W^\mp Z$ vertex
is generated only at the one-loop  level \cite{Mendez:1990epa}.
Unfortunately, the situation is not  much better  in  realistic composite models, 
due to  the custodial $\SO(3)$ that plays a crucial role. To analyze this,  
we will use  the CCWZ construction of  the $\SO(4)/\SO(3)$ $\sigma$-model effective Lagrangian obtained after EWSB, in full analogy with the discussion of $Zb\bar b$ in section~\ref{situation}. 
Let us start with   operators involving only the custodial triplets ${\cal D}_\mu^a$, ${\cal E}_\mu^a$ and   $H^a$. 
%and possibly the external gauge field strengths   $W_{\mu\nu}$ though the latter necessarily leads to effects suppressed by a power of $\partial^2_\mu/m_\rho^2$.
At the two derivative level the only object one can write is
\begin{equation}
(\partial^\mu+{\cal E}^\mu) H^a\,  {\cal D}_\mu^a\, ,
\end{equation}
%where the covariant derivative $D_\mu$ involves 
where ${\cal E}_\mu$  contains both NGB and gauge fields, upon weak gauging of the global symmetry. The above term however induces a kinetic mixing between the electroweak gauge bosons and the heavy Higgs triplet $H^a$. As a consequence this term will be eliminated by the suitable (physical) gauge choice for which no such mixing exists. Another possible two derivative operator ${\cal D}_\mu^a{\cal D}^{b \mu} H^c\epsilon_{abc}$ vanishes by Bose symmetry. It is also similarly easy to deal with operators that involve $H^a$ and two powers of ${\cal D}_\mu^a$, plus a number of covariant derivatives acting on them. These would correspond to effects that are genuinely associated to the strong sector, and thus only involve the longitudinally polarized vectors. Zooming on the trilinear interactions we can replace ${\cal D}_\mu^a=\partial_\mu G^a$, where $G^a$ is the triplet of SM NGB fields. Then, integrating by parts, and using the lowest order equation of motion $\Box G^a=0$ such operators can always be written as
\begin{equation}
	H^a (\partial_\mu\dots\partial_\nu G^b) (\partial^\mu\dots\partial^\nu G^c)\, .
\end{equation}
Again the contraction with $\epsilon_{abc}$, which would lead to a singlet, vanishes. This means that the only contribution to $H^\pm W^\mp Z$  can only  come from the terms we neglected in the above procedure: custodial  breaking terms (from the NGB equations of motion) and terms  explicitly involving  gauge fields. The latter survive only to the extent that the gauge field configuration is not gauge equivalent to a NGB field, so they must necessarily involve the gauge field strength and therefore will lead to effects  suppressed by  powers of $\partial^2_\mu/m^2_\rho$.
We conclude then that the contributions to $H^\pm W^\mp Z$ should involve either custodial breaking spurions or gauge field strengths or both. Before considering these other effects, we must point out that for symmetric cosets, like $\SO(6) / \SO(4)\times \SO(2)$, there are no interaction terms involving an odd number of NGB.
In those cases it is  trivial to realize that $H^\pm W^\mp Z$ is not enhanced by  pure $\sigma$-model terms involving  the eaten NGBs. The proof we gave in this paragraph is however more general, as it solely relys on $\SO(3)$ invariance.

Let us now consider terms with  field strengths.  The lowest-order terms involve just one field strength, and  we find two such terms 
\begin{equation}
{\cal O}_B=\frac{1}{m_\rho^2}D_\mu H^a {\cal D}_\nu^a B^{\mu\nu}\qquad ,\qquad {\cal O}_W=\frac{1}{m_\rho^2}D_\mu H^a {\cal D}_\nu^b W^{c\mu\nu}\epsilon_{abc}\, .
\label{zwhextra}
\end{equation}
One further   question concerns the order at which these terms arise  in our $\SO(6)/\SO(4)\times \SO(2)$ cosets. To investigate that, we must work with the $\SO(6)/\SO(4)\times \SO(2)$ NGB fields ${\cal D}^{\alpha i}_\mu$. In  doing so one is easily convinced that, because of Bose symmetry, only the first operator survives
\begin{equation}\label{eqHWZBmunu}
\frac{1}{m_\rho^2} \epsilon^{\alpha\beta}{\cal D}^{\alpha i}_\mu{\cal D}^{\beta i}_\nu B^{\mu\nu}\, .
\end{equation}
Of course this operator is only generated  if the strong sector breaks $C_2$. Notice also that $B_{\mu\nu}$, from the point of view of the strong sector, plays the role of the field strength of the weakly gauged $\U(1)_X$. In that sense the presence of $C_X$ charge conjugation within the strong sector would forbid that term. The above operator \eq{eqHWZBmunu} 
contributes to both $H^+\to W^+ Z$ and to the more interesting $H^+\to W^+ \gamma$. 
Unfortunately,  assuming minimal coupling \cite{Giudice:2007fh} 
(as it is the case in five-dimensional  realization of these models), the coefficient of the operators (\ref{zwhextra}) is further suppressed by $g_\rho^2/16\pi^2$, and the phenomenological relevance
of these decay modes is very limited.

%In   section \ref{almost}  we shall comment on the potential phenomenological relevance of these channels.

Let us now consider effects induced by the custodial-breaking spurions. We have two of them, the top proto-Yukawa $y_L$ and the gauge coupling $g'$. As already discussed in section~\ref{secExplicitModels}, only combinations that are invariant under the additional rephasing of the external fields can enter in the strong-sector Lagrangian. These are respectively $\Upsilon_L$ of  eq.~(\ref{iu}) that contains a     $\SO(3)$ triplet, $w^a$, with VEV along $a=3$, and $\Gamma^\pm_{g'}$ of eq.~(\ref{gp})   that contains  a  $\SO(3)$ triplet and a quintuplet, $w^{ab}$ (a traceless symmetric tensor), with VEV along the 33 component. In classifying the operators $CP$ plays a crucial role.
For this purpose it is useful to recall, as discussed in section~\ref{secExplicitModels}, that on the  bosonic fields $CP$ reduces to just parity times $C_1$,  a 180 degree rotation in the 1-3 plane of O(4) defined in eq. \eqref{eqC1definition}. In view of that,  $w^3$ and $w^{33}$ are respectively $C_1$ odd and $C_1$ even. Operators with odd powers of $w^a$ will break $CP$ \footnote{We stress that this breaking would be due to the strong sector. This is because $w^a={\rm Im}(\Upsilon _L)$ is odd under complex conjugation, so that $w_3$ is even under the combined action of complex conjugation and $C_1$, which is precisely $CP$ on the spurion $-$see eqs. \eqref{c1tr} and \eqref{eqC1Pgauge}.}
. Now at the two derivative level we find the following terms
\begin{eqnarray}
{\cal O}_3&=&\frac{v}{f^2}\frac{w^a}{16\pi^2} {\cal D}_\mu^a H^b{\cal D}^{b\mu}\ ,\qquad
{\cal O}_4=\frac{v}{f^2}\frac{w^a}{16\pi^2} H^a{\cal D}_\mu^b {\cal D}^{b\mu}\, ,\\
{\cal O}_5&=&\frac{v}{f^2}\frac{w^{ab}}{16\pi^2} H^c{\cal D}_\mu^a {\cal D}^{d\mu}\epsilon^{bcd}\, ,
\end{eqnarray}
as well as a  term of the same form as ${\cal O}_5$ but with $w^{ab}$ replaced by $w^a w^b/g_\rho^2$. For $y_L\sim Y_t$ that term is however subleading. 
${\cal O}_{3,4}$ can only be generated if $CP$ is broken by the strong sector. They lead to the $CP$-odd coupling
\begin{equation}\label{eqHWZCPodd}
	i Z_\mu \left( H^+W^{-\, \mu} - H^-W^{+\, \mu} \right)\, , 
\end{equation}
together with different combinations of $A^0W_\mu^+W^{-\mu}$ and  $A^0Z_\mu Z^\mu$. We recall that our conventions on $CP$ are $H^\pm,W_\mu^\pm\to H^\mp,W_\mu^\mp$ and $Z_\mu\to-Z_\mu$.
On the other hand ${\cal O}_{5}$ respects $CP$ and will be generically present leading just to the trilinear 
\begin{equation}\label{eqHWZCPeven}
	Z_\mu  \left( H^+W^{-\, \mu} - H^-W^{+\, \mu} \right)\, .
	\end{equation}
Of course all these terms require explicit breakdown of $C_2$ and a relevant question concerns the possibility to generate them in the $\SO(6)/\SO(4)\times \SO(2)$ models. We have explicitly checked that only  ${\cal O}_5$ is generated in these models, its avatar being, using the notation of section~\ref{secExplicitModels},
\begin{equation}\label{eqHWZgploop}
	\frac{g'^2}{16\pi^2}\frac{m_\rho^2}{g_\rho^2}(\Gamma_{g'}^+)^{\alpha i\beta j} \epsilon^{\alpha\gamma}{(\cal D_\mu)}^{\gamma i}{(\cal D^\mu)}^{\beta j}\, .
\end{equation}
Notice that only the $T_3^R$ part of $g'$ contributes.

\subsection{Phenomenology of $\SO(6)/\SO(4)\times \SO(2)$  Models}
We here focus on the explicit examples described in section~\ref{secExplicitModels} corresponding to the $\SO(6)/\SO(4)\times \SO(2)$ coset.
In that case we have
\begin{equation}
	\ca	=	-\dfrac{1}{2}
	\ ,\ \
	\cb	=	0
	\ ,\ \
	\cc	=	-\dfrac{1}{2}
	\ ,\ \
	\cd	=	 0
	\ ,\ \
	\ce	=	0\, ,
\label{coe64}
\end{equation}
where $\cb$
and $\ce$ are zero at leading order  due to the accidental $C_2$ symmetry of the coset, and 
$\cd$ vanishes because of the absence, at leading order, of an operator involving the $\SO(4)$ Levi-Civita tensor.
From eq.~(\ref{coe64}) we find that in these models only the  first two sum rules of eq.~(\ref{sumrules}) are modified,
with  the ratios of eq.~(\ref{ratios}) not being altered. Therefore a precise  determination of the coupling of the 
$CP$-even  Higgses, $h^0_1$ and $h^0_2$, to    $V^aV^a$ or $V^aH^a$ will be needed to study 
the two first sum rules of  eq.~(\ref{sumrules}), and possibly unravel the composite nature of the Higgs bosons.
Alternatively,  one could measure the growth with energy of the $V^aV^a$ 
scattering amplitude, that proceeds as in eqs.~(4.14)-(4.16) of 
ref.~\cite{Giudice:2007fh}  with $c_H=-2c_1=1$.
The Higgs couplings  to  fermions are  modified by corrections of order $v^2/f^2$ 
arising from  eq.~(\ref{fermionrules}). The coefficients $c_t^{(i)}$ depend on the specific representations
of the operator to which the SM fermions couple and are model dependent.  
For the models of  section~\ref{secC2model} and \ref{secC1Pmodel} we find respectively
\begin{equation}
	c_{t}^{(1)}	= - \frac{1}{2}
	\ ,\ \
	c_{t}^{(2)}	= c_{t}^{(3)} = \frac{i}{4} \tan \theta
	\ ,\ \
%	c_{t}^{(3)}	= \frac{i}{3} \tan \theta
%	\ ,\ \
	c_{t}^{(4)}	= 0
	\, ,
\end{equation}
and
\begin{equation}
	c_{t}^{(1)}	= - \frac{1}{2}
	\ ,\ \
	c_{t}^{(2)}	= c_{t}^{(3)}= c_{t}^{(4)} = 0
%	\ ,\ \
%	c_{t}^{(3)}	=	.....
%	\ ,\ \
%	c_{t}^{(4)}	=	.....
	\, ,
\end{equation}
where $c_t^{(2)} = c_t^{(3)}$ due to the $\SO(4)$ symmetry of the coset.
Again, notice that we need to measure the coupling of both $CP$-even Higgs
to fermions in order to establish deviations from  the sum rules of  eq.~(\ref{fermionrules}).

Besides the above features, the phenomenology of the explicit models discussed in section~\ref{secExplicitModels} 
will be very similar to that of  an elementary 2HDM.
The main characteristic  will be the approximate custodial symmetry of the Higgs potential
and the smallness of $\theta_h$, as we explore in the following examples.

\subsubsection{Composite Inert Higgs}

The $C_2$-invariant model of section~\ref{secC2model} corresponds to an inert Higgs scenario,
whose  phenomenology has been extensively studied in the literature starting with \cite{Barbieri:2006dq}. We can identify  $H_1$ and $H_2$, defined in the basis \eq{smartbasis2},
respectively with  $H_\uno$ and $H_\due$ defined in  section~\ref{secC2model}.
As described there, the  lightest Higgs, now called $h^0$, is $C_2$-even and  behaves as  the SM Higgs with a mass ranging from 150  to  250 GeV for $N=10$ to $N=3$.
The other  Higgs bosons, $H^0$, $A^0$ and $H^\pm$, are $C_2$-odd and 
their masses fulfill the approximate relations
\beq
m_{H^+}^2\simeq  m_{A^0}^2\ ,\ \ \ \   m_{H^0}^2\simeq \left(1-\frac{\xi}{3}\right) m_{A^0}^2\, .
\eeq
The lightest $C_2$-odd  Higgs is the neutral $H^0$ that  is stable with  a mass that can be as low as $m_{H^0}\sim 500 \GeV$  by taking  $N\simeq 10$.
The $C_2$-odd Higgses can  only be   pair produced at the LHC 
%The inert Higgs has to be produced in pairs at colliders. At the LHC, the leading production processes would be 
through the processes $q q' \rightarrow \gamma, Z^*, W^{\pm *} \rightarrow H^aH^b$, 
with cross-sections   below $\textrm{fb}$ for masses $\sim 500 \GeV$.
% Subleading production processes, arising from the composite nature of the inert Higgs and its large coupling to top quarks, are gluon fusion $gg \rightarrow H_2^a H_2^a$ (through a top quark loop) or W fusion $qq' \rightarrow H_2^a H_2^b$. Both of these processes receive $1/f^2$ suppressed contributions from the strong sector, which are important in the high energy collision regime. 
 The strong degeneracy of $H^+$ and $A^0$ implies  that  the decay channel  $H^+\rightarrow W^+A^0$, or vice versa, $A^0\rightarrow W^+H^-$,  cannot proceed, and  the main decay channels are $H^+\rightarrow W^+H^0$ and $A^0\rightarrow ZH^0$ where $H^0$
 being stable escapes from the detector giving missing energy.

The $C_2$ symmetry could explicitly be broken,  as discussed in the last model of section~\ref{secC2model}.
In this case, however,  we saw
that approximate accidental $ C_1P$ and $C_2$ (and custodial) symmetry of the scalar potential
force   the   VEV  of $H_\due$ to be small (see \eq{v2tadpole1}  and (\ref{v2tadpole2})).
We can then still  approximately identify
$H_\uno$ and $H_\due$ respectively with $H_1$ and $H_2$. We also found that the breaking of $ C_1P\cdot C_2$
is further suppressed with respect to that of $C_2$ in the scalar potential.
Notice that
$ C_1P\cdot C_2$
acts as a  $CP$ symmetry under which $A^0$ is $CP$-even and $H^0$ is $CP$-odd.
The main consequence  of the explicit $C_2$ breaking is that now
$H_2$ couples to fermions according to eq.~(\ref{yukawasc2}),
implying  that 
$H^0,A^0$ and $H^+$
can be single produced  by gluon fusion,
decaying  mainly  into tops and bottoms. 
Their decay into  gauge bosons  is suppressed by the small VEV of  $H_2$, and 
we estimate that the corresponding branching ratios
are always smaller than 1$\%$.

\subsubsection{Almost Inert Higgs}
\label{almost}

In the $C_1P$-invariant model of section~\ref{secC1Pmodel},
the  $C_2$ symmetry is preserved by the top Yukawa coupling,
but violated by the coupling to the bottom and the tau, since $\tilde \theta$ defined in 
\eq{tildetheta} is a free parameter.
Furthermore, $C_2$ is accidentally preserved in the Higgs potential at order  $y_L^2$.
To go from the Higgs doublet basis $(H_\uno, H_\due)$  of \eq{2HDMV}, 
to the basis $(H_1,H_2)$ of eq.~(\ref{smartbasis2}), where only one Higgs doublet gets a VEV,
we must perform a rotation  of order $m^2_{12}/m^2_{22}\sim y_L^2/g_\rho^2$ that generates a 
contribution to $a_t$ of this order. Therefore  we find   a 2HDM with the following properties:\\

\noindent $\displaystyle 1)\ \ a_t\sim \frac{y^2_L }{g_\rho^2}\ , \ \  \ a_{b,\tau}=\tan\tilde\theta_{d,l}\, ,  \text{where $a_{b,\tau}$ is the equivalent of $a_t$ for the bottom quark and the tau lepton.}$
\noindent $\displaystyle 2)\ \    \theta_h\sim \frac{y^2_L }{g_\rho^2}\xi \sim  a_t\xi\ll 1\, ,\ 
\text{implying  that  $h^0,H^0$ are approximately mass-eigenstates.}$\\
\noindent $\displaystyle 3)\ \    m^2_{h^0}\sim  \xi m^2_{H^0}\sim \frac{N_c y^2_Lg_\rho^2v^2}{16\pi^2}\, .$\\
\noindent $\displaystyle 4)\ \    m^2_{H^+}\simeq  m_{A^0}^2  \ ,\     m_{A^0}^2\simeq m_{H^0}^2-\frac{1}{3} ( m_{h^0}^2  + \xi m_{H^0}^2)\, .$\\
\noindent $\displaystyle 5)\ \ \lambda_{H^0h^0h^0}\sim \frac{y_L^2}{g_\rho^2}\frac{m_{H^0}^2v}{f^2} \sim a_t\frac{ m_{H^0}^2v}{f^2}\, ,\  \text{where  $\lambda_{H^0h^0h^0}$  is the Higgs trilinear coupling in the potential.}$\\
%\begin{align*}
%&1)\ \ a_t\sim \frac{y^2_L }{g_\rho^2}\ , \ \  \ a_{b,\tau}=\tan\tilde\theta_{d,l}\, ,  \text{where $a_{b,\tau}$ is the equivalent of $a_t$ for the bottom quark and the tau lepton.}\\
%&2)\ \    \theta_h\sim \frac{y^2_L }{g_\rho^2}\xi \sim  a_t\xi\ll 1\, ,\ \text{implying  that  $h^0,H^0$ are approximately mass-eigenstates.}\\
%&3)\ \    m^2_{h^0}\sim  \xi m^2_{H^0}\sim \frac{N_c y^2_Lg_\rho^2v^2}{16\pi^2}\, .\\
%&4)\ \    m^2_{H^+}\simeq  m_{A^0}^2  \ ,\     m_{A^0}^2\simeq m_{H^0}^2-\frac{1}{3} ( m_{h^0}^2  + \xi m_{H^0}^2)\, . \\
%&5)\ \ \lambda_{H^0h^0h^0}\sim \frac{y_L^2}{g_\rho^2}\frac{m_{H^0}^2v}{f^2} \sim a_t\frac{ m_{H^0}^2v}{f^2}\, ,\  \text{where  $\lambda_{H^0h^0h^0}$  is the Higgs trilinear coupling in the potential.}
%\lambda_{H^0h^0h^0}\sim \frac{a_t m_{H^0}^2v}{f^2}\, ,\  \text{where  $\lambda_{H^0h^0h^0}$  is the Higgs trilinear coupling in the potential.}
%\end{align*}

\begin{figure}
\centering
\epsfig{file=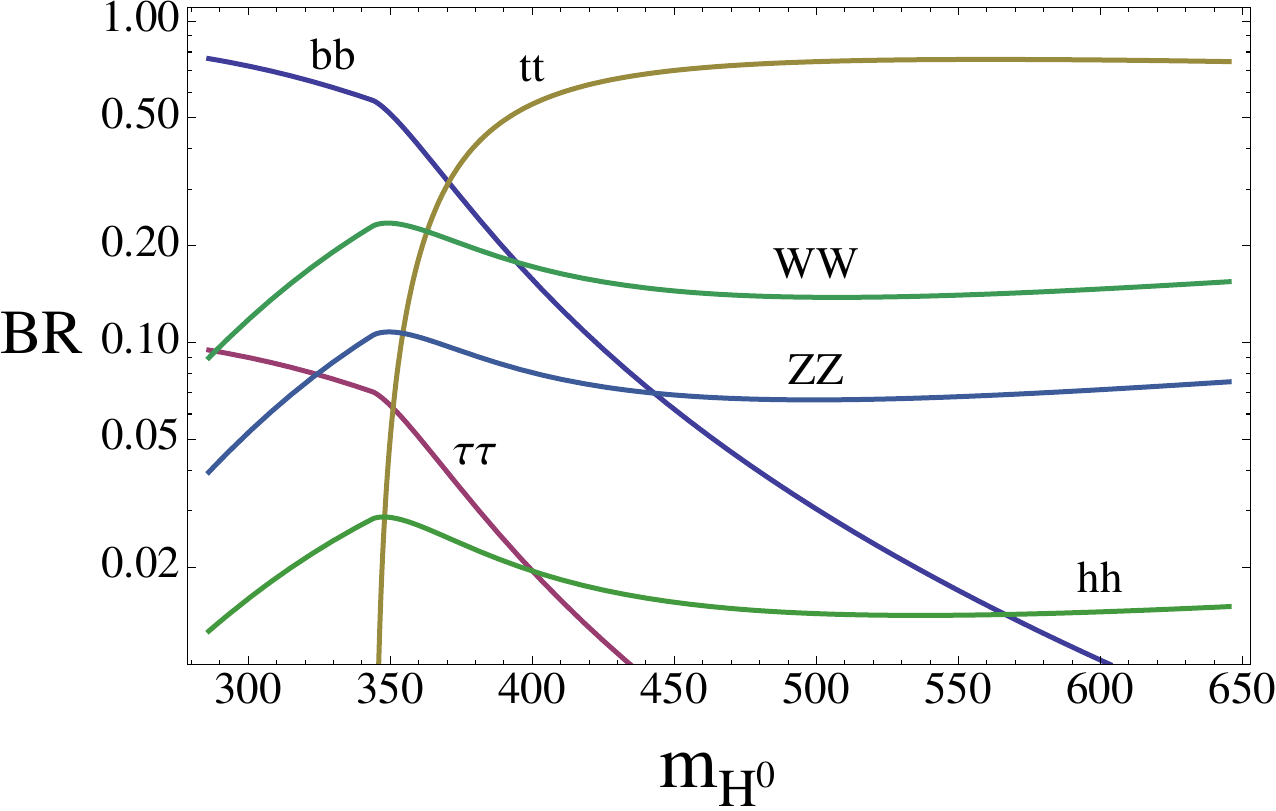,height=200pt}
\caption{Branching ratios for $H^0$ in the almost inert Higgs model.}\label{figH1}
\end{figure}

\begin{figure}
\centering
%\hspace{-300pt}
\epsfig{file=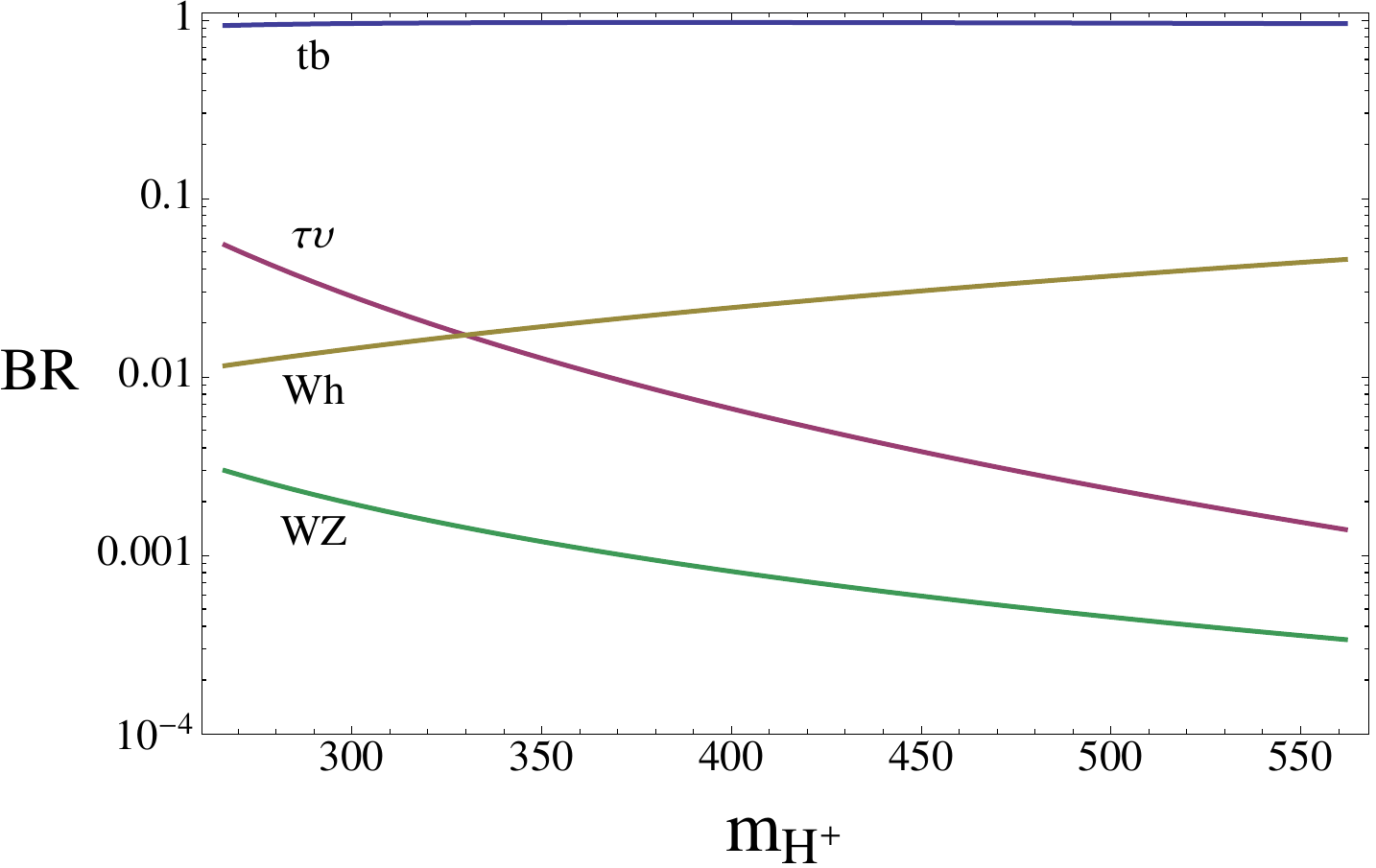,height=150pt}\ 
%\hspace{30pt}
\ \epsfig{file=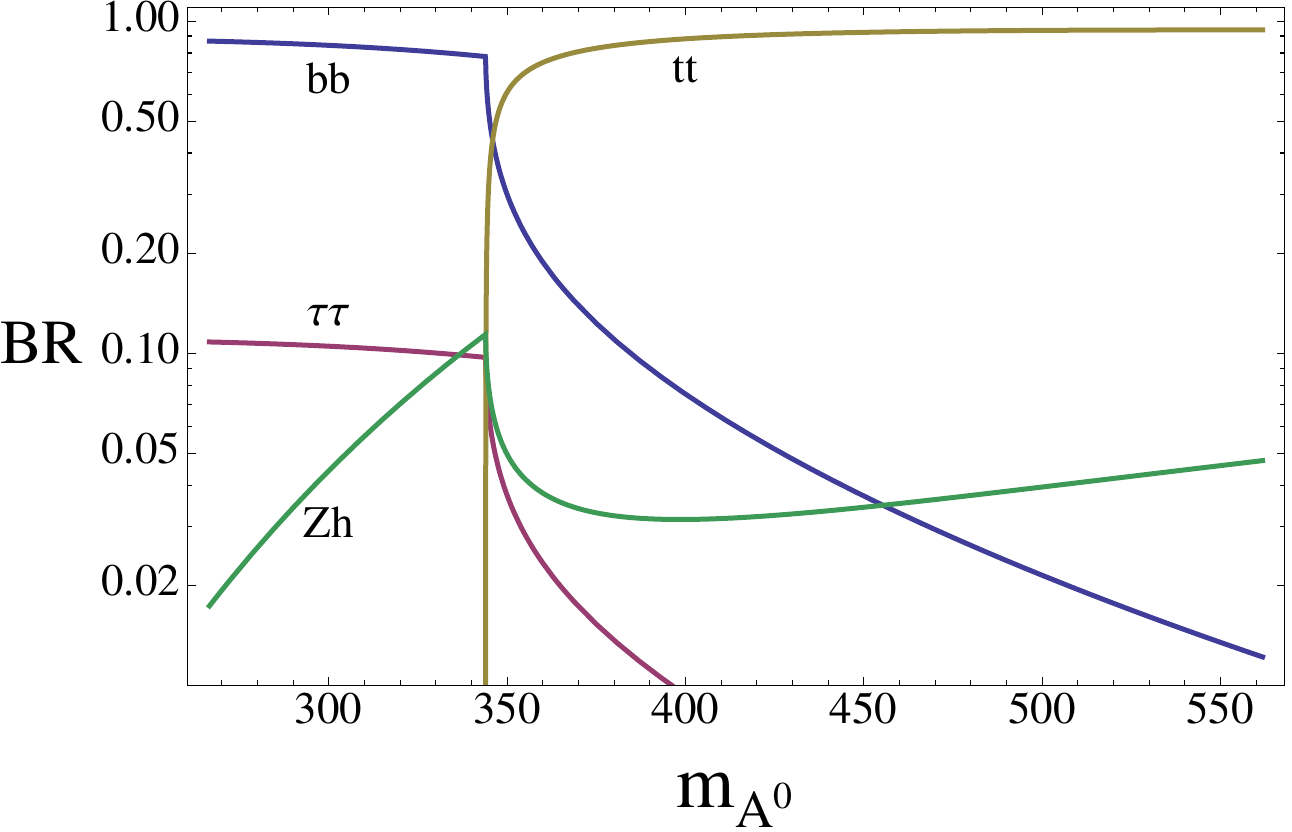,height=150pt}
%\vspace{1cm}
\caption{Branching ratios for $H^+$ and $A^0$   in the almost inert Higgs model.}\label{figH2}
\end{figure}

From the above we can calculate the production cross-sections and branching ratios for the Higgs bosons.
We focus on the values  $N\sim 8$, $\xi\sim 0.25$ and $\tan\tilde\theta_{d,l}\sim 1$, taking  
$y_L$ ranging from $Y_t\sim 1$ to $\sqrt{Y_tg_\rho}\sim 2.1$.
For these values of the parameters, the lightest Higgs $h^0$, that  behaves  as a SM Higgs, has  a mass ranging from  150 to 300 GeV, while
$m_{H^0}$ ranges from 300 to 650 GeV. The Higgs doublet $H_2$  couples to the top with a coupling proportional to $Y_t a_t$ that 
takes a value between  $0.05$ and $0.2$. This implies that  single Higgs production for $H^0$, $A^0$ and $H^+$
via gluon fusion is suppressed by $Y_t^2 a_t^2$ with respect to that of the SM Higgs. 
Vector boson fusion for $H^0$ is also very  small since it is suppressed by $\theta_h^2$.
The cross-section for the double-production $H^+H^0$ is  of few fb
for Higgs masses around $300$ GeV.
The decay channels  of the Higgs bosons depend strongly on their masses.
We show in Fig.~\ref{figH1} the branching ratio for  $H^0$ as a function of its mass.
Notice that for low mass values $H^0$ decays mainly into bottoms,
but as its mass increases the channel into gauge bosons becomes sizable. 
The decay to tops dominates whenever it is kinematically allowed.
It is important to remark that the fact that the mass splitting between $H^0$ and
$A^0,H^+$ goes as $\xi m_{H^0}$ implies that
the  decay channels  $H^0\rightarrow WH^+,ZA^0$ are only open for large values 
of the $H^0$ mass. For the mass values given in Fig.~\ref{figH1} these decay channels are always close, but we must emphasize 
that this is very sensitive to the value of $\xi$ and $m_h^2$.
In Fig.~\ref{figH2} we also show the branching ratios of $H^+$ and   $A^0$.
It is worth mentioning the   branching ratio for  $H^+\rightarrow WZ/\gamma$ that 
in this model  can reach values    $ \sim 0.01$, much larger than in a renormalizable 2HDM where it is induced by top 
loops and takes a value $\sim 10^{-4}$ for $m_{H^+}\sim 300$ GeV \cite{Mendez:1990epa}.
In our model this decay width arises  mainly from the operator
\eqref{eqHWZBmunu} that gives
\begin{equation}
	\Gamma(H^\pm\to W^\pm Z/\gamma) \sim \frac {g'^4}{8\pi} \frac {v^2m_{H^+}^3}{m_\rho^4}\, .
\end{equation}
As we increase $N$ and $g_\rho$ becomes smaller,
we have that $a_t$ and $\theta_h$ increase.  
This  makes the Higgs easier to be detected. 
For example, for $N\sim 10$ we  can have $h^0$ and $H^0$  with  masses around 170  and 350 GeV respectively, decaying 
both mainly into  gauge bosons. 

\section{Summary}

In this paper we have considered the construction and the broad phenomenology of models with a composite Higgs sector featuring two light scalar doublets. 
The possibility that the Higgs dynamics is determined by light PNGB's 
from some strong sector is a plausible one. The minimal model, based 
on ${\textrm{SO}}(5)/{\textrm{SO}}(4)$, has indeed been extensively studied in recent years. However, there seems to be no a priori, theoretical or experimental, 
reason to avoid  considering less minimal options. Mapping out the structure and phenomenology of non-minimal options is thus a potentially useful thing to do. 
Our study represents one step in that direction.

Our construction is largely based on the use of symmetries. It is  a  fact that additional symmetries are typically needed to meet experimental constraints in extensions of the SM. In particular, already in the simple renormalizable 2HDM,  in order to control FCNC, either an additional discrete symmetry or minimal flavor violation (that is flavor $\SU(3)^5$ selection rules under specific assumptions on the sources of breaking) are invoked. That gives rise  to respectively the type I-II and type III models. On the other hand, in composite Higgs models and in technicolor, an approximate ${\textrm{SO}}(4)$ symmetry of the strong sector must be assumed in order to control the corrections to $\widehat T$.
Our study shows that in order to have a phenomenologically acceptable composite 2HDM it is enough to postulate the strong sector
is invariant under $\SO(4)$ times a discrete symmetry. Two possibilities are given for the latter, either $C_2$, that is a $Z_2$ symmetry distinguishing the two doublets, or  $CP$, that we call $C_1P$ given its specific action on the strong sector. These discrete symmetries are, maybe surprisingly, essential to control $\widehat T$, while Higgs-mediated FCNC can be tamed
by the non-linear symmetry $G$ of the strong sector combined with either $C_2$ or flavor minimality assumptions similar
but different from MFV. Models based on ${\textrm{SO}}(6)/{\textrm{SO}}(4)\times {\textrm{SO}}(2)$ and ${\textrm{SU}}(5)/{\textrm{SU}}(4)\times \U(1)$ can realize our
scenario, but we have studied in detail only explicit realizations of the first possibility. 

In the spirit of ref.~\cite{Giudice:2007fh} we have only constructed a low-energy effective description of our models. We see no obstacle to ``UV completing'' our models into a warped compactification, but we do not want to be tied to that perspective. The underlying hypothesis of our scenario is that it  UV completes into a 4D CFT possessing the right set of operators, most notably  fermions with scaling dimension $d\sim 5/2$, in order to implement the 
partial-compositeness paradigm. Our  construction is based on two  assumptions. The first one is that  the strong sector is broadly characterized by a mass scale $m_\rho$ and a coupling strength $g_\rho$. The second assumption is that the global symmetry $G$ is only broken by the (linear) coupling of SM fields to strong sector operators. While the coupling to vector bosons is fully fixed by gauge invariance, more freedom exists in the fermionic sector. To fully specify the model we must choose the quantum numbers of the operators that mix with the SM fermions, in particular to the top quark. In this paper we have focussed on two possibilities. In the first class of models, that we indicate by $\{{\bf 6},{\bf 6}\}$, both left-  and right-handed fermions couple to operators in the ${\bf 6}$ of ${\textrm{SO}}(6)$. In the second class, indicated by $\{{\bf 20'},{\bf 1}\}$, the left and right-handed fermions couple to respectively a ${\bf 20'}$ and to a total singlet.

Under the above assumptions, our methodology to derive the low energy effective Lagrangian relies on the CCWZ formalism and makes broad use of all spurionic symmetries.  We find it easier in the CCWZ language, compared to the 
approach based on the linear Higgs field $\Sigma$ that breaks $G\to H$,  to count and classify the independent invariants at any given order.  Our effective Lagrangian is organized as an expansion in derivatives and in powers
of the $G$ breaking spurions, the most relevant ones being the top quark proto-Yukwas $y_L$ and $y_R$, and the gauge couplings $g$ and $g'$. One crucial result of our analysis is the emergence of accidental symmetries at the lowest orders in the expansion. On one hand, these accidental symmetries can  help the models to meet the phenomenological constraints. On the other hand they provide smoking guns for the whole scenario.

%One first accidental symmetry that emerges in our analysis up to the order relevant for all experimental tests is the $Z_2$ parity in ${\textrm{O}}(4)$, 
%defined by ${\textrm{O}}(4)={\textrm{SO}}(4)\times P_{LR}$. 
%This symmetry crucially protects the $Z\bar b b$ vertex form receiving large corrections. The analysis in sect. \ref{situation}
%nicely shows that this happy accident happens  in all cosets  ${\textrm{SO}}(N)/ {\textrm{SO}}(4)\times {\textrm{SO}}(N-4)$ and left handed fermions mixing to 
%a $(2,2)$, regardless of $P_{LR}$ being 
%exact. In particular it happens in the minimal composite Higgs and in the composite 2HDM we consider in this paper.
%The fact that this symmetry is simply accidental, was not appreciated before. Thanks to $P_{LR}$ a larger value of $y_L$
%and compatibly a larger value of $m_h$ can be achieved, without a stark contradiction of the EWPT.

One accidental symmetry that emerges in our analysis up to the order relevant for all experimental tests is the $Z_2$ parity in ${\textrm{O}}(4)$, 
defined by ${\textrm{O}}(4)={\textrm{SO}}(4)\times P_{LR}$. 
This symmetry crucially protects the $Z\bar b b$ vertex form receiving large corrections. 
This happy accident, as the analysis of section~\ref{situation} shows, depends on the choice of the strong sector's global group 
and on the representation to which the left-handed fermions mix. In particular this happens in the MCHM with 
fermions in the ${\mathbf5}$ and in the composite 2HDM we consider in this paper.
The fact that this symmetry is simply accidental, and needs not to be respected by the fundamental strong sector's dynamics, 
was not appreciated before. Thanks to $P_{LR}$ a larger value of $y_L$
and compatibly a larger value of $m_h$ can be achieved, without a stark contradiction with EWPT.

More specifically to our ${\textrm{SO}}(6)/{\textrm{SO}}(4)\times {\textrm{SO}}(2)$ models, we also find that $C_2$, $C_1P$ and 
${\textrm{SO}}(3)$ can arise as accidental symmetries of subsectors in the low-energy effective action or simply at lowest order. 
In particular the leading $O(y^2)$ contribution to the scalar potential is invariant under $C_2\times C_1P \times {\textrm{SO}}(3)$ under the 
weak assumptions that the strong sector and the top proto-Yukawa respects either $C_2$ or $C_1P$. Indeed in the $\{{\bf 6},{\bf 6}\}$ model,
this result holds even when $C_2$ is maximally broken by the top proto-Yukawa.  Then around the generic vacuum of this model, the Higgs bosons self-interactions and their 
coupling to vector bosons, respects $C_2$ to a good approximation, implying that one of the Higgs doublets, say $H_2$, is quasi-inert. However the interactions of both $H_1$ 
and $H_2$ to the top maximally break $C_2$. The resulting signal is that $A,H^\pm, H$ can be singly produced in gluon fusion
and decay dominanty to $t,b$ quarks, with only a tiny branching ratio to vector bosons and to $h$.  
In the case that $C_2$ is completely preserved, we obtain a composite inert Higgs model.
In the  $\{{\bf 20'},{\bf 1}\}$ model, when $C_1P$ is a symmetry of the strong sector and of the proto-Yukawas, one has that $C_2$ arises as an accidental symmetry of the 
lowest-order effective action. In this case the second Higgs genuinely behaves like a quasi-inert doublet, including   its couplings to fermions. It is therefore mostly doubly produced, 
even though the underlying strong dynamics maximally breaks $C_2$. In that case to reveal the accident and learn about the structure of the theory one would need to observe 
$C_2$ violation in the production of heavy resonances, and not just the PNGBs.

The presence of ${\textrm{SO}}(3)$ symmetry at leading order in the Higgs potential can be easily understood by noticing that $y_L^2$ can be decomposed into a  triplet plus a singlet under 
${\textrm{SO}}(3)$. Given that $h$, $H$ are singlets and $A$, $H^\pm$ form a triplet, $H^a$, and given that the neutrals $h$, $H$, $A$ cannot mix by $C_2$ and $C_1P$, one readily realizes 
no mass term involving the triplet in $y_L^2$ can be written. The degeneracy $m_A\simeq m_{H^\pm}$ is thus one indirect but robust prediction of the composite 2HDM. This  is only broken by small effects of order $g'^2$ and $y_L^4$.
On top of ${\textrm{SO}}(3)$ invariance, in the specific models we considered there are additional predictions for the Higgs mass spectrum,  arising from the limited number of independent structures at leading order. For instance, in the $\{{\bf 20'},{\bf 1}\}$ model the leading  $O(y^2)$ potential is determined by three unknown coefficients associated to the three invariants one can write. There is thus one relation among the four parameters $\xi,\, m_h,\, m_H,\, m_A$, given by eq.~(\ref{prediction201}).  Notice that one always has $m_H>m_A$.
In the $\{{\bf6},{\bf6}\}$ model there are only two parameters at leading order, so that one has in principle
an additional prediction at leading $O(y^2)$. However, that just amounts to $m_h^2\ll m_H^2/\xi$ corresponding to the fact that in order to have electroweak symmetry breaking it is necessary to tune $m_h^2$ to be $O(y^4)$ rather than $O(y^2)$. The other prediction is given by eq.~(\ref{splitt}). Notice that in contrast to the $\{{\bf 20'},{\bf1}\}$ model,  here we always have $m_H<m_A$. 

Aside of the above restricted structure which is  mostly a  consequence of model building constraints, the genuine predictions
of a composite 2HDM model reside in $O(\xi)$ deviation in the Higgs couplings with respect to the elementary case, in the 
growth with energy of scattering amplitudes  involving scalars and longitudinally polarized vectors, and eventually in the production of strongly coupled resonances. Using our effective Lagrangian we have given a general parametrization of the first two classes of effects. Our results generalize the composite Lagrangian of ref.~\cite{Giudice:2007fh}.
Putting those effects in evidence at the LHC will be difficult unless $\xi$ is somewhat large, likely  above $\sim 0.2$.
In the composite 2HDM the situation does not seem easier. Indeed given the rich set of relations and sum rules 
implied in the renormalizable 2HDM (see section~\ref{elementary}), one may have naively expected more dramatic effects when turning on non-renormalizable couplings. In particular $g_{H^+W^-Z}$, which vanishes at tree level in the renormalizable case, was a candidate to potentially large effects. However it turns out that in any realistic composite 2HDM this coupling is also suppressed because of symmetry reasons, as we have discussed in detail. Therefore one does not get dramatic enhancements of the branching ratio for $H^\pm \to WZ$. Nevertheless it is worth noticing that in the $\bf \{20',1\}$ model the branching ratios for $H^\pm \to WZ/\gamma$ are significantly enhanced over the renormalizable case, basically due to  the accidental $C_2$ symmetry in the coupling with fermions. That symmetry  is instead broken at $O(g'^2)$  in the coupling to vectors.

The conclusion of our study is that realistic composite 2HDM  can  reasonably  be constructed, showing a procedure that can be used 
for other composite models with richer  Higgs structure.
The main signatures that can 
be extracted at the LHC concern the structure of the spectrum, mainly its peculiar $\SO(3)$ invariance, that can be used to distinguish them from supersymmetric models.
Beyond that, we have outlined  a rich pattern of deviations from  the renormalizable 2HDM. The study of those effects
belongs to the worthy motivations of a Linear Collider.

\vskip3cm
%%%%%%%%%%%%%%%%%%%%%%%%%%%%%%%%%%%%%%%%%%%%%%%%
\noindent{\bf\large Acknowledgments} 

We would like to thank R.Contino and G.Panico for discussions.
The work of JM, RR and AW has been partially supported by the 
Swiss National Science Foundation under contracts 200021-125237 and  200021-116372.
The work of AP and JS has been partly supported by CICYT-FEDER-FPA2008-01430,
2009SGR894, AP2006-03102, and ICREA Academia program.
%%%%%%%%%%%%%%%%%%%%%%%%%%%%%%%%%%%%%%%%%%%%%%%%%%%%%%%%%%%%%%%%%%%%%%%%%%%%%%%%%%%
\vfill
\pagebreak
\appendix

\section{Two Higgs $\SO(4)$-invariant derivative interactions}\label{appCustSigmaModel}

We want to classify all the dimension-six $\SO(4)$-invariant operators with two derivatives, which are
of three different kinds: 
\begin{enumerate}
\item 
Operators with  $\Box$ of the form
\beq
\left(\Phi^\alpha_i \; \Box\Phi^\beta_i \right)\,\left(\Phi^\gamma_i \; \Phi^\delta_i \right)\,,
\eeq
with $\alpha = \widehat{1}, \widehat{2}$ and analogously for $\beta$, $\gamma$ and $\delta$. There are $12$ of them.

\item 
Operators in which an $\SO(4)$ singlet is formed by contracting one $\Phi^\alpha$ field with one derivative $\partial_\mu\Phi^\alpha$. Those 
can be written in terms of four objects
\bea
d_\mu^\uno&\equiv& \frac{1}{2}\partial_\mu\left(\Phi^\uno\right)^2
\,,\nonumber\\
d_\mu^\due&\equiv& \frac{1}{2}\partial_\mu\left(\Phi^\due\right)^2
\,,\nonumber\\
d_\mu^{\uno \due}&\equiv& \frac{1}{2}\partial_\mu\left(\Phi^\uno\cdot\Phi^\due\right)
\,,\nonumber\\
c_\mu&\equiv& \Phi^\uno\partial_\mu\Phi^\due\,-\,\Phi^\due\partial_\mu\Phi^\uno
\,,
\eea
by forming all possible Lorentz vector products. However, since $\partial^\mu c_\mu =0$, the only non-vanishing contraction of 
$c_\mu$ is with itself, so that we have $1+6=7$ independent invariants in this class.
Operators in which an $\SO(4)$ singlet  is formed by contracting two derivatives $\partial_\mu\Phi^\alpha$ can obviously be 
rewritten as the previous ones by integration by parts.

\item
The $P_{LR}$-odd operator
\beq
\frac{c_{\epsilon}}{f^2} 
\epsilon^{ijkl} \Phi^\uno_i \Phi^\due_j D_\mu \Phi^\uno_k D_\mu \Phi^\due_l\,.
\eeq

\end{enumerate}
All the $12$ operators of the first class can be eliminated by the $12$ field redefinitions of the form
\beq
\Phi^\alpha_i\rightarrow\Phi^\alpha_i\,+A\,\Phi^\beta_i\left(\Phi^\gamma\cdot\Phi^\delta\right)\,,
\label{redef}
\eeq
so that we are left with the following operators
\begin{equation}
\renewcommand\arraystretch{1.7}\begin{array}{l}
{\cal L}_{6d}=	c_{H_1}\,		(d^\uno)_\mu (d^\uno)^\mu+
c_{H_2}\,		(d^\due)_\mu (d^\due)^\mu		+
	c_{H_{12}}\,	(d^{\uno \due})_\mu (d^{\uno \due})^\mu	+
	c_{H_1H_2}\,	(d^\uno)_\mu (d^\due)^\mu		\\
+	c_{H_1H_{12}}\,	(d^\uno)_\mu (d^{\uno \due})^\mu		+
	c_{H_2H_{12}}\,	(d^\due)_\mu (d^{\uno \due})^\mu		+
	c_T\,		c_\mu c^\mu		+
	c_{\epsilon}\, 
	\epsilon^{ijkl} \Phi^\uno_i \Phi^\due_j D_\mu \Phi^\uno_k D_\mu \Phi^\due_l\,.
\end{array}
\label{indops}
\end{equation}
As also explained in the text, it might seem that one 
of the field redefinition in eq.~(\ref{redef}) (in particular, 
$\Phi^\uno\rightarrow\Phi^\due\,+A\,\Phi^\uno\;\Phi^\uno\cdot\Phi^\uno$) cannot be 
performed if willing to remain in a basis where $\Phi^\due$ does not take VEV.
The VEV of $\Phi^\due$ induced by this field redefinition, however, can 
always be eliminated by performing a further redefinition, which consists 
in an $\SO(2)$ rotation in the $(\Phi^\uno,\Phi^\due)$ plane. By this second rotation the 
operators in eq.~(\ref{indops}) merely rotate into each other. 

Finally for the $\SO(6)/\SO(4) \times \SO(2)$ coset studied above 
the coefficients 
are given by
\begin{equation}\begin{array}{rclcrcl}
	c_{H_1}		= \frac{1}{2}
	\ ,\
	c_{H_2}		= \frac{1}{2}
	\ ,\
	c_{H_{12}} 	= 1
	\ ,\
	c_{H_1 H_2}	= 0
	\ ,\
	c_{H_1 H_{12}}	= 0
	\ ,\
	c_{H_1 H_{12}}	= 0
	\ ,\
	c_T			= -\frac{1}{4}
	\ ,\
	c_{\epsilon}	= 0\, .
\end{array}\end{equation}

\end{document}